\documentclass[10pt,a4paper,fleqn]{scrreprt}
\pdfoutput=1
\usepackage[utf8]{inputenc}
\usepackage[english]{babel}
\usepackage{amsmath}
\usepackage{amsfonts}
\usepackage{amssymb}
\usepackage{braket}
\usepackage[]{graphicx}
\usepackage{bbold}
\usepackage{appendix}
\usepackage{siunitx}
\usepackage{mathtools}
\usepackage{pbox}
\usepackage{hyperref}
\usepackage{authblk}
\usepackage{subcaption}
\usepackage{url}
\usepackage{chngcntr}

\usepackage[
backend=bibtex,
firstinits=true,
maxbibnames=15,
sorting=none,
url=false,
doi=true,
]{biblatex}

\bibliography{References}

\renewbibmacro{in:}{}
\AtEveryBibitem{%
  \ifboolexpr{test {\ifentrytype{article}}}
    {\clearfield{issue}}
    {}%
}
\AtEveryBibitem{%
  \ifboolexpr{test {\ifentrytype{article}} and not test {\iffieldundef{doi}}}
    {\clearfield{number}}
    {}%
}
\AtEveryBibitem{%
  \ifboolexpr{test {\ifentrytype{article}} and not test {\iffieldundef{doi}}}
    {\clearfield{issue}}
    {}%
}
\DeclareFieldFormat[article]{volume}{\mkbibbold{#1}}
\DeclareFieldFormat{postnote}{#1}
\DeclareFieldFormat{multipostnote}{#1}
\DeclareFieldFormat{pages}{#1}

\renewbibmacro*{volume+number+eid}{%
  \setunit*{\addcomma\space}
  \printfield{volume}%
  \setunit*{\addcomma\space}
  \printfield{number}%
  \setunit{\addcomma\space}%
  \printfield{eid}}

\DeclareFieldFormat[article]{volume}{\bibstring{volume}~#1}
\DeclareFieldFormat[article]{number}{\bibstring{number}~#1}

\renewbibmacro*{journal+issuetitle}{%
  \usebibmacro{journal}%
  \setunit*{\addspace}%
  \iffieldundef{series}
    {}
    {\newunit
     \printfield{series}%
     \setunit{\addspace}}%
  \usebibmacro{volume+number+eid}
  \newunit}

\DeclareBibliographyDriver{article}{%
  \usebibmacro{bibindex}%
  \usebibmacro{begentry}%
  \usebibmacro{author/translator+others}%
  \setunit{\labelnamepunct}\newblock
  \usebibmacro{title}%
  \newunit
  \printlist{language}%
  \newunit\newblock
  \usebibmacro{byauthor}%
  \newunit\newblock
  \usebibmacro{bytranslator+others}%
  \newunit\newblock
  \printfield{version}%
  \newunit\newblock
  \usebibmacro{journal+issuetitle}%
  \newunit
  \usebibmacro{byeditor+others}%
  \newunit
  \usebibmacro{note+pages}%
  \setunit{\addspace}
  \usebibmacro{issue+date}
  \setunit{\addcolon\space}
  \usebibmacro{issue}
  \newunit\newblock
  \iftoggle{bbx:isbn}
    {\printfield{issn}}
    {}%
  \newunit\newblock
  \usebibmacro{doi+eprint+url}%
  \newunit\newblock
  \usebibmacro{addendum+pubstate}%
  \setunit{\bibpagerefpunct}\newblock
  \usebibmacro{pageref}%
  \usebibmacro{finentry}}

\DeclareFieldFormat[article]{volume}{\mkbibbold{#1}}

\counterwithout{footnote}{chapter}

\DeclareMathOperator{\Cov}{Cov}
\DeclareMathOperator{\Tr}{Tr}
\DeclareMathOperator{\diag}{diag}

\usepackage[left=2cm,right=2cm,top=2cm,bottom=2cm]{geometry}

\newcommand{\eqcolon}{\mathrel{\resizebox{\widthof{$\mathord{=}$}}{\height}{ $\!\!=\!\!\resizebox{1.2\width}{0.8\height}{\raisebox{0.23ex}{$\mathop{:}$}}\!\!$ }}}
\newcommand{\coloneq}{\mathrel{\resizebox{\widthof{$\mathord{=}$}}{\height}{ $\!\!\resizebox{1.2\width}{0.8\height}{\raisebox{0.23ex}{$\mathop{:}$}}\!\!=\!\!$ }}}

\usepackage{array}

\newcommand{\randp}{\mathcal{P}}
\newcommand{\randq}{\mathcal{Q}}

\title{Continuous-Variable Quantum Key Distribution with Gaussian Modulation -- The Theory of Practical Implementations}
\date{}

\author[*1,3]{\large Fabian Laudenbach}
\author[1]{\large Christoph Pacher}
\author[2]{\large Chi-Hang Fred Fung}
\author[2]{\large Andreas Poppe}
\author[2]{\large Momtchil Peev}
\author[1]{\large Bernhard Schrenk}
\author[1]{\large Michael Hentschel}
\author[3]{\large Philip Walther}
\author[1]{\large Hannes H\"{u}bel}
\affil[1]{\normalsize Security \& Communication Technologies, Center for Digital Safety \& Security,\protect \\AIT Austrian Institute of Technology GmbH, Donau-City-Str. 1, 1220 Vienna, Austria}
\affil[2]{\normalsize Optical and Quantum Laboratory, Munich Research Center,\protect \\Huawei Technologies Düsseldorf GmbH, Riesstr. 25-C3, 80992 Munich, Germany}
\affil[3]{\normalsize Quantum Optics, Quantum Nanophysics \& Quantum Information, Faculty of Physics,\protect \\University of Vienna, Boltzmanngasse 5, 1090 Vienna, Austria \protect \\ \quad}
\affil[*]{\normalsize mail: fabian.laudenbach@ait.ac.at}

\begin{document}

\maketitle

\begin{abstract}
\textbf{Abstract.} Quantum key distribution using weak coherent states and homodyne detection is a promising candidate for practical quantum-cryptographic implementations due to its compatibility with existing telecom equipment and high detection efficiencies. However, despite the actual simplicity of the protocol, the security analysis of this method is rather involved compared to discrete-variable QKD. In this article we review the theoretical foundations of continuous-variable quantum key distribution (CV-QKD) with Gaussian modulation and rederive the essential relations from scratch in a pedagogical way. The aim of this paper is to be as comprehensive and self-contained as possible in order to be well intelligible even for readers with little pre-knowledge on the subject. Although the present article is a theoretical discussion of CV-QKD, its focus lies on practical implementations, taking into account various kinds of hardware imperfections and suggesting practical methods to perform the security analysis subsequent to the key exchange. Apart from a review of well-known results, this manuscript presents a set of new original noise models which are helpful to get an estimate of how well a given set of hardware will perform in practice.
\end{abstract}

\tableofcontents

\chapter{Introduction}

Continuous-variable quantum key distribution (CV-QKD) with Gaussian modulation (GM) of quantum coherent states is regarded as an auspicious contender for deployment in widely-used applications. The aim of this tutorial is to provide an overview of the protocol, its security against eavesdropping and some aspects of its experimental implementation. We try to derive the fundamental relations from scratch in most detail while assuming as little pre-knowledge on the topic as possible. In this regard, the present article might serve as an introduction for beginners in the field or as a reference for experimentalists who want to deepen their theoretical understanding of CV-QKD. In addition to a pedagogical review of long-established relations, we present a detailed derivation of our own analytical noise models. The purpose of these models is to link the rather abstract formalism of continuous-variable quantum information theory to the specifications on actual hardware datasheets.

Quantum key distribution~\cite{gisin2002rmp,scarani2009security} is a method to generate a secret key between two distant parties, Alice and Bob, based on transmitting non-orthogonal quantum states. After the transmission and measurement of these quantum states, Alice and Bob exchange classical messages and perform post-processing to generate a secure key. In order to prevent a man-in-the-middle attack, Alice and Bob need to authenticate these classical messages in advance (so strictly speaking QKD is a key-growing protocol).

QKD was first introduced with single photons acting as information carrier~\cite{bennet1984quantum, ekert1991quantum}, sometimes referred to as \emph{discrete-variable} (DV) QKD. The exchanged quantum states are encoded into the polarisation, phase or time bin of the transmitted qubits and the secret key is established upon detection of the individual photons. With a delay of fifteen years after the first DV-QKD protocol, QKD with \emph{continuous variables} was introduced as a promising alternative. Firstly proposed with discrete~\cite{ralph1999, hillery2000squeezed, reid2000cv} and Gaussian~\cite{cerf2001quantum} encoding of squeezed states, the concept was soon developed further to Gaussian-modulated CV-QKD with coherent states~\cite{grosshans2002continuous, grosshans2003gmcv, grosshans2003virtual, weedbrook2004quantum}. The advantage of CV-QKD over DV-QKD lies in the efficient, high-rate and cost-effective detection using homodyne receivers as opposed to single-photon counters; the advantage of CV-QKD with \emph{coherent} states over \emph{squeezed}-state protocols lies mainly in the avoidance of the technologically challenging generation of squeezed light. A comprehensive theoretical overview of CV-QKD (including squeezed-state protocols) is provided in~\cite{garcia2007quantum}.

By now, many experimental demonstrations underlined the feasibility of CV-QKD with coherent states~\cite{qi2007experimental, lodewyck2007quantum, fossier2009field, jouguet2013experimental, huang2015high, qi2015generating, soh2015self, huang2016long, kleis2017continuous, zhang2017continuous, laudenbach2017pilot, wang2018high}. The experimental performance of these works has to be evaluated in the context of the various settings and assumptions that the respective authors applied and are therefore difficult to compare. After all, a CV-QKD setup is primarily appraised by the secure-key rate that can be achieved over a given transmission distance. The key rate, however, as elaborated in the present article, crucially depends on a number of assumptions, e.g.\ on the power of a potential eavesdropper, the reconciliation efficiency, the consideration of finite-key effects, the classification of so-called trusted noise, the security of a given modulation alphabet and others. Moreover, different experimental demonstrations exhibit different degrees of vulnerability to side-channel attacks. For example, a few papers so far successfully demonstrated long-distance CV-QKD with a non-zero key over a channel length of $80$ to $\SI{100}{km}$ \cite{jouguet2013experimental, huang2016long}. On the other hand, in these implementations the local oscillator (a strong laser signal required to measure the quantum signals) is transmitted to the receiver over the quantum channel and is therefore accessible to possible manipulations by an eavesdropper -- a deficiency which opens a number of severe security loopholes~\cite{haseler2008testing, huang2013quantum, ma2013wavelength, qin2013saturation, jouguet2013preventing, ma2013local, huang2014quantum, qin2016quantum}. This is why recent works are focusing on different methods to avoid the transmission of the local oscillator and to generate it at the receiver lab instead (often referred to as ``\emph{true} local oscillator'' or ``\emph{local} local oscillator'' [LLO])~\cite{qi2015generating, soh2015self, huang2015high, kleis2017continuous, laudenbach2017pilot, marie2017self, wang2018high}. Some of the latest papers demonstrated the feasibility of a secure-key rate of a few Mbit/s over a channel length of some tens of kilometres using a LLO scheme~\cite{laudenbach2017pilot, wang2018high}.

A good overview of the state-of-the-art in GM CV-QKD and the security of various protocols is given in~\cite{Diamanti2015}, however that review does not provide any detailed calculations. While discrete-variable QKD (photon counting instead of homodyne detection) has been proven secure against an eavesdropper Eve who has infinite classical and/or quantum computational power and memory,\footnote{All reported attacks against QKD attacked imperfections in the implementation, i.e.\ deviations from the theoretical model.} security proofs for CV-QKD with Gaussian modulation of coherent states are, up to now, less advanced. Unconditional security against the strongest (the so-called coherent) attack is established only for the cases of infinitely or unpractically long keys \cite[Table 1]{Diamanti2015} (see Section~\ref{sec_asymptotic}). However, a recent work suggests a proof yielding non-vanishing secure-key rates even for practical block sizes \cite{leverrier2017security}. In order not to undermine the introductory purpose of this tutorial, the present work is restricted to the less complex security analysis of keys with infinite length. Still, the formalism derived in this article remains valid also when finite-key effects are taken into account, albeit of course under some adequate extensions.

In this article we first introduce the information-theoretic notions of security and describe under which assumptions a QKD protocol can be considered secure (Section~\ref{ch_security}). Subsequently, in Section~\ref{ch_protocol}, we describe the Gaussian-modulated coherent-state protocol. The very fundamental notions of transmittance and noise are introduced in Section~\ref{ch_transmittancenoise}. We then discuss the covariance matrix with respect to Alice and Bob in Section~\ref{ch_covariancemat}, which is used to calculate Eve's information (the Holevo information) in Section~\ref{ch_holevo}. The mutual information between Alice and Bob which represents the amount of information that can be transmitted from one party to the other is calculated in Section~\ref{ch_SNR}. These two informational quantities are needed to compute the final asymptotic secret-key rate. In Section~\ref{ch_parameterest} we describe how an estimate on the secret-key rate can be obtained in practice. In Section~\ref{ch_noise}, we analyse the various noise sources in realistic implementations of the protocol. The noise models derived in this section allow to get a preview on the experimental performance of a CV-QKD system by simulating the impact of certain of hardware choices~\cite{laudenbach2016cvsim}. Finally, in Section~\ref{ch_experimentalimplications} we discuss some experimental implications of the basic relations and noise models derived in this article. While Sections~\ref{ch_security}--\ref{ch_holevo}, the better part of Section~\ref{ch_parameterest} as well as the entire appendix represent an introduction to and derivation of well-known established relations in CV-QKD, the noise models and their evaluation for practical implementations (Sections~\ref{ch_noise} and \ref{ch_experimentalimplications}) represent our own original research.

\chapter{Notions of Security and Secure-Key Rate} \label{ch_security}

Generally, in our assumptions we allow Eve to have full access to the quantum channel which she is free to control and manipulate. She can monitor the public channel but cannot intervene in the conversation between Alice and Bob which introduces the important requirement of an \emph{authenticated channel}. For her eavesdropping attack, Eve is allowed to prepare arbitrary ancillary states that she gets to interact with the transmitted signal states and subsequently performs measurements on. Very importantly, she might be in possession of a quantum memory which allows her to store her states and perform her measurement at a later time according to what she learned during the classical post-processing~(see Section~\ref{sec_post-processing}). The actual degree of information-theoretic security of a given QKD protocol depends strongly on the assumed technological capabilities a potential eavesdropper might have. Classified by her powers, we distinguish three different types of eavesdropping attacks (i.e.\ attempts to obtain information on the secret key) that are typically considered in security proofs~\cite{scarani2009security}:

\begin{description}
    \item [individual attack] Eve performs an independent and identically distributed (i.i.d.) attack on all signals, i.e.\ she prepares separable ancilla states each of which interacts individually with one signal pulse in the quantum channel. The states are stored in a quantum memory until the end of the sifting procedure (but until \emph{before} the post-processing step) and subsequently measured independently from one another.
    \item [collective attack] Eve performs an i.i.d.\ attack with separable ancilla states, stores her state in a quantum memory and performs an optimal \emph{collective} measurement on all quantum states at any later time (in particular, \emph{after} post-processing).
    \item [coherent attack] The most general attack where no (i.i.d.) assumption is made. In particular, Eve may prepare an optimal global ancilla state whose (possibly mutually dependent) modes interact with the signal pulses in the channel and are then stored and collectively measured after the classical post-processing.
\end{description}
In addition, security can be proven either in the \emph{asymptotic limit} in which an infinite number of symbols is transmitted or for the case of a finite number of transmitted symbols. Naturally, the study of the asymptotic limit does not directly apply to any realistic system. Nonetheless, it is useful to analyse since it provides an upper bound on the corresponding non-asymptotic result and because it can typically be derived more easily.

Initial security proofs focused on individual attacks~\cite{grosshans2003virtual, grosshans2004individual}, followed by collective attacks~\cite{grosshans2005collective}. Further efforts to prove the security against coherent attacks (the strongest possible attack) have been eased by two ideas: The first is the reduction of coherent attacks to collective attacks using the de~Finetti representation theorem for infinite dimensions~\cite{renner2009definetti}. The second is the optimality of Gaussian attacks~\cite{wolf2006extremality, garcia-patron2006gaussianattacks, navascues2006gaussianattacks} which states that for a given covariance matrix, the state that minimises the key rate is Gaussian. Subsequently, the security of Gaussian-modulated CV-QKD against coherent attacks was proved based on computing covariance matrices assuming Gaussian channels~\cite{leverrier2010proof, lodewyck2007quantum, pirandola2008proof}. While these proofs only hold in the asymptotic limit, the security of CV-QKD with finite block sizes is much more involved and, to this date, a vital field of ongoing research~\cite{leverrier2010finite, leverrier2013security, ruppert2014long, thearle2016estimation}.

A non-asymptotic \emph{composable}\footnote{\label{FN-comp}Composability is an important property of cryptographic protocols. A protocol is called universally composable if its outcome is (almost) indistinguishable from that of an ideal protocol. Any protocol that uses a composable QKD key is (almost) as secure as if it would use an ideal key instead. See \cite{MullerQuade2009,Portmann2014} for a discussion of composable security of quantum key distribution.} security proof against collective attacks was first introduced for continuous-variable quantum key distribution with squeezed-states~\cite{furrer2012continuous}, finally followed by a proof for coherent-state CV-QKD~\cite{Leverrier2014}. Unfortunately, the author notes that both the usual de~Finetti method and the post-selection method (another method to derive security against coherent attacks from security against collective attacks, see \cite{Christandl2009}) fall short on providing useful finite-size key estimates against coherent (general) attacks. However, a more recent result seems to overcome this problem by exploitation of a new kind of Gaussian de~Finetti reduction \cite{leverrier2017security}.

\section{Asymptotic Security against Collective Attacks} \label{sec_asymptotic}

To avoid a lot of complications and in order to keep the text at an introductory level, we will consider only the asymptotic secure-key rate in the case of collective attacks in the remainder of this tutorial. Since it comes almost for free, we will, however, take the non-asymptotic behaviour of the information reconciliation algorithm into account.

In general, the secure-key rate $K$ is given by

\begin{align} \label{eq_K}
K= f_{\text{sym}} \cdot r,
\end{align}
where $f_{\text{sym}}$ is the symbol rate (in units of symbols/s) and $r$ is the secret fraction (i.e.\ key rate per symbol; in bits/symbol)~\cite{scarani2009security}. The asymptotic secret fraction for an CV-QKD system with ideal post-processing in the case of collective attacks is given by the Devetak-Winter formula \cite{Devetak2005,scarani2009security}:

\begin{align}
    r_\textrm{coll}^\textrm{asympt}\ge I_{AB} - \chi.
\end{align}
It is lower-bounded by the difference between \emph{the mutual information} $I_{AB}$ (in bits/symbol) between Alice and Bob and a bound on Eve's possible information on the key, the \emph{Holevo information} $\chi$. The actual representation of $\chi$ will depend on the error-correction protocol carried out by Alice and Bob (see Section~\ref{sec_post-processing}). When Bob corrects his data according to information he receives from Alice and Alice's data remains unmodified, we speak of \emph{direct reconciliation} and the asymptotic secret fraction reads

\begin{align}
    r_\textrm{coll,DR}^\textrm{asympt}\ge I_{AB} - \chi_{EA},
\end{align}
where $\chi_{EA}$ represents Eve's potential information on Alice's key. In the opposite case of \emph{reverse reconciliation} (Bob's key is unmodified, Alice corrects her key according to Bob's) the secret fraction reads

\begin{align} \label{eq_rRR}
    r_\textrm{coll,RR}^\textrm{asympt} \ge I_{AB} - \chi_{EB},
\end{align}
where $\chi_{EB}$ is the Holevo bound representing Eve's possible information on Bob's key. In the remainder of this review we will restrict our analysis to the case of reverse reconciliation. Equation~\eqref{eq_rRR} is valid for an ideal system; however, if we take into account that information reconciliation does not operate at the (asymptotic) Shannon limit, that a fraction of blocks (frames) will fail to reconcile, and that a fraction of the symbols is used for error estimation we obtain the asymptotic secure-key fraction of a practical CV-QKD system \cite{lodewyck2007quantum}:\footnote{See \cite{johnson2017problem} for a critical discussion of this equation.}

\begin{align} \label{eq_r}
    r_\textrm{coll}^\textrm{asympt}\ge (1-\text{FER})(1-\nu)(\beta I_{AB}-\chi_{EB}).
\end{align}
Here $\text{FER} \in [0,1]$ and $\beta \in [0,1]$ represent the frame-error rate and the efficiency of information reconciliation, respectively, and $\nu \in [0,1]$ is the fraction of the symbols which has to be disclosed in order to estimate the entries of the covariance matrix. The efficiency $\beta$ measures how closely an information reconciliation method approaches the theoretical limit. Note that the inequation~\eqref{eq_r} describes a \emph{lower bound} for the inherent secret fraction because $\chi_{EB}$ represents an \emph{upper bound}. In actual implementations of CV-QKD protocols, however, we always assume Eve to perform an optimal attack and therefore maximise $\chi_{EB}$ right to its upper bound. Therefore, the achievable secret fraction $r$ is defined by the lower bound of the above inequation and the left and right side become equal.

If a binary error-correcting code with code rate $R$ is used, the relation

\begin{equation}
R =\beta I_{AB}
\end{equation}
holds. If an error-correcting code over an alphabet of size $q$ is used, every symbol of the code corresponds to $\log_2 q$ bits (typically $q$ is an integer power of 2) and the previous relation can be generalized to

\begin{equation}
    R \log_2 q =\beta I_{AB}.
\end{equation}
Within the subsequent chapters we will derive in detail the mutual information $I_{AB}$ and the Holevo information $\chi_{EB}$ addressing the scenario of coherent-state CV-QKD with Gaussian modulation and reverse reconciliation.

\chapter{Basics of a Coherent-State Protocol} \label{ch_protocol}

Any QKD protocol consists of two subsequent phases: first the preparation, transmission and measurement of non-orthogonal quantum states in order to distribute the raw key, followed by classical post-processing in which Alice and Bob perform sifting (reconciling the measurement bases, if required by the protocol), error correction and privacy amplification. The transmission of the quantum states is performed through a possibly insecure quantum channel~\cite{pirandola2009direct, wilde2013quantum, gyongyosi2018survey, imre2012advanced} whereas the error-correction is conducted using an authenticated classical channel.

The next sections describe in which way these steps are carried out in coherent-state CV-QKD with Gaussian modulation.

\section{Gaussian Modulation}

\begin{figure}
\centering
\includegraphics[width=0.8\linewidth]{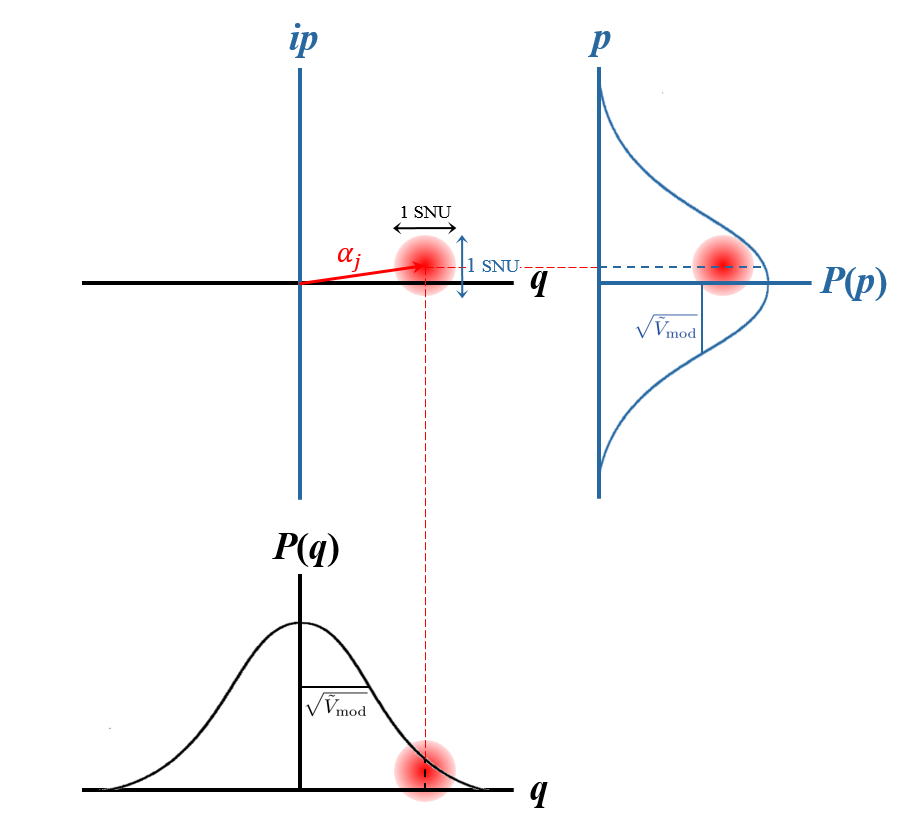}
\caption{Phase-space representation of Gaussian modulation. The coherent state $\alpha_{j}$ represents one of many subsequent states, randomly prepared by Alice. During preparation both quadratures obey the same Gaussian probability distribution with variance $\tilde{V}_{\text{mod}}$, illustrated by the graphs on the bottom left ($q$) and top right ($p$). As for any coherent state, the quadrature operators both carry an inherent variance of one shot-noise unit.}
\label{gaussianmodulation}
\end{figure}

In a Gaussian-modulated scenario \cite{grosshans2002continuous} Alice prepares displaced coherent states with quadrature components $q$ and $p$ that are realisations of two independent and identically distributed (i.i.d.) random variables $\randq$ and $\randp$. The random variables $\randq$ and $\randp$ obey the same zero-centred normal distribution

\begin{equation} \label{eq_gaussdistribution}
    \randq \sim \randp \sim \mathcal{N} (0,\tilde{V}_{\text{mod}})
\end{equation}
where $\tilde{V}_{\text{mod}}$ is referred to as \emph{modulation variance} (cf. Figure~\ref{gaussianmodulation}). Alice prepares a sequence $\ket{\alpha_{1}},\dots \ket{\alpha_{j}}, \dots, \ket{\alpha_{N}}$ of displaced coherent states

\begin{align}
\ket{\alpha_{j}}=\ket{q_{j}+ip_{j}},
\end{align}
obeying the usual eigenvalue equation:

\begin{subequations}
\begin{align}
\hat{a} \ket{\alpha_{j}} & = \alpha_{j} \ket{\alpha_{j}} , \\
\frac{1}{2} \left( \hat{q}+i\hat{p} \right) \ket{\alpha_{j}} & = (q_{j}+ip_{j}) \ket{\alpha_{j}} ,
\end{align}\\
\end{subequations}
where $\hat{a}=\frac{1}{2}(\hat{q} + i \hat{p})$ is the annihilation operator and $\hat{q}$ and $\hat{p}$ are the quadrature operators, defined in the framework of shot-noise units\footnote{See Appendix~\ref{ch_units} for a definition of shot-noise units and a direct comparison to other conventions.} (SNU). Each individual state $\ket{\alpha_j}$ has a mean photon number of

\begin{align}
    \braket{n_{j}}=\braket{\alpha_j|\hat n|\alpha_j} = |\alpha_{j}|^{2}=q_{j}^{2}+p_{j}^{2}.
\end{align}
Considering that $q_j$ and $p_j$ are sampled from the distribution in \eqref{eq_gaussdistribution}, the mean photon number of the state \emph{ensemble} that Alice prepares reads

\begin{align}\label{eq:expectn}
\braket{n}= \braket{\randq^{2}}+\braket{\randp^{2}} = 2 \tilde{V}_{\text{mod}}.
\end{align}
The variance of the quadrature operators applied to the state ensemble relates to the modulation variance by\footnote{For a derivation see the footnote in Appendix~\ref{ch_coherent}.}

\begin{equation}\label{eq:variancequad}
V(\hat{q}) = V(\hat{p}) \eqcolon V = 4 \tilde{V}_{\text{mod}} + 1 \eqcolon V_{\text{mod}} + 1 ,
\end{equation}
where $\tilde{V}_{\text{mod}}$ represents the modulation variance of the quadrature \emph{components} $q$ and $p$, and $V_{\text{mod}}= 4 \tilde{V}_{\text{mod}}$ is the modulation variance of the quadrature \emph{operators} $\hat{q}$ and $\hat{p}$. Note that even in the case of $\tilde{V}_{\text{mod}}=V_{\text{mod}}=0$, the quadrature operators still carry a variance of $V_{0}=1$ due to the uncertainty relation. $V_{0}$ is usually referred to as shot noise and it can be simply added to the modulation variance because of mutual stochastic independence of the modulation and the shot noise. Combining \eqref{eq:expectn} and \eqref{eq:variancequad}, the mean photon number in terms of the quadrature operators' variance reads

\begin{equation} \label{nvsVmod}
\braket{n}=\frac{1}{2}(V-1) = \frac{1}{2} V_{\text{mod}}.
\end{equation}
After preparation of each coherent state Alice transmits $\ket{\alpha_j}$ to Bob through a Gaussian quantum channel. Bob uses homodyne or heterodyne detection (see Section~\ref{sec_homhet}) to measure the eigenvalue of either one or both of the quadrature operators.

\section{Classical Post-Processing} \label{sec_post-processing}

Here we briefly mention the individual steps during the classical data post-processing (see e.g.~\cite{scarani2009security}) that transform Alice's modulation data and Bob's measurement results into a universally composable (see footnote~\ref{FN-comp} on page~\pageref{FN-comp}) secure key.

\begin{description}
    \item [(Sifting)] In some variants of CV-QKD Alice and Bob select the bases which they use to prepare and measure states, resp., by using independently and uniformly generated random bits. In these cases the sifting step eliminates all (uncorrelated) signals where different bases have been used for preparation and measurement. In variants of CV-QKD where Alice and Bob use both bases simultaneously no sifting is performed.
    \item [Parameter Estimation] After transmitting a sequence of states Alice and Bob will reveal and compare a random subset of the data that was sent and the corresponding measurements. This comparison allows them to estimate the total transmission and excess noise (see Section~\ref{ch_transmittancenoise}) of the channel by which they are able to compute their mutual information $I_{AB}$ and bound Eve's information $\chi$. 
If $\chi$ turns out to be greater than $\beta I_{AB}$ the protocol aborts at this point. 
    \item [Information Reconciliation] Otherwise, if $\beta I_{AB} > \chi$, Alice and Bob will perform information reconciliation which is a form of error correction. One-way information reconciliation where one party sends information on her key to the other party can be carried out in two different ways: Either Bob corrects his bits according to Alice's data (\emph{direct reconciliation}) or Alice corrects her bits according to Bob's data (\emph{reverse reconciliation})~\cite{grosshans2003virtual}. In the case of forward reconciliation, for a total transmittance of $T_{\text{tot}}<0.5$ ($\approx$ -3 dB), Eve potentially has more information on what Alice prepared than Bob has, hence no secret key can be distilled (assuming that Eve can use the entire loss for her own benefit). This \SI{3}{dB} loss limit can be overcome by using reverse reconciliation, where Bob sends the correction information to Alice who thereupon corrects her bit string according to Bob's. In this scenario Bob's data is primary, and since Alice's information on Bob's measurement results is always greater than Eve's, the mutual information $I_{AB}$ can remain greater than $\chi$ for any total transmission $T$ (of course, the lower $T$ is the more critical will the excess noise $\xi$ become). 
	    
For CV-QKD with Gaussian modulation different reconciliation schemes have been proposed. Two important schemes are slice reconciliation~\cite{VanAssche2004reconGauss} and multidimensional reconciliation~\cite{Leverrier2008multidim}. Both schemes can employ low-density parity-check (LDPC) codes~\cite{Richardson2008MCT} for the actual error correction. In the reverse reconciliation scheme one or several LDPC codes are used to calculate a compressed version of Bob's data. This compressed version is then transmitted over a classical channel to Alice and fed into her decoder. Since 2011, the state-of-the art of LDPC codes for CV-QKD in the high-loss (low SNR) regime are multi-edge-type (MET) LDPC codes~\cite{Jouguet2011errorcorrection}. 
    \item [Confirmation] After information reconciliation Alice and Bob perform a confirmation step using a family of (almost) universal hash functions~\cite{carterwegman1979} to bound the probability that error correction has failed: Alice or Bob choose with uniform probability one particular hash function from the family and transmits the choice to the partner. Both apply that hash function to their key to obtain a hash value. Subsequently, Alice and Bob exchange and compare their hash values. If the hash values are different the keys must be different and they abort; if the hash values are equal they continue and know that they have obtained an upper bound on the probability that the keys are not identical. This error probability depends on the length of the hash values and of the type of hashing functions used. 
    \item [Privacy Amplification] After successful confirmation Alice and Bob will share the same bit string with very high probability. However, Eve has a certain amount of information on the key. 
        In order to reduce Eve's probability to successfully guess (a part of) the key to a tolerable value, Alice and Bob will perform a privacy amplification protocol by applying a seeded randomness extractor (algorithm) to their bit strings. Again a family of universal hash functions is typically used for that purpose. 
    \item [Authentication]
To avoid a man-in-the-middle attack by Eve (cf.~\cite{Pacher2016} for a related discussion) Alice and Bob need to authenticate their classical communication using a family of strongly universal hash functions.
\end{description}

\chapter{Transmittance and Noise} \label{ch_transmittancenoise}

Apart from the specifications on signal modulation, any CV-QKD system is primarily characterised by the transmittance (channel transmission, detection efficiency, coupling losses) and the noise. Both undermine the experimental performance, lowering the signal-to-noise ratio and raising the advantage of a possible eavesdropper. Unlike in this present article, many references include the transmittance $T$ into the definition of noise, summarising transmission and noise in one noise parameter $\Xi$.\footnote{In most references these noise parameters are represented by the symbol $\chi$ instead of $\Xi$. However, we decided to deviate from this convention in order to avoid confusion with the Holevo information $\chi_{EA}$ and $\chi_{EB}$.} Literature often distinguishes between channel noise $\Xi_{\text{ch}}$ and detection noise $\Xi_{\text{det}}$. The channel noise reads \cite{scarani2009security}

\begin{equation}
\Xi_{\text{ch}} = \frac{1-T_{\text{ch}}}{T_{\text{ch}}}+\xi_{A},
\end{equation}
decomposing $\Xi_{\text{ch}}$ into a term caused by channel losses $(1-T_{\text{ch}})/T_{\text{ch}}$ and the so-called \emph{excess noise} $\xi_{A}$, an additional contribution to the noise variance of the quadrature operators (on top of the inherent vacuum noise $V_{0} \stackrel{\text{SNU}}{=} 1$). The subscript $A$ indicates that $\xi_{A}$ is referring to the channel input (hence to Alice), just like $\Xi_{\text{ch}}$ itself. The excess noise may originate from multiple sources such as from imperfect modulation, Raman scattering, quantisation, phase fluctuations and so on. Assuming all excess-noise sources to be stochastically independent from one another, they can be summed up due to the additivity of variances:

\begin{equation}
\xi_{A} = \xi_{\text{modul},A} + \xi_{\text{Raman},A} + \xi_{\text{quant},A} + \xi_{\text{phase},A} + \dots
\end{equation}
Very similarly to the channel noise, the detection noise is represented as \cite{scarani2009security}

\begin{equation}
\Xi_{\text{det}} = \frac{1-\eta_{\text{det}}}{\eta_{\text{det}}}+\frac{\nu_{\text{el}}}{\eta_{\text{det}}}.
\end{equation}
Again, this expression for the noise is composed of losses $(1-\eta_{\text{det}})/\eta_{\text{det}}$ and an additional variance $\nu_{\text{el}}$, this time caused by the electronic noise of the detector. Conversely to the channel noise, the detection noise $\Xi_{\text{det}}$ is referring to the channel \emph{output}, hence to Bob's lab. The total noise is given by the sum:

\begin{equation}
\Xi=\Xi_{\text{ch}}+\frac{1}{T_{\text{ch}}}\Xi_{\text{det}} ,
\end{equation}
where the detection noise is divided by $T_{\text{ch}}$ in order for it to refer to the channel input just like $\Xi_{\text{ch}}$ and, consequentially, $\Xi$. After transmission of Alice's signal through a lossy and noisy channel, Bob will measure a total quadrature variance of \cite{scarani2009security}

\begin{equation}
V_{B}=T_{\text{ch}}\eta_{\text{det}}(V+\Xi).
\end{equation}
Decomposing $\Xi$ into its constituents yields

\begin{align} \label{eq_varB}
V_{B} & =T_{\text{ch}}\eta_{\text{det}} \left(V+\Xi_{\text{ch}}+\frac{1}{T_{\text{ch}}}\Xi_{\text{det}} \right) \notag \\
& = T_{\text{ch}}\eta_{\text{det}}\left( V+ \frac{1-T_{\text{ch}}}{T_{\text{ch}}}+\xi_{A} +\frac{1}{T_{\text{ch}}} \left( \frac{1-\eta_{\text{det}}}{\eta_{\text{det}}}+\frac{\nu_{\text{el}}}{\eta_{\text{det}}} \right) \right) \notag \\
& = T_{\text{ch}}\eta_{\text{det}} V -T_{\text{ch}}\eta_{\text{det}}+T_{\text{ch}}\eta_{\text{det}}\xi_{A} + 1 + \nu_{\text{el}} \notag \\
& \eqcolon T V - T + T \xi_{A} + 1 + \nu_{\text{el}} \notag \\
& = T (V -1) + T\xi_{A} + 1 + \nu_{\text{el}} ,
\end{align}
where we defined $T \coloneq T_{\text{ch}}\eta_{\text{det}}$. Since $\nu_{\text{el}}$ represents a noise variance in shot-noise units and can be assumed to be stochastically independent from the other noise sources, we can treat it as just another constituent of the excess noise, $\nu_{\text{el}} \eqcolon \xi_{\text{det}}$, which allows us to write

\begin{align}
& T\xi_{A}+\xi_{\text{det}} \notag \\
= & T \left( \xi_{A}+\frac{1}{T}\xi_{\text{det}} \right) \notag \\
\eqcolon & T \xi_{\text{tot},A}
\end{align}
Unlike most of the literature on CV-QKD, in the present article, we have the excess noise defined at the place where it is observed by measurement, hence at the channel output in Bob's lab. We therefore introduce the definition

\begin{equation}
\xi \coloneq \xi_{\text{tot},B}=T \xi_{\text{tot},A}.
\end{equation}
This and the integration of $\nu_{\text{el}}$ into the excess noise lead to a reexpression of Bob's variance \eqref{eq_varB} to\footnote{This relation is rigorously derived, using the covariance-matrix formalism, in Section~\ref{sec_covmatchannelnoise}.}

\begin{align}
V_{B} & =  T (V -1) + 1 + \xi \notag \\
& = T V_{\text{mod}} + 1 + \xi ,
\end{align}
where, in a realistic setup, the transmittance $T$ comprises not only channel- and detection losses but also the coupling efficiency:

\begin{equation}
T=T_{\text{ch}} \cdot \eta_{\text{coup}} \cdot \eta_{\text{det}},
\end{equation}
and the excess noise is the sum of the variances of all possible noise sources:

\begin{equation}
\xi = \xi_{\text{modul}} + \xi_{\text{Raman}} + \xi_{\text{quant}} + \xi_{\text{phase}} + \xi_{\text{det}} + \xi_{\text{RIN}} + \xi_{\text{CMRR}} + \dots
\end{equation}
In Section~\ref{ch_noise} we derive analytic models for the various kinds of excess-noise constituents.

\chapter{Covariance Matrix} \label{ch_covariancemat}

Bosonic multi-mode states can conveniently be described by \emph{covariance matrices} whose diagonal entries represent the variances of the quadrature operators and whose off-diagonal terms represent the mutual covariance functions of two quadratures.\footnote{Find more details on the covariance-matrix formalism in Appendix~\ref{ch_gaussianquinfo}.} Knowledge about the covariance matrix $\Sigma$ and its transformations is helpful to understand the behaviour of the particular bosonic modes and their correlations. However, in CV-QKD the knowledge of $\Sigma$ is not only helpful but vital in order to compute the Holevo information $\chi$.

In the protocol discussed in this article, Alice prepares displaced coherent states whose quadrature components are measured by Bob after transmission through a Gaussian channel (prepare-and-measure, PM). This is equivalent to a scenario where Alice prepares a two-mode squeezed vacuum state (TMSVS), performs a measurement of both quadratures of one mode and sends the other mode to Bob (entanglement-based, EB), as for example explained in \cite{grosshans2003virtual}. In shot-noise units the TMSVS is described by the covariance matrix\footnote{See Appendix~\ref{sec_covmatinit} for a derivation of the covariance matrix of a TMSVS.}

\begin{equation}
\Sigma_{AB}=
\bordermatrix{        & \hat{q}_{A} & \hat{p}_{A} & \hat{q}_{B} & \hat{p}_{B} \cr
                \hat{q}_{A} & V &  0  & \sqrt{V^{2}-1} & 0\cr
                \hat{p}_{A} & 0  &  V & 0 & -\sqrt{V^{2}-1}\cr
                \hat{q}_{B} & \sqrt{V^{2}-1} & 0 & V & 0\cr
                \hat{p}_{B} & 0  &   -\sqrt{V^{2}-1}  & 0 & V}
=
\begin{pmatrix}
V \mathbb{1}_{2} & \sqrt{V^{2}-1} \sigma_{z} \\ 
\sqrt{V^{2}-1} \sigma_{z} & V \mathbb{1}_{2}
\end{pmatrix},
\end{equation}
where $\sigma_{z}$ is the Pauli matrix $\left(\begin{smallmatrix}1&0\\0&-1\end{smallmatrix}\right)$ and $V$ is the variance of the quadrature operators:

\begin{subequations}
\begin{align}
V(\hat{q}) & =\braket{\hat{q}^{2}}-\braket{\hat{q}}^{2}, \\
V(\hat{p}) & =\braket{\hat{p}^{2}}-\braket{\hat{p}}^{2}.
\end{align}
\end{subequations}
In our case the expectation values of the operators are zero and both quadratures $\hat{q}$ and $\hat{p}$ carry the same variance (as opposed to a squeezed state)

\begin{equation}
V \coloneqq V(\hat{q})=V(\hat{p})=\braket{\hat{q}^{2}}=\braket{\hat{p}^{2}}.
\end{equation}
This variance can be understood as the sum of the actual variance Alice is modulating with and the vacuum shot-noise variance:

\begin{equation}
V=V_{\text{mod}}+V_{0} \stackrel{\text{SNU}}{=} V_{\text{mod}}+1 .
\end{equation}
Hence the covariance matrix can be rewritten as

\begin{equation}
\Sigma_{AB}=
\begin{pmatrix}
(V_{\text{mod}}+1) \mathbb{1}_{2} & \sqrt{V_{\text{mod}}^{2}+2V_{\text{mod}}} \sigma_{z} \\ 
\sqrt{V_{\text{mod}}^{2}+2V_{\text{mod}}} \sigma_{z} & (V_{\text{mod}}+1) \mathbb{1}_{2}
\end{pmatrix} .
\end{equation}
The variance relates to the mean photon number per pulse as follows:\footnote{Derived in Appendix~\ref{ch_coherent}.}

\begin{align}
\braket{\hat{n}_{A}} & = \frac{1}{4} \left( \braket{\hat{q}^{2}}+\braket{\hat{p}^{2}} \right)-\frac{1}{2} \notag \\
& = \frac{1}{2} (V-1) = \frac{1}{2} V_{\text{mod}} .
\end{align}
After transmission of Bob's mode and influence of the various noise sources the covariance matrix transforms to\footnote{See Appendix~\ref{sec_covmatchannelnoise} for a derivation of this matrix.}

\begin{equation} \label{eq_covmathom}
\Sigma_{AB}=
\begin{pmatrix}
V \mathbb{1}_{2} & \sqrt{T(V^{2}-1)} \sigma_{z} \\ 
\sqrt{T(V^{2}-1)} \sigma_{z} & (T[V-1]+1+\xi) \mathbb{1}_{2}
\end{pmatrix}
=
\begin{pmatrix}
(V_{\text{mod}}+1) \mathbb{1}_{2} & \sqrt{T(V_{\text{mod}}^{2}+2V_{\text{mod}})} \sigma_{z} \\ 
\sqrt{T(V_{\text{mod}}^{2}+2V_{\text{mod}})} \sigma_{z} & (TV_{\text{mod}}+1+\xi) \mathbb{1}_{2}
\end{pmatrix} .
\end{equation}
Note that the matrix \eqref{eq_covmathom} represents a quadrature modulation according to a continuous Gaussian distribution $\randq \sim \randp \sim \mathcal{N}(0,\tilde{V}_{\text{mod}})$. Other modulation alphabets like phase-shift keying (PSK) or quadrature amplitude modulation (QAM) are not discussed in this review. Those alphabets are represented by their own covariance matrices which distinguish from \eqref{eq_covmathom} only in the off-diagonal term, replacing $ \sqrt{T(V_{\text{mod}}^{2}+2V_{\text{mod}})}$ by an appropriate covariance function of $A$ and $B$.

\section{Homodyne and Heterodyne Detection} \label{sec_homhet}

\begin{figure}
\centering
\subcaptionbox{}
    [\linewidth]{\includegraphics[width=\linewidth]{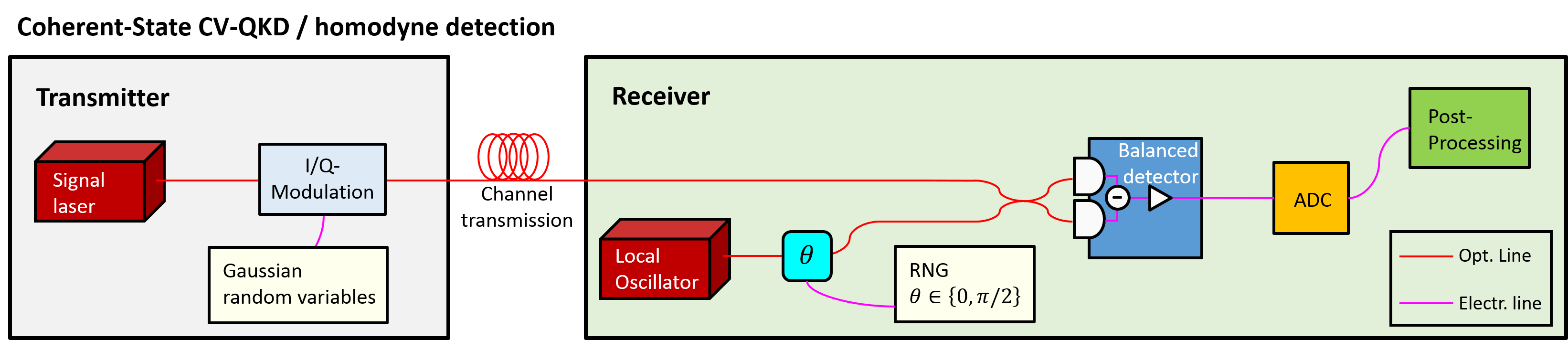}}
\subcaptionbox{}
    [\linewidth]{\includegraphics[width=\linewidth]{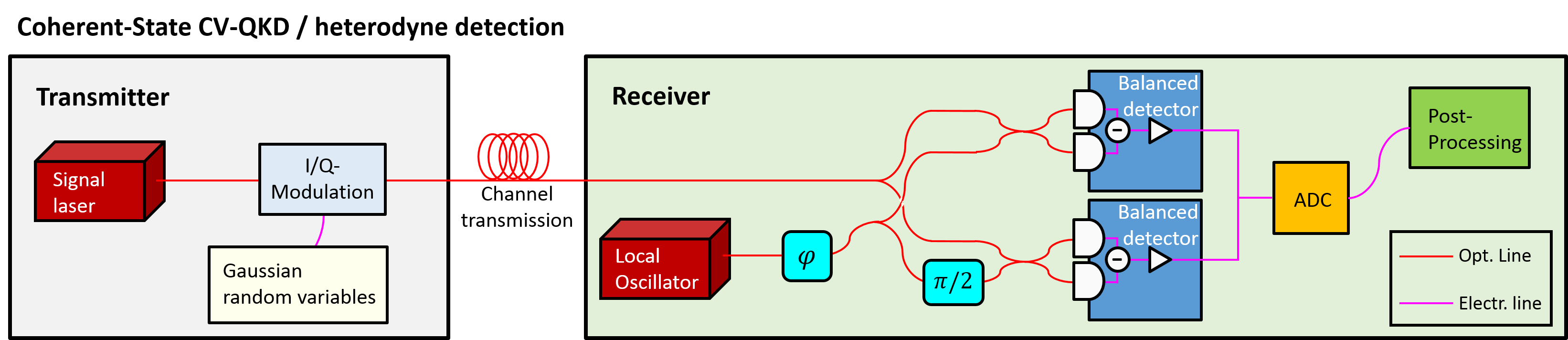}}
\caption{Distinction between the (a) homodyne- and (b) heterodyne-detection protocol. In the first case a random-number generator (RNG) is used to select the phase of the local oscillator: $0$ or $\pi/2$ to measure $q$ or $p$, respectively. Only one homodyne detector is used, measuring one quadrature at a time. In the case of heterodyne detection the quantum signal is split using a balanced beamsplitter. One arm is used to measure $q$, the other one -- after a LO-phase shift of $\pi/2$ -- to measure $p$.}
\label{homohetero}
\end{figure}

During the key transmission phase Bob receives noisy coherent states with a given symbol rate $f_{\text{sym}}$. These subsequent modes each carry randomly modulated quadrature components $q$ and $p$. The quadratures can be measured using the technique of homodyne detection where the signal mode is mixed with a reference laser (the local oscillator, LO) at a balanced beamsplitter. Depending on the relative phase $\theta$ of signal mode and LO, the photon number difference at the output ports of the BS is proportional to either the $q$- or $p$ quadrature:\footnote{See Appendix~\ref{ch_homodynedet} for a derivation of the homodyne-detection law.}

\begin{align}
\Delta \hat{n} = |\alpha_{\text{LO}}| ( \hat{q} \cos{\theta} + \hat{p} \sin{\theta} ).
\end{align}
Depending on the protocol, Bob can either measure one quadrature component at a time by randomly selecting between $\theta=0$ and $\theta=\pi/2$ for each incoming mode, or he measures \emph{both} quadratures of each mode simultaneously, which is sometimes referred to as ``no-switching protocol''~\cite{weedbrook2004quantum}. The two variations of the protocol are illustrated in Figure~\ref{homohetero}. For the second approach Bob will separate the incoming states with a balanced BS and then measure the two quadratures by homodyne detection on each half of the signal -- one with $\theta=0$ to measure $q$ and one with $\theta=\pi/2$ to measure $p$. Although this convention is misleading in more than just one way, we are speaking of \emph{homodyne} detection when Bob measures only one quadrature component at a time and of \emph{heterodyne} detection when Bob uses a BS and \emph{two} homodyne detectors to measure both quadrature components simultaneously.\footnote{Note that in the telecommunication jargon, \emph{heterodyne detection} denotes a differing optical frequency of transmitted signal and local oscillator.} Using heterodyne instead of homodyne detection will double the mutual information for each symbol for the price of additional \SI{3}{dB} loss introduced by the heterodyning BS.

Splitting Bob's mode for heterodyne detection will transform the covariance matrix to\footnote{This covariance matrix describing a heterodyne measurement setup is in detail derived in Appendix~\ref{sec_bobhet}.}

\begin{equation}
\Sigma_{AB} = 
\bordermatrix{
& A & B_{1} & B_{2} \cr
A      & (V_{\text{mod}}+1) \mathbb{1}_{2} & \sqrt{\frac{T}{2}(V_{\text{mod}}^{2}+2V_{\text{mod}})} \sigma_{z} & -\sqrt{\frac{T}{2}(V_{\text{mod}}^{2}+2V_{\text{mod}})}\sigma_{z} \cr
B_{1}  & \sqrt{\frac{T}{2}(V_{\text{mod}}^{2}+2V_{\text{mod}})} \sigma_{z}  &  (\frac{T}{2}V_{\text{mod}}+1+\frac{\xi}{2})\mathbb{1}_{2} & -\frac{1}{2}(TV_{\text{mod}}+\xi)\mathbb{1}_{2} \cr
B_{2}  & -\sqrt{\frac{T}{2}(V_{\text{mod}}^{2}+2V_{\text{mod}})}\sigma_{z} & -\frac{1}{2}(TV_{\text{mod}}+\xi)\mathbb{1}_{2} & (\frac{T}{2}V_{\text{mod}}+1+\frac{\xi}{2})\mathbb{1}_{2} } .
\end{equation}
Note that Bob now has two modes $B_1$, $B_2$ since he split his signal into half. The covariance matrix between Alice's mode and one of Bob's two modes is then represented as

\begin{equation} \label{eq_covmathet}
\Sigma_{AB_{1,2}} = 
\begin{pmatrix}
(V_{\text{mod}}+1) \mathbb{1}_{2} & \pm \sqrt{\frac{T}{2}(V_{\text{mod}}^{2}+2V_{\text{mod}})} \sigma_{z} \\
\pm \sqrt{\frac{T}{2}(V_{\text{mod}}^{2}+2V_{\text{mod}})} \sigma_{z}  &  (\frac{T}{2}V_{\text{mod}}+1+\frac{\xi}{2})\mathbb{1}_{2} 
\end{pmatrix},
\end{equation}
which is the equivalent to the case of homodyne detection \eqref{eq_covmathom} up to a factor of $1/2$ at transmission $T$ and excess noise $\xi$. The above matrix illustrates nicely the impact of heterodyne detection on Bob's measurement outcome. However, even in case of heterodyne detection, the Holevo information can already be derived from \eqref{eq_covmathom}, i.e.\ the matrix describing the TMSVS \emph{before} action of Bob's measurement apparatus (see Section~\ref{ch_holevo}).

\section{Prepare-and-Measure vs. Entanglement-Based Protocols}

In practical implementations of coherent-state CV-QKD protocols, Alice will, instead of generating two-mode squeezed states, prepare coherent states using $q/p$ modulation according to a given probability distribution with variance $V_{\text{mod}}$. However, for the sake of a simplified security analysis, it is convenient to assume that Alice and Bob are sharing an entangled TMSVS to which Eve holds a purification. The equivalence of both pictures, prepare-and-measure (PM) and entanglement-based (EB), is discussed in the present section. Assume Alice prepares her coherent state with quadrature operators $\hat{q}$ and $\hat{p}$ according to a probability distribution with variance $V_{A}=V_{\text{mod}}$ (or equivalently, quadrature components $q$ and $p$ with variance $V_{\text{mod}}/4$). Due to the minimal uncertainty of 1 shot-noise unit, described by a normal distribution $\mathcal{N}(0,1)$, Bob will receive coherent states with variance $V_{B}=V_{\text{mod}}+1$ (neglecting for now transmission losses and excess noise):

\begin{align}
\hat{q}_{B} & \sim \hat{q}_{A} + \mathcal{N}(0,1), \notag \\
V_{B} & = V(\hat{q}_{A} + \mathcal{N}(0,1))=V(\hat{q}_{A})+1=V_{\text{mod}}+1.
\end{align}
Assuming the quadratures to be zero-centred ($\braket{\hat{q}}=\braket{\hat{p}}=0$), the covariance of Alice's and Bob's data reads

\begin{align}
\Cov(\hat{q}_{A},\hat{q}_{B})=\braket{\hat{q}_{A}\hat{q}_{B}} = \braket{\underbrace{(\hat{q}_{B}-\hat{q}_{A})}_{\mathcal{N}(0,1)}\hat{q}_{A}} + \braket{\hat{q}_{A}^{2}} =0 + \braket{\hat{q}_{A}^{2}} = V_{A}=V_{\text{mod}}.
\end{align}
This allows us to write down the covariance matrix of a prepare-and-measure protocol:

\begin{align}
\Sigma^{\text{PM}}=
\begin{pmatrix}
V_{A} \mathbb{1}_{2} & \Cov(A,B) \mathbb{1}_{2} \\
\Cov(A,B) \mathbb{1}_{2} & V_{B} \mathbb{1}_{2} .
\end{pmatrix}
=
\begin{pmatrix}
V_{\text{mod}} \mathbb{1}_{2} & V_{\text{mod}} \mathbb{1}_{2} \\
V_{\text{mod}} \mathbb{1}_{2} & (V_{\text{mod}}+1) \mathbb{1}_{2} .
\end{pmatrix}
\end{align}
Suppose now that Alice and Bob share a two-mode squeezed vacuum state with variance $V$:

\begin{align}
\Sigma^{\text{EB}}=
\begin{pmatrix}
V \mathbb{1}_{2} & \sqrt{(V^{2}-1)} \sigma_{z} \\ 
\sqrt{(V^{2}-1)} \sigma_{z} & V \mathbb{1}_{2}
\end{pmatrix} .
\end{align}
Alice, in order to collapse Bob's mode into a coherent state, will perform a measurement of both quadratures of her mode. Therefore she has to split her mode into half using a balanced beamsplitter on whose output ports she will determine $\hat{q}$ and $\hat{p}$ respectively. The covariance matrix is then represented as

\begin{align}
\Sigma^{'\text{EB}}=
\begin{pmatrix}
\frac{V+1}{2} \mathbb{1}_{2} & \frac{V-1}{2} \mathbb{1}_{2} & \sqrt{\frac{1}{2}(V^{2}-1)} \sigma_{z} \\ 
\frac{V-1}{2} \mathbb{1}_{2} & \frac{V+1}{2} \mathbb{1}_{2} & \sqrt{\frac{1}{2}(V^{2}-1)} \sigma_{z} \\
\sqrt{\frac{1}{2}(V^{2}-1)} \sigma_{z} & \sqrt{\frac{1}{2}(V^{2}-1)} \sigma_{z} &  V \mathbb{1}_{2}
\end{pmatrix} .
\end{align}
Since both Alice's modes are exactly equivalent, we reduce the matrix to the familiar $4\times4$ dimensions, representing only one of Alice's modes and Bob's mode:

\begin{align} \label{eq_SigmaEB}
\Sigma^{''\text{EB}}=
\begin{pmatrix}
\frac{V+1}{2} \mathbb{1}_{2} & \sqrt{\frac{1}{2}(V^{2}-1)} \sigma_{z} \\
\sqrt{\frac{1}{2}(V^{2}-1)} \sigma_{z} &  V \mathbb{1}_{2}
\end{pmatrix} .
\end{align}
If Alice rescales her measurement operators according to the transformations

\begin{subequations}
\begin{align}
\hat{q}_{A} & \longrightarrow  \sqrt{\frac{2(V-1)}{V+1}} \hat{q}_{A}, \\
\hat{p}_{A} & \longrightarrow -\sqrt{\frac{2(V-1)}{V+1}} \hat{p}_{A},
\end{align}
\end{subequations}
the variance $V_{A}$ and the covariance $\Cov(A,B)$ will change accordingly:

\begin{subequations}
\begin{align}
V(\hat{q}_{A}) & \longrightarrow  \frac{2(V-1)}{V+1} V(\hat{q}_{A}), \notag \\
V(\hat{p}_{A}) & \longrightarrow  \frac{2(V-1)}{V+1} V(\hat{p}_{A}), \notag \\
\Cov(\hat{q}_{A},\hat{q}_{B}) & \longrightarrow \sqrt{\frac{2(V-1)}{V+1}} \Cov(\hat{q}_{A},\hat{q}_{B}), \\
\Cov(\hat{p}_{A},\hat{p}_{B}) & \longrightarrow -\sqrt{\frac{2(V-1)}{V+1}} \Cov(\hat{p}_{A},\hat{p}_{B}).
\end{align}
\end{subequations}
This will transform the covariance matrix \eqref{eq_SigmaEB} as follows:

\begin{align}
\Sigma^{''\text{EB}} \longrightarrow &
\begin{pmatrix}
\frac{2(V-1)}{V+1} \frac{V+1}{2} \mathbb{1}_{2} & \sqrt{\frac{2(V-1)}{V+1}} \sqrt{\frac{1}{2}(V^{2}-1)} \mathbb{1}_{2} \\ 
\sqrt{\frac{2(V-1)}{V+1}} \sqrt{\frac{1}{2}(V^{2}-1)} \mathbb{1}_{2} & V \mathbb{1}_{2}
\end{pmatrix} \notag \\
= &
\begin{pmatrix}
(V-1) \mathbb{1}_{2} & (V-1) \mathbb{1}_{2} \\ 
(V-1) \mathbb{1}_{2} & V \mathbb{1}_{2}
\end{pmatrix} \notag \\
= &
\begin{pmatrix}
V_{\text{mod}} \mathbb{1}_{2} & V_{\text{mod}} \mathbb{1}_{2} \\ 
V_{\text{mod}} \mathbb{1}_{2} & (V_{\text{mod}}+1) \mathbb{1}_{2}
\end{pmatrix} = \Sigma^{\text{PM}}.
\end{align}
So by rescaling the data obtained by her heterodyne measurement, Alice can simulate a prepare-and-measure scenario without Bob or Eve taking any notice of it. In other words, the prepare-and-measure scenario is equivalent to the entanglement-based version after Alice has rescaled her measurement outcomes by a factor of $\kappa=\pm \sqrt{2(V-1)/(V+1)}$. Conversely, in the experimentally more realistic scenario where Alice actively modulates the coherent states instead of measuring one mode of a TMSVS, she will rescale the values of the prepared quadratures with $\kappa^{-1}$ in order to simulate an entanglement-based scenario:

\begin{subequations}
\begin{align}
\hat{q}_{A}^{\text{EB}} & =  \sqrt{\frac{V+1}{2(V-1)}} \hat{q}_{A}^{\text{PM}}, \\
\hat{p}_{A}^{\text{EB}} & = -\sqrt{\frac{V+1}{2(V-1)}} \hat{p}_{A}^{\text{PM}}.
\end{align}
\end{subequations}
This will of course transform the prepare-and-measure matrix to the entanglement-based one for which the security analysis is significantly more simple.

\begin{align}
\Sigma^{\text{PM}} = &
\begin{pmatrix}
(V-1) \mathbb{1}_{2} & (V-1) \mathbb{1}_{2} \\ 
(V-1) \mathbb{1}_{2} & V \mathbb{1}_{2}
\end{pmatrix}
\notag \\
\longrightarrow &
\begin{pmatrix}
\frac{1}{\kappa^{2}}(V-1) \mathbb{1}_{2} & \frac{1}{\kappa}(V-1) \mathbb{1}_{2} \\ 
\frac{1}{\kappa}(V-1) \mathbb{1}_{2} & V \mathbb{1}_{2}
\end{pmatrix}
\notag \\
= &
\begin{pmatrix}
\frac{V+1}{2} \mathbb{1}_{2} & \sqrt{\frac{1}{2}(V^{2}-1)} \sigma_{z} \\
\sqrt{\frac{1}{2}(V^{2}-1)} \sigma_{z} &  V \mathbb{1}_{2}
\end{pmatrix} = \Sigma^{\text{EB}} .
\end{align}
Rescaling Alice's values is therefore a crucial procedure in order to properly determine the Holevo bound and has to be taken into account for the parameter-estimation procedure, as will be described in Section~\ref{sec_covmatestimation}.

\chapter{Signal-to-Noise Ratio and Mutual Information} \label{ch_SNR}

The signal-to-noise ratio (SNR) is simply described by

\begin{equation}
\text{SNR}=\frac{P_{S}}{P_{N}},
\end{equation}
where $P_{S}$ is the total signal power and $P_{N}$ is the total noise power arriving at the channel output. As seen in \eqref{eq_covmathom} and \eqref{eq_covmathet}, the variance of Bob's quadratures is

\begin{equation}
V_{B} = V(\hat{q}_{B}) = V(\hat{p}_{B}) = \frac{T}{\mu} V_{\text{mod}} +1+\frac{\xi}{\mu},
\end{equation}
where $T$ is the total transmittance (including coupling- and detection efficiency) and $\mu=1$ in case of homodyne detection and $\mu=2$ in case of heterodyne detection. The damped modulation variance corresponds to the signal whereas vacuum noise 1 and excess noise $\xi$ add up to the total noise, hence

\begin{equation} \label{eq_SNR}
\text{SNR}=\frac{\frac{1}{\mu} T V_{\text{mod}}}{1+\frac{1}{\mu} \xi} .
\end{equation}
The mutual information between Alice and Bob (and hence the maximal code rate) is given by

\begin{equation} \label{eq_IAB}
I_{AB} = \frac{\mu}{2} \log_{2} (1+\text{SNR}) = \frac{\mu}{2} \log_{2}  \left( 1 + \frac{\frac{1}{\mu} T V_{\text{mod}}}{1+\frac{1}{\mu} \xi} \right) .
\end{equation}
As the equation indicates, measuring both quadratures at the same time seemingly doubles the mutual information: $\log_{2}(1+\text{SNR})$ instead of $\log_{2}(1+\text{SNR})/2$. This advantage is, however, compromised by the fact that the SNR is decreased by roughly a factor of 1/2 (not exactly 1/2 since also the excess noise $\xi$ will be halved by the heterodyning beamsplitter, as shown in \eqref{eq_SNR}).

\section{Relation to Normalised SNR}

Say, $n$ raw bits are needed to transmit $k$ information bits. Then $k < n$ and the fraction of the two defines the code rate:

\begin{equation}
R=\frac{k}{n}.
\end{equation}
So when $E_{r}$ and $E_{b}$ are the energies required in order to transmit a raw bit and information bit respectively, then they two are related to each other by the code rate as well:

\begin{equation}
E_{r}=R E_{b}.
\end{equation}
Rewriting the SNR in terms of energy per bit yields

\begin{equation}
\text{SNR}=\frac{P_{S}}{P_{N}} = \frac{E_{r}/\tau}{N_{0}/2\cdot B} = 2 \frac{E_{r}}{N_{0}} = 2R \frac{E_{b}}{N_{0}},
\end{equation}
where $N_{0}$ is the spectral noise density, $\tau$ is the pulse duration of signal and LO, $B$ is the frequency bandwidth and we assumed $\tau B \approx 1$. Using the above relation and \eqref{eq_IAB}, we obtain an implicit function for the code rate:

\begin{equation}
R = \beta I_{AB} = \beta \frac{\mu}{2} \log_{2}  \left( 1 + 2 R \frac{E_{b}}{N_{0}} \right).
\end{equation}
Here $0\le \beta<1$ is the efficiency that the error correcting code with rate $R$ achieves for the mutual information $I_{AB}$.

\chapter{Holevo Information} \label{ch_holevo}

The upper bound on an eavesdropper's information, the Holevo information $\chi_{EB}$, (assuming reverse reconciliation) reads \cite{weedbrook2012gaussian}

\begin{equation}
\chi_{EB} = S_{E} - S_{E|B},
\end{equation}
where $S_{E}$ is the von Neumann entropy \eqref{eq_vonneumann} of the state accessible to Eve for collective measurement and $S_{E|B}$ is the von Neumann entropy of the same state after a projective (homodyne or heterodyne) measurement has been performed by Bob. The entropies $S_{E}$ and $S_{E|B}$ are determined by the symplectic eigenvalues\footnote{See Appendix~\ref{sec_vonneumann} for a definition of the von Neumann entropy and symplectic eigenvalues.} of the covariance matrices describing these states. The actual quantum state from which $S_{E}$ are $S_{E|B}$ derived depends on which kind of eavesdropping attack is assumed. For a simplified security analysis it is very convenient to assume that Eve holds a purification of Alice's and Bob's mutual state $\rho_{AB}$. Fortunately there don't need to be made any kinds of assumptions on \emph{what purification} exactly Eve holds. This is due to the freedom-in-purifications theorem~\cite{nielsen2010quantum} which states that for any two ancillary states $\rho_{E_{1}}$ and $\rho_{E_{2}}$ which purify $\rho_{AB}$ such that $\rho_{ABE_{1}}$ and $\rho_{ABE_{2}}$ are both pure, there exists a unitary transformation $U_{E_{1}}$ acting on $\rho_{E_{1}}$ which transforms $\rho_{ABE_{1}}$ into $\rho_{ABE_{2}}$:

\begin{equation}
(\mathbb{1}_{AB} \otimes U_{E_{1}}) \rho_{ABE_{1}} (\mathbb{1}_{AB} \otimes  U_{E_{1}}^{\dagger}) = \rho_{ABE_{2}} .
\end{equation}
Since the von Neumann entropy is invariant under unitary transformations:

\begin{equation}
S(\rho)=S(U \rho U_{\dagger}),
\end{equation}
any purification of $\rho_{AB}$ that Eve may possess will result in the same entropy and hence the same Holevo information $\chi_{EB}$.

The next section will discuss the entangling cloner attack which represents a concrete example of how Eve can use a purification of $\rho_{AB}$ in order to gain information on the key. Section~\ref{sec_purification} presents a convenient way to compute the Holevo bound $\chi_{EB}$ regardless of the actual attack carried out by Eve. Although computed from entirely different covariance matrices, the resulting $\chi_{EB}$ in the next two sections will -- due to the freedom in purifications -- turn out exactly the same.

\section{Entangling Cloner Attack}

This attack is nicely described in \cite{weedbrook2012continuous}. Suppose Alice generates a two-mode squeezed vacuum state with variance $V$ and covariance matrix

\begin{equation}
\Sigma_{AB} =
\begin{pmatrix}
V \mathbb{1}_{2} & \sqrt{V^{2}-1} \sigma_{z} \\ 
\sqrt{V^{2}-1} \sigma_{z} & V \mathbb{1}_{2}
\end{pmatrix}.
\end{equation}
She keeps one mode to perform heterodyne detection on it (which is equivalent to preparing a coherent state) and sends the other one to Bob through a Gaussian channel. Eve generates a TMSVS herself with variance $W$ and covariance matrix

\begin{equation}
\Sigma_{E_{1}E_{2}} =
\begin{pmatrix}
W \mathbb{1}_{2} & \sqrt{W^{2}-1} \sigma_{z} \\ 
\sqrt{W^{2}-1} \sigma_{z} & W \mathbb{1}_{2}
\end{pmatrix},
\end{equation}
so the overall covariance matrix reads

\begin{equation} \label{eq_SigmaABEE_0}
\Sigma_{A B E_{1} E_{2}} =
\begin{pmatrix}
V \mathbb{1}_{2} & \sqrt{V^{2}-1} \sigma_{z} & 0 & 0 \\ 
\sqrt{V^{2}-1} \sigma_{z} & V \mathbb{1}_{2} & 0 & 0 \\
0 & 0 & W \mathbb{1}_{2} & \sqrt{W^{2}-1} \sigma_{z} \\ 
0 & 0 & \sqrt{W^{2}-1} \sigma_{z} & W \mathbb{1}_{2}
\end{pmatrix} .
\end{equation}
Now suppose Eve replaces the quantum channel with a lossless channel in which she inserts a beamsplitter with transmission $T$. The beamsplitter mixes Bobs mode $B$ with one of Eve's modes $E_{1}$. Eve keeps one of the outputs to herself and passes the other one to Bob. The symplectic beamsplitter operator BS acting on the modes $B$ and $E_{1}$ of the state \eqref{eq_SigmaABEE_0} is represented as\footnote{See Appendix~\ref{sec_symplop} for more details on the transformation laws of Gaussian states.}

\begin{equation}
\text{BS}_{B E_{1}} =
\begin{pmatrix}
\mathbb{1}_{2} & 0 & 0 & 0 \\ 
0 & \sqrt{T} \mathbb{1}_{2} & \sqrt{1-T} \mathbb{1}_{2} & 0 \\
0 & -\sqrt{1-T} \mathbb{1}_{2} & \sqrt{T} \mathbb{1}_{2} &0 \\ 
0 & 0 & 0 & \mathbb{1}_{2}
\end{pmatrix} .
\end{equation}
Action of the beamsplitter operator on the total state shared between Alice, Bob and Eve yields

\begin{align} \label{eq_entclon_total}
\Sigma_{A B E_{1} E_{2}}' & = \text{BS}_{B E_{1}} \Sigma_{A B E_{1} E_{2}} \text{BS}_{B E_{1}}^{T} \notag \\
& \hspace{-60pt} =
\left( \begin{array}{cccc}
 V\mathbb{1}_{2} & \sqrt{T} \sqrt{V^2-1} \sigma_{z} & -\sqrt{1-T} \sqrt{V^2-1} \sigma_{z} & 0 \\
 \sqrt{T} \sqrt{V^2-1} \sigma_{z} & (T V+[1-T] W)\mathbb{1}_{2} & \sqrt{T(1-T)} (W-V) \mathbb{1}_{2} & \sqrt{1-T} \sqrt{W^2-1} \sigma_{z} \\
 -\sqrt{1-T} \sqrt{V^2-1} \sigma_{z} & \sqrt{T(1-T)} (W-V) \mathbb{1}_{2}  & ([1-T] V+T W)\mathbb{1}_{2} & \sqrt{T} \sqrt{W^2-1} \sigma_{z} \\
 0 & \sqrt{1-T} \sqrt{W^2-1} \sigma_{z} & \sqrt{T} \sqrt{W^2-1} \sigma_{z} & W\mathbb{1}_{2} \\
\end{array}
\right) .
\end{align}
Given the above total state, Alice's and Bob's substate reads

\begin{equation}
\Sigma_{AB}' =
\begin{pmatrix}
V \mathbb{1}_{2} & \sqrt{T(V^{2}-1)} \sigma_{z} \\ 
\sqrt{T(V^{2}-1)} \sigma_{z} & (T V+[1-T] W) \mathbb{1}_{2}
\end{pmatrix}.
\end{equation}
When Eve chooses the variance of her TMSVS to be

\begin{equation}
W=\frac{\xi}{1-T}+1
\end{equation}
we obtain the covariance matrix

\begin{equation}
\Sigma_{AB}' =
\begin{pmatrix}
V \mathbb{1}_{2} & \sqrt{T(V^{2}-1)} \sigma_{z} \\ 
\sqrt{T(V^{2}-1)} \sigma_{z} & (T [V-1]+1+\xi) \mathbb{1}_{2}
\end{pmatrix} ,
\end{equation}
which coincides exactly with \eqref{eq_covmathom}, a covariance matrix that represents a transmission through a lossy channel and excess noise $\xi$ \emph{without} the presence of an eavesdropper. In order to obtain Eve's information we compute the symplectic eigenvalues of the covariance matrix describing her substate of \eqref{eq_entclon_total}:

\begin{equation} \label{eq_entclonE}
\Sigma_{E_{1}E_{2}}' =
\left(
\begin{array}{cc}
 ([1-T] V+T W ) \mathbb{1}_{2}& \sqrt{T(W^2-1)} \sigma_{z} \\
 \sqrt{T(W^2-1)} \sigma_{z} & W \mathbb{1}_{2} \\
\end{array}
\right) .
\end{equation}
When a matrix is of the form

\begin{equation}
\begin{pmatrix}
 a \mathbb{1}_{2} & c \sigma_{z} \\
 c \sigma_{z} & b \mathbb{1}_{2} \\
\end{pmatrix} ,
\end{equation}
its symplectic eigenvalues can be obtained by \cite{weedbrook2012gaussian}

\begin{equation} \label{eq_symplEV1}
\nu_{1,2}=\frac{1}{2} \left(z \pm [b-a] \right)
\end{equation}
with

\begin{equation} \label{eq_symplEV2}
z=\sqrt{(a+b)^{2}-4c^{2}} .
\end{equation}
These two eigenvalues can be used to calculate the entropy $S_{E}$. In order to obtain the conditional entropy $S_{E|B}$ based on Bob's measurement, we need to find the symplectic eigenvalues of Eve's covariance matrix after Bob has performed a (homodyne or heterodyne) measurement on his mode.

\subsection{Homodyne Detection}

The covariance matrix shared between Eve and Bob reads

\begin{align}
\Sigma_{B E_{1} E_{2}}'= 
\bordermatrix{ 
  & B & E_{1} & E_{2} \cr
B & (T V+[1-T] W)\mathbb{1}_{2} & \sqrt{T(1-T)} (W-V) \mathbb{1}_{2} & \sqrt{1-T} \sqrt{W^2-1} \sigma_{z} \cr
E_{1} & \sqrt{T(1-T)} (W-V) \mathbb{1}_{2}  & ([1-T] V+T W)\mathbb{1}_{2} & \sqrt{T} \sqrt{W^2-1} \sigma_{z} \cr
E_{2} & \sqrt{1-T} \sqrt{W^2-1} \sigma_{z} & \sqrt{T} \sqrt{W^2-1} \sigma_{z} & W\mathbb{1}_{2} \cr
} .
\end{align}
According to \eqref{eq_partialhom}, a homodyne measurement of Bob's mode will transform Eve's substate to\footnote{See Appendix~\ref{sec_partial} for detailed considerations on partial measurements on $N$-mode Gaussian states.}

\begin{equation}
\Sigma_{E|B} = \Sigma_{E} - \frac{1}{V_{B}} \Sigma_{C} \Pi  \Sigma_{C}^{T} ,
\end{equation}
where $\Sigma_{E}$ is given by \eqref{eq_entclonE}, $V_{B}=T V+[1-T] W = T(V-1) + 1 +\xi$ and

\begin{equation}
\Sigma_{C} = 
\begin{pmatrix}
\sqrt{T(1-T)} (W-V) \mathbb{1}_{2} \\
\sqrt{1-T} \sqrt{W^2-1} \sigma_{z}
\end{pmatrix}
\eqcolon
\begin{pmatrix}
X \mathbb{1}_{2} \\
Y \sigma_{z}
\end{pmatrix} .
\end{equation}
So we obtain

\begin{align}
\Sigma_{E|B} & = \Sigma_{E} - \frac{1}{V_{B}}
\begin{pmatrix}
X^{2} & 0 & XY & 0 \\
0 & 0 & 0 & 0 \\
XY & 0 & Y^{2} & 0 \\
0 & 0 & 0 & 0 
\end{pmatrix} \notag \\
& = 
\left(
\begin{array}{cccc}
 \frac{V W}{T (V-W)+W} & 0 & \frac{\sqrt{T} V \sqrt{W^2-1}}{T (V-W)+W} & 0 \\
 0 & -T V+V+T W & 0 & -\sqrt{T} \sqrt{W^2-1} \\
 \frac{\sqrt{T} V \sqrt{W^2-1}}{T (V-W)+W} & 0 & \frac{V W T-T+1}{T V-T W+W} & 0 \\
 0 & -\sqrt{T} \sqrt{W^2-1} & 0 & W \\
\end{array}
\right) .
\end{align}
The symplectic eigenvalues of this matrix can be obtained by finding the eigenvalues of $\tilde{\Sigma}_{E|B}=i\Omega\Sigma_{E|B}$ (where $\Omega$ is defined as in \eqref{eq_Omega}) and taking their modulus.

\subsection{Heterodyne Detection}

When Bob performs heterodyne measurement on his mode, Eve's state transforms according to \eqref{eq_partialhet}

\begin{equation}
\Sigma_{E|B} = \Sigma_{E} - \frac{1}{V_{B}+1} \Sigma_{C} \Sigma_{C}^{T} ,
\end{equation}
where $\Sigma_{E}$, $V_{B}$ and $\Sigma_{C}$ are defined as in the previous section. Straightforward calculation yields the matrix

\begin{align}
\Sigma_{E|B} & =
\left(
\begin{array}{cccc}
 \frac{-T V+W V+V+T W}{T V-T W+W+1} & 0 & \frac{\sqrt{T} (V+1) \sqrt{W^2-1}}{T (V-W)+W+1} & 0 \\
 0 & \frac{-T V+W V+V+T W}{T V-T W+W+1} & 0 & -\frac{\sqrt{T} (V+1) \sqrt{W^2-1}}{T (V-W)+W+1} \\
 \frac{\sqrt{T} (V+1) \sqrt{W^2-1}}{T (V-W)+W+1} & 0 & \frac{V W T-T+W+1}{T V-T W+W+1} & 0 \\
 0 & -\frac{\sqrt{T} (V+1) \sqrt{W^2-1}}{T (V-W)+W+1} & 0 & \frac{V W T-T+W+1}{T V-T W+W+1} \\
\end{array}
\right) \notag \\
& \eqcolon
\begin{pmatrix}
 a \mathbb{1}_{2} & c \sigma_{z} \\
 c \sigma_{z} & b \mathbb{1}_{2} \\
\end{pmatrix} ,
\end{align}
whose symplectic eigenvalues can be computed using \eqref{eq_symplEV1} and \eqref{eq_symplEV2}.

\section{Universal Analysis of Purification Attacks} \label{sec_purification}

Making no assumptions on how a specific attack might be carried out concretely, all we presume is Eve holding a purification of Alice's and Bob's shared quantum state $\rho_{AB}$. This means that the total state $\rho_{\text{tot}}=\rho_{ABE}$ can be represented as a pure state

\begin{equation}
\rho_{\text{tot}} = \ket{\psi} \bra{\psi} ,
\end{equation}
which, when Eve's subspace is traced out, will deliver Alice's and Bob's mixed state:

\begin{equation}
\Tr_{E} (\rho_{\text{tot}}) = \rho_{AB} .
\end{equation}
The von Neumann entropy of a mixed state $\rho=\sum_{i} p_{i} \ket{i}\bra{i}$ is defined as \cite{nielsen2010quantum}

\begin{equation}
S=-\sum_{i} p_{i} \log(p_{i}).
\end{equation}
We consider $\ket{\psi}$ a bipartite state with Alice's and Bob's joint substate making up one part and Eve's one the other. Any pure bipartite state can be written in terms of a Schmidt decomposition, hence

\begin{equation}
\ket{\psi} = \sum_{i} \sqrt{\lambda_{i}} \ket{i}_{AB} \otimes \ket{i}_{E} .
\end{equation}
Tracing out either of the subsystems then yields

\begin{subequations}
\begin{align}
& \Tr_{E}  = \rho_{AB} = \sum_{i} \lambda_{i} \ket{i}_{AB}\bra{i}_{AB}, \\
& \Tr_{AB} = \rho_{E}  = \sum_{i} \lambda_{i} \ket{i}_{E} \bra{i}_{E} .
\end{align}
\end{subequations}
The von Neumann entropy depends only on the components $\lambda_{i}$ which are -- thanks to the Schmidt representation of $\ket{\psi}$ -- coinciding for $\rho_{AB}$ and $\rho_{E}$. Therefore the assumption that Eve holds a purification of $\rho_{AB}$ leads to the fact that her von Neumann entropy coincides with the one shared by Alice and Bob:

\begin{equation}
S_{E}=S_{AB}=-\sum_{i} \lambda_{i} \log{\lambda_{i}} ,
\end{equation}
which drastically simplifies the computation of the Holevo information:

\begin{align}
\chi_{EB} & = S_{E} - S_{E|B} \notag \\
& = S_{AB} - S_{A|B} .
\end{align}
The entropy $S_{AB}$ can be obtained by \eqref{eq_vonneumann}, using Alice's and Bob's covariance matrix \eqref{eq_covmathom} and its symplectic eigenvalues $\nu_{1}$ and $\nu_{2}$, obtained by \eqref{eq_symplEV1} and \eqref{eq_symplEV2}. However, similar to the case of an entangling cloner attack, the computation of $S_{A|B}$ is a bit more involved and depends on whether Bob performs a homodyne or heterodyne measurement.

\subsection{Homodyne Detection}

Using the covariance matrix

\begin{equation} \label{eq_purificcov}
\Sigma_{AB} =
\begin{pmatrix}
V \mathbb{1}_{2} & \sqrt{T(V^{2}-1)} \sigma_{z} \\ 
\sqrt{T(V^{2}-1)} \sigma_{z} & (T [V-1]+1+\xi) \mathbb{1}_{2}
\end{pmatrix} \eqcolon
\begin{pmatrix}
a \mathbb{1}_{2} & c \sigma_{z} \\ 
c \sigma_{z} & b \mathbb{1}_{2}
\end{pmatrix}
\end{equation}
and the relation describing the effect of a partial homodyne measurement \eqref{eq_partialhom}

\begin{equation}
\Sigma_{A|B} = \Sigma_{A} - \frac{1}{V(q_{B})} \Sigma_{C} \Pi_{q}  \Sigma_{C}^{T} ,
\end{equation}
where $\Sigma_{A}=a \mathbb{1}_{2}$, $\Sigma_{C}=c\sigma_{z}$ and $V(q_{B})=V_{B}=b$, we obtain

\begin{align}
\Sigma_{A|B} =
\begin{pmatrix}
a & 0 \\
0 & a
\end{pmatrix}
-\frac{1}{b}
\begin{pmatrix}
c^{2} & 0 \\
0 & 0
\end{pmatrix} =
\begin{pmatrix}
a-\frac{c^{2}}{b} & 0 \\
0 & a
\end{pmatrix} .
\end{align}
Again, the symplectic eigenvalues of $\Sigma_{A|B}$ are given by the modulus of the eigenvalues of $\tilde{\Sigma}_{A|B}=i\Omega\Sigma_{A|B}$. We obtain

\begin{equation}
\tilde{\Sigma}_{A|B} = 
i
\begin{pmatrix}
0 & 1 \\
-1 & 0
\end{pmatrix}
\begin{pmatrix}
a-\frac{c^{2}}{b} & 0 \\
0 & a
\end{pmatrix} 
=
i
\begin{pmatrix}
0 & a \\
-a+\frac{c^{2}}{b} & 0
\end{pmatrix} .
\end{equation}
The eigenvalue $\nu_{3}$ of this matrix can be derived obtained in the well-known fashion:

\begin{align}
\det \left( \tilde{\Sigma}_{A|B} - \nu_{3} \mathbb{1}_{2} \right) & = \det
\begin{pmatrix}
-\nu_{3} & ia \\
i \left( -a+\frac{c^{2}}{b} \right) & -\nu_{3}
\end{pmatrix}
\notag \\
& =  \nu_{3}^{2} + a \left( -a+\frac{c^{2}}{b} \right) \stackrel{!}{=} 0 .
\end{align}
And therefore

\begin{equation}
\nu_{3} = \sqrt{a \left(a-\frac{c^{2}}{b} \right)} .
\end{equation}

\subsection{Heterodyne Detection}

In case of heterodyne detection we use \eqref{eq_partialhet}

\begin{equation}
\Sigma_{A|B} = \Sigma_{A} - \frac{1}{V_{B}+1} \Sigma_{C} \Sigma_{C}^{T} ,
\end{equation}
where again $\Sigma_{A}=a \mathbb{1}_{2}$, $\Sigma_{C}=c\sigma_{z}$, $V_{B}=b$ and $a$, $b$ and $c$ are defined as in \eqref{eq_purificcov}. Using these definitions we obtain

\begin{align}
\Sigma_{A|B} = 
\begin{pmatrix}
a & 0 \\
0 & a
\end{pmatrix}
- \frac{1}{b+1}
\begin{pmatrix}
c^{2} & 0 \\
0 & c^{2}
\end{pmatrix}
= \left( a-\frac{c^{2}}{b+1} \right) \mathbb{1}_{2}.
\end{align}
The eigenvalue is computed in analogous fashion to the previous section and yields

\begin{equation}
\nu_{3} = a-\frac{c^{2}}{b+1} .
\end{equation}

\section{Strict and Loose Security Assumptions}

Consider the assumption that an eavesdropper might have access to the quantum channel but \emph{not} to the instruments in Bob's lab (``trusted-device scenario'', see~\cite{usenko2016trusted} for a detailed discussion). This would be beneficial for the key rate since not the entire noise and transmission are attributed to Eve. For example, detection noise and quantisation noise, as well as the detection efficiency might not contribute to the Holevo information (although they of course still influence Bob's measurement results and therefore the SNR and mutual information $I_{AB}$). Decomposing $\xi$ and $T$ into constituents associated to the channel and the receiver yields

\begin{subequations}
\begin{align}
\xi & =\xi_{\text{ch}}+\xi_{\text{rec}}, \\
T & =T_{\text{ch}} \cdot T_{\text{rec}},
\end{align}
\end{subequations}
where $\xi_{\text{rec}}$ and $T_{\text{rec}}$ might be composed as follows:

\begin{subequations}
\begin{align}
\xi_{\text{rec}} & = \xi_{\text{det}} + \xi_{\text{CMRR}} + \xi_{\text{ADC}}, \\
T_{\text{rec}} & = \eta_{\text{det}} \cdot \eta_{\text{coup}}.
\end{align}
\end{subequations}
This allows us to rewrite the covariance matrix \eqref{eq_purificcov} as

\begin{equation}
\Sigma_{AB}^{\text{strict}} =
\begin{pmatrix}
V \mathbb{1}_{2} & \sqrt{T_{\text{ch}} \cdot T_{\text{rec}}(V^{2}-1)} \sigma_{z} \\ 
\sqrt{T_{\text{ch}} \cdot T_{\text{rec}}(V^{2}-1)} \sigma_{z} & (T_{\text{ch}} \cdot T_{\text{rec}} [V-1]+1+\xi_{\text{ch}}+\xi_{\text{rec}}) \mathbb{1}_{2}
\end{pmatrix} .
\end{equation}
This matrix will be used to compute $\chi_{EB}$ under the strict assumptions that Eve has access not only to the channel but also to Bob's apparatus. Therefore, the entire noise and transmittance are attributed to Eve. However, under the trusted-device scenario, the covariance matrix would be modified to

\begin{equation}
\Sigma_{AB}^{\text{loose}} =
\begin{pmatrix}
V \mathbb{1}_{2} & \sqrt{T_{\text{ch}}(V^{2}-1)} \sigma_{z} \\ 
\sqrt{T_{\text{ch}} (V^{2}-1)} \sigma_{z} & (T_{\text{ch}} [V-1]+1+\xi_{\text{ch}}) \mathbb{1}_{2}
\end{pmatrix} .
\end{equation}
It is easy to show that this representation of the covariance matrix under loose assumptions is equivalent to the one often used in CV-QKD literature \cite{lodewyck2007quantum}:

\begin{align} \label{eq_varBliterature}
\Sigma_{AB}^{\text{loose}} & =
\begin{pmatrix}
V \mathbb{1}_{2} & \sqrt{T_{\text{ch}}(V^{2}-1)} \sigma_{z} \\ 
\sqrt{T_{\text{ch}} (V^{2}-1)} \sigma_{z} & (T_{\text{ch}} [V+\Xi_{\text{ch}}]) \mathbb{1}_{2}
\end{pmatrix} ,
\end{align}
which uses the notation introduced in Section~\ref{ch_transmittancenoise}:

\begin{subequations}
\begin{align}
\Xi & =\Xi_{\text{ch}}+\frac{1}{T_{\text{ch}}}\Xi_{\text{det}}, \\ 
\Xi_{\text{ch}} & = \frac{1-T_{\text{ch}}}{T_{\text{ch}}}+\xi_{\text{ch},A}, \\
\Xi_{\text{rec}} & = \frac{1-T_{\text{rec}}}{T_{\text{rec}}}+\frac{\xi_{\text{rec}}}{T_{\text{rec}}}.
\end{align}
\end{subequations}
Bob's variance in \eqref{eq_varBliterature} can be multiplied out to obtain

\begin{align}
V_{B}^{\text{loose}} & =T_{\text{ch}} \left( V+\frac{1-T_{\text{ch}}}{T_{\text{ch}}}+\xi_{\text{ch},A} \right) \notag \\
& = T_{\text{ch}} (V-1) + 1 + T_{\text{ch}} \xi_{\text{ch},A} \notag \\
& = T_{\text{ch}} (V-1) + 1 + \xi_{\text{ch}} ,
\end{align}
where we used this article's convention that the parameter $\xi$ refers to the channel output:

\begin{equation}
\xi=\xi_{B}=T\xi_{A} .
\end{equation}
Note that in addition to the trusted-receiver assumption discussed in this section, one might tend to classify the preparation noise as trusted as well. After all, it originates in Alice's lab from her well-known devices. This assumption, however, was shown to open the door to severe side-channel attacks~\cite{derkach2016preventing, derkach2017continuous} and is therefore not further considered in this article.

\chapter{Practical Parameter Estimation} \label{ch_parameterest}

The estimation of the channel parameters is an important task in any QKD protocol. In this work we avoid security problems related to attacks against a co-transmitted local oscillator (see e.g.~\cite{jouguet2013preventing, kunz2015robust}) and concentrate on the estimation procedure for the case of a true or local local oscillator (LLO).

\section{Introduction}

A lower bound to the asymptotic secret-key rate of a CV-QKD protocol, according to \eqref{eq_K} and \eqref{eq_r}, reads:

\begin{equation}
K=f_{\text{sym}}(1-\text{FER})(1-\nu)(\beta I_{AB}-\chi_{EB}),
\end{equation}
where $f_{\text{sym}}$ is the symbol rate, FER represents the frame-error rate, $\nu$ is the key fraction disclosed for parameter estimation, $\beta$ is the post-processing efficiency, $I_{AB}$ represents the mutual information between Alice and Bob (upper-bounded by the channel capacity/Shannon limit) and $\chi_{EB}$ is the Holevo bound, i.e. an upper bound to the mutual information between Eve and Bob 
The mutual information between Alice and Bob $I_{AB}$ depends uniquely on the signal-to-noise ratio, which is given by~\eqref{eq_SNR}

\begin{equation} \label{eq_SNR2}
\text{SNR}=\frac{\frac{1}{\mu}TV_{\text{mod}}}{1+\frac{1}{\mu}\xi},
\end{equation}
where T is the total channel transmittance, $V_{\text{mod}}$ is the modulation variance, $\xi$ is the total excess noise and $\mu=1$ for homodyne and $\mu=2$ for heterodyne detection. The Holevo bound is derived from the covariance matrix~\eqref{eq_covmathom}

\begin{align} \label{eq_SigmaEB2}
\Sigma_{AB}=
\begin{pmatrix}
V_{A} \mathbb{1}_{2} & \Cov(A,B) \sigma_{z} \\
\Cov(A,B) \sigma_{z} & V_{B} \mathbb{1}_{2}
\end{pmatrix} = 
\begin{pmatrix}
(V_{\text{mod}}+1) \mathbb{1}_{2} & \sqrt{T(V_{\text{mod}}^{2}+2V_{\text{mod}})} \sigma_{z} \\
\sqrt{T(V_{\text{mod}}^{2}+2V_{\text{mod}})} \sigma_{z} & (TV_{\text{mod}} + 1 +\xi)\mathbb{1}_{2}
\end{pmatrix},
\end{align}
where $\sigma_{z}$ is the Pauli-$z$ matrix. In order to estimate a secret-key rate from a given measurement of $N$ displaced states in phase space, the SNR and Holevo bound need be obtained from the measurement results, as derived in Section~\ref{sec_covmatestimation}.

However, prior to the actual parameter estimation, the measurement apparatus needs to be calibrated in order for Bob's experimental results to be mapped from voltages to shot-noise units (SNU). A possible procedure for this conversion is presented in the following section.

\section{Calibration of the Measurement Setup}

We describe briefly how to obtain the proper conversion factor $\phi$ which relates experimental measurement data to the language of Gaussian quantum information, hence shot-noise units. In order to keep track of possibly varying calibration parameters, this procedure should be carried out repeatedly during the key-exchange phase.

The total variance of the quadrature operator $\hat{q}$ at the receiver side is, according to \eqref{eq_covmathom} and \eqref{eq_covmathet},

\begin{equation}
V(\hat{q}_{B})=\frac{T}{\mu}V_{\text{mod}}+1+\frac{\xi}{\mu}.
\end{equation}
We divide the excess noise into one part originating from the quantum channel $\xi_{\text{ch}}$ and one part related to the receiver $\xi_{\text{rec}} = \xi_{\text{det}} +\xi_{\text{ADC}}$. With this we reexpress the above equation as

\begin{equation} \label{VB}
V(\hat{q}_{B})=\frac{T}{\mu}V_{\text{mod}}+1+\frac{\xi_{\text{ch}}}{\mu} + \frac{\xi_{\text{rec}}}{\mu}.
\end{equation}
However, what Bob actually measures is a voltage variance:

\begin{equation} \label{VU}
V(U)=\phi V(\hat{q}_{B}),
\end{equation}
where $\phi$ is the conversion factor specific to the measurement setup in units of $\text{V}^{2}/\text{SNU}$. The maximum likelihood estimator (MLE) for the voltage variance is given by

\begin{equation} \label{VU2}
V(U)=\braket{U^{2}}-\braket{U}^{2}=\frac{1}{N}\sum_{i=1}^{N} U_{i}^{2}-\left( \frac{1}{N}\sum_{i=1}^{N} U_{i} \right)^{2},
\end{equation}
where $U_{i}$ are the voltages measured during $N$ samples. In the current framework of an asymptotic analysis we ignore the fact that the actual variance might deviate from the MLE and refer to~\cite{Leverrier2014} for the detailed estimation procedure that involves the calculation of confidence intervals. 

According to our model, the factor $\phi$ can be approximated by

\begin{equation}
\phi\approx P_{\text{LO}} \ \rho^{2} g^{2} B hf,
\end{equation}
where $P_{\text{LO}}$ is the local-oscillator power, $\rho$ is the PIN diodes' responsivity (in A/W), $g$ is the total amplification of the receiver (in V/A), $B$ is the electronic bandwidth (in Hz), $h$ is Planck's constant and $f$ is the optical frequency. However, instead of relying on this model, the factor $\phi$ must be determined experimentally. In order to do so, Bob will disconnect the signal input ($TV_{\text{mod}}=\xi_{\text{ch}}=0$) and measure the quadratures of a vacuum state instead. This reduces \eqref{VB} and \eqref{VU} to

\begin{subequations}
\begin{align}
V(\hat{q}_{B}) & =1+\frac{\xi_{\text{rec}}}{\mu},  \qquad \text{for } TV_{\text{mod}}=\xi_{\text{ch}}=0 \\
V(U) & = \phi+\phi \frac{\xi_{\text{rec}}}{\mu}, \\
& =: \phi+N_{\text{rec}}.  \qquad \text{for } TV_{\text{mod}}=\xi_{\text{ch}}=0  \label{VU3}
\end{align}
\end{subequations}
Note that $\phi$ scales linearly with the local-oscillator power while the receiver noise scales with the inverse $P_{\text{LO}}$, as seen in \eqref{eq_xidet} and \eqref{eq_xiADC}. Therefore the product of the two is constant with respect to $P_{\text{LO}}$:

\begin{align}
\phi & \propto P_{\text{LO}} , \\
\xi_{\text{rec}} & \propto \frac{1}{P_{\text{LO}}} , \\
\frac{\partial N_{\text{rec}}}{\partial P_{\text{LO}}} & = \frac{\partial(\phi \xi_{\text{rec}})}{\partial P_{\text{LO}}} = 0.
\end{align}
So in the case of $P_{\text{LO}}=0$, the conversion factor $\phi$ will become zero but $N_{\text{rec}}$ will stay unaffected since it is independent from $P_{\text{LO}}$. Therefore, when not only the signal input but also the local oscillator is disconnected \eqref{VU3} becomes

\begin{equation}
V(U) = N_{\text{rec}} \qquad \text{for } P_{\text{LO}}=TV_{\text{mod}}=\xi_{\text{ch}}=0 ,
\end{equation}
which can easily be obtained by measurement, using \eqref{VU2}. Now for a given voltage variance $V(U)$, obtained with a given non-zero local-oscillator power $P_{\text{LO}}$, we can rewrite \eqref{VU3} as follows:

\begin{equation}
\phi=V(U) - N_{\text{rec}},
\end{equation}
which is the voltage-square measure of exactly one unit of shot noise (still provided that only a vacuum input is measured, hence $TV_{\text{mod}}=0$). For the subsequent parameter estimation (as described in the following two sections) Bob will divide his the measured voltages, representing $q$ and $p$, by $\sqrt{\phi}$ and any computed voltage variance by $\phi$, so all his data will be given in shot-noise units.

\section{Estimation of the Covariance Matrix} \label{sec_covmatestimation}

The Holevo bound and the characteristic parameters $T$ and $\xi$ are determined by an estimation of the covariance matrix $\Sigma$. This procedure requires for Alice and Bob to reveal a certain fraction of their raw key, say $n$ samples which will later be omitted and not used for the generation of the secure key. In practical CV-QKD Alice prepares a sequence of randomly distributed coherent states which are transmitted to Bob which is referred to as prepare-and-measure (PM) scenario. On the other hand, the entire security analysis is based on the distribution of two-mode squeezed vacuum states, referred to as entanglement-based (EB) picture. Therefore, in order to satisfy the requirements of a sound security analysis, Alice and Bob have to rescale their experimental quadrature values, transferring their prepare-and-measure covariance matrix $\Sigma^{\text{PM}}$ to an entanglement-based version $\Sigma^{\text{EB}}$. The prepare-and-measure covariance matrix, after transmission through a lossy ($T$) and noisy ($\xi$) channel into Bob's lab who performs a homodyne ($\mu=1$) or heterodyne ($\mu=2$) measurement, is represented as~\eqref{eq_covmathet}

\begin{align}
\Sigma^{\text{PM}} = &
\begin{pmatrix}
V_{\text{mod}} \mathbb{1}_{2} & \sqrt{\frac{T}{\mu}}V_{\text{mod}} \mathbb{1}_{2} \\ 
\sqrt{\frac{T}{\mu}} V_{\text{mod}} \mathbb{1}_{2} & \frac{T}{\mu} V_{\text{mod}} + 1 + \frac{\xi}{\mu} \mathbb{1}_{2}
\end{pmatrix} \eqcolon
\begin{pmatrix}
 a^\text{PM} \mathbb{1}_{2} & c^\text{PM} \mathbb{1}_{2} \\
 c^\text{PM} \mathbb{1}_{2} & b^\text{PM} \mathbb{1}_{2} \\
\end{pmatrix} .
\end{align}
On the other hand, the covariance matrix of a TMSVS, after transmission of one mode, looks as follows:

\begin{align}
\Sigma^{\text{EB}} & =
\begin{pmatrix}
 V \mathbb{1}_{2} & \sqrt{T}\sqrt{V^{2}-1} \sigma_{z} \\
 \sqrt{T}\sqrt{V^{2}-1} \sigma_{z} & (T[V-1]+1+\xi) \mathbb{1}_{2} \\
\end{pmatrix} \notag \\
& =
\begin{pmatrix}
 (V_{\text{mod}}+1) \mathbb{1}_{2} & \sqrt{T}\sqrt{V_{\text{mod}}^{2}+2V_{\text{mod}}} \sigma_{z} \\
 \sqrt{T}\sqrt{V_{\text{mod}}^{2}+2V_{\text{mod}}} \sigma_{z} & (TV_{\text{mod}}+1+\xi) \mathbb{1}_{2} \\
\end{pmatrix} \notag \\
& \eqcolon
\begin{pmatrix}
 a^\text{EB} \mathbb{1}_{2} & c^\text{EB} \sigma_{z} \\
 c^\text{EB} \sigma_{z} & b^\text{EB} \mathbb{1}_{2} \\
\end{pmatrix} .
\end{align}
Alice determines the coefficients $a^\text{EB}$, $b^\text{EB}$ and $c^\text{EB}$ which she will use to compute the Holevo bound $\chi_{EB}$ (see Section~\ref{ch_holevo}):

\begin{enumerate}
\item The coefficient $a^\text{EB}$ is easily obtained, since the modulation variance is well known ahead of the raw-key transmission:

\begin{equation}
a^\text{EB}=V_{\text{mod}}+1.
\end{equation}

\item The computation of $b^\text{EB}$ is straightforward as well, since it is distinguished from $b^\text{PM}$ only in the case of heterodyne detection where the factor $\mu = 2$ needs to be taken into account.

\begin{align}
b^\text{EB}=\mu b^\text{PM} - \mu + 1,
\end{align}
where $b^{\text{PM}}$, the variance of Bob's quadratures in SNU, is either announced by Bob or computed by Alice using the $n$ disclosed samples. The above transformation will leave $b^{\text{PM}}$ unchanged in case of homodyne detection ($\mu=1$) and change $b^{\text{PM}}=T V_{\text{mod}}/2 + 1 + \xi/2$ to $b^{\text{EB}}=TV_{\text{mod}}+1+\xi$ in case of heterodyne detection ($\mu=2$).

\item For the computation of $c^\text{EB}$, Alice's and Bob's values need to rescale their values as follows in order to adhere to the entanglement-based picture:

\begin{subequations}
\begin{align}
\hat{q}_{A}^{\text{EB}} & =  \sqrt{\frac{V+1}{V-1}} \hat{q}_{A}^{\text{PM}}= \sqrt{\frac{V_{\text{mod}}+2}{V_{\text{mod}}}} \hat{q}_{A}^{\text{PM}}, \\
\hat{p}_{A}^{\text{EB}} & = -\sqrt{\frac{V+1}{V-1}} \hat{p}_{A}^{\text{PM}}= -\sqrt{\frac{V_{\text{mod}}+2}{V_{\text{mod}}}} \hat{p}_{A}^{\text{PM}}, \\
\hat{q}_{B}^{\text{EB}} & = \sqrt{\mu} \ \hat{q}_{B}^{\text{PM}}, \\
\hat{p}_{B}^{\text{EB}} & = \sqrt{\mu} \ \hat{p}_{B}^{\text{PM}},
\end{align}
\end{subequations}
where the conversion factor $\phi$ had been used to convert Bob's measurements from volts to SNU. This allows us to write

\begin{align}
c^{\text{EB}}=\braket{\hat{q}_{A}^{\text{EB}} \hat{q}_{B}^{\text{EB}}}= \sqrt{\frac{V_{\text{mod}}+2}{V_{\text{mod}}}} \sqrt{\mu} \braket{\hat{q}_{A}^{\text{PM}} \hat{q}_{B}^{\text{PM}}} .
\end{align}
A quick check is sufficient to confirm this result:

\begin{align}
& \sqrt{\frac{V_{\text{mod}}+2}{V_{\text{mod}}}} \sqrt{\mu} \braket{\hat{q}_{A}^{\text{PM}} \hat{q}_{B}^{\text{PM}}} \notag \\
= & \sqrt{\frac{V_{\text{mod}}+2}{V_{\text{mod}}}} \sqrt{\mu} \ c^{\text{PM}} \notag \\
= & \sqrt{\frac{V_{\text{mod}}+2}{V_{\text{mod}}}} \sqrt{\mu} \ \sqrt{\frac{T}{\mu}}V_{\text{mod}} \notag \\
= & \sqrt{T}\sqrt{V_{\text{mod}}^{2}+2V_{\text{mod}}} = c^{\text{EB}} .
\end{align}
So in order to obtain $c^{\text{EB}}$, Alice will use the disclosed fraction of $n$ samples to compute the mean values $\braket{q_{A}^{\text{PM}}q_{B}^{\text{PM}}}$ and $\braket{p_{A}^{\text{PM}}p_{B}^{\text{PM}}}$ and multiply the result with $\sqrt{\mu(V_{\text{mod}}+2) / V_{\text{mod}}}$.

\end{enumerate}
The knowledge of the three coefficients $a^{\text{EB}}$, $b^{\text{EB}}$ and $c^{\text{EB}}$ is sufficient to compute the Holevo information $\chi_{EB}$. However, if required, the parameters $T$ and $\xi$ can quickly be obtained by

\begin{align}
T & =  \frac{ \left( c^{\text{EB}} \right)^{2}}{ V_{\text{mod}}^{2}+2V_{\text{mod}}} \notag \\
& = \mu \left( \frac{ c^{\text{PM}}}{V_{\text{mod}}} \right)^{2}
\end{align}
and

\begin{align}
\xi & =V_{B}-TV_{\text{mod}}-1 \notag \\
& = b^{\text{EB}}-TV_{\text{mod}}-1.
\end{align}
The signal-to-noise ratio (and therefore the mutual information $I_{AB}$) can be either be determined using the parameters $T$ and $\xi$, or alternatively, the total variance and the conditional variance of Bob's data:

\begin{align}
\text{SNR}=\frac{\frac{T}{\mu}V_{\text{mod}}}{1+\frac{\xi}{\mu}}=\frac{V_{B}-(1+\frac{\xi}{\mu})}{1+\frac{\xi}{\mu}}=\frac{V_{B}-V_{B|A}}{V_{B|A}} = \frac{V_{B}}{V_{B|A}}-1,
\end{align}
where $V_{B}=\frac{T}{\mu}V_{\text{mod}}+1+\frac{\xi}{\mu}$ is the total variance of Bob's measurement in one basis and $V_{B|A} = V(\hat{q}_{B}|A)=1+\frac{\xi}{\mu}$ is the conditional variance, i.e. the variance of Bob's measurements conditioned on Alice's data. It is equivalent to the variance of the difference $\sqrt{T/ \mu} \ \hat{q}_{A}-\hat{q}_{B}$:

\begin{align}
V_{B|A} & = \left\langle \left( \sqrt{T/ \mu} \ \hat{q}_{A}-\hat{q}_{B} \right)^{2} \right\rangle - \left\langle \sqrt{T/ \mu} \ \hat{q}_{A}-\hat{q}_{B} \right\rangle^{2} \notag \\
& =  \left\langle \left( \sqrt{T/ \mu} \ \hat{q}_{A}-\hat{q}_{B} \right)^{2} \right\rangle - \left( \sqrt{T/ \mu} \ \left\langle \hat{q}_{A} \right\rangle - \left\langle \hat{q}_{B} \right\rangle \right)^{2} \notag \\
& = \left\langle \left( \sqrt{T/ \mu} \ \hat{q}_{A}-\hat{q}_{B} \right)^{2} \right\rangle \notag \\
& = \frac{1}{n} \sum_{i=1}^{n} \left( \sqrt{T/ \mu} \ \hat{q}_{A,i}-\hat{q}_{B,i} \right)^{2} ,
\end{align}
where we used $\braket{\hat{q}_{A}} = \braket{\hat{q}_{B}} = 0$. Note that in the case of discrete modulation (e.g. QPSK or 16-QAM, as opposed to continuous Gaussian modulation reviewed in this article) the conditional variance is easier to obtain since it corresponds to the variance of the subset of samples that are associated to one and the same (discrete) modulation symbol.

\chapter{Noise Models} \label{ch_noise}

In addition to the obligatory shot-noise, various experimental imperfections will contribute to the total noise and undermine the system's performance.\footnote{Note that the augmentation of excess noise is not the only way for experimental imperfections to manifest. For example, when the transmitted quantum states are generated digitally (as it is usually the case in the prepare-and-measure scheme) they deviate from a truly continuous Gaussian distribution due to the finite bit resolution of any digital-to-analog converter. The effects of the discretised Gaussian modulation go beyond the mere occurrence of quantisation noise. In fact, since the standard security analysis of CV-QKD assumes continuous Gaussian states, this imperfection has to be taken into account on a more fundamental level~\cite{leverrier2011continuous, jouguet2012analysis}.} We refer to this additional noise as excess noise $\xi$, describing a variance of the quadrature operators and given in shot-noise units. The constituents of $\xi$ may originate from noisy detection, from a noisy preparation~\cite{filip2008continuous, usenko2010feasibility, weedbrook2010quantum} due to intensity fluctuations of the used lasers or imperfect modulation, from quantisation of the measurement results, from Raman scattering caused by classical channels in the fibre and others. We assume all these noise sources to be stochastically independent which makes the variances they cause to the quadratures additive:

\begin{equation}
\xi=\xi_{\text{det}} + \xi_{\text{RIN}} + \xi_{\text{quant}} + \xi_{\text{Ram}} + \dots
\end{equation}
In the present chapter we are providing mathematical models for some of the possible noise components. These models may serve as guidance for the design of actual CV-QKD implementations as they provide estimates of the impact that particular hardware choices (e.g.\ lasers, modulators, detectors, analog-to-digital converters, etc.) on the total noise and, consequently, the expected secure-key rate. Section~\ref{ch_experimentalimplications} discusses some practical conclusions and identifies the most crucial hardware specifications based on the noise models derived in this chapter.

\section{Relative Intensity Noise -- Signal}

Power fluctuations of a laser are quantified by the so-called relative intensity noise (RIN), which is given by

\begin{equation}
\text{RIN}=\frac{(\delta p)^{2}}{\braket{P}^{2}} ,
\end{equation} 
where $\braket{P}$ represents the average optical power and $(\delta p)^{2}$ the spectral density of the power-fluctuation, given in units of $[\text{W}^{2}/\text{Hz}]$. Therefore, the absolute power variance, given a bandwidth $B$, can be written as

\begin{equation} \label{VP_RIN}
V(P)=(\delta P)^{2}=(\delta p)^{2} B=\text{RIN}{\braket{P}^{2}} B.
\end{equation}
In order to obtain a measure for how the RIN of the signal laser contributes to the excess noise, we first investigate in which way a certain laser-power variance $V(P)$ influences the variance of the quadrature operator $V(\hat{q})$:

\begin{align}
V(P) & = V\left( \frac{hf}{\tau}\braket{n} \right) = \frac{h^{2}f^{2}}{\tau^{2}} V\left( \braket{n} \right) = \frac{h^{2}f^{2}}{\tau^{2}} V(\hat{n})= \frac{h^{2}f^{2}}{\tau^{2}} V(\hat{a}^{\dagger}\hat{a}) \notag \\
& = \frac{h^{2}f^{2}}{16 \tau^{2}} V((\hat{q}-i\hat{p})(\hat{q}+i\hat{p})) \notag \\
& = \frac{h^{2}f^{2}}{16 \tau^{2}} V(\hat{q}^{2}+\hat{p}^{2}-2) \notag \\
& = \frac{h^{2}f^{2}}{16 \tau^{2}} V(\hat{q}^{2}+\hat{p}^{2}) \notag \\
& = \frac{h^{2}f^{2}}{16 \tau^{2}} \left( V(\hat{q}^{2})+V(\hat{p}^{2}) \right),
\end{align}
where we used \eqref{expect_n} in the second and third line. Assuming variance symmetry of the quadrature operators, $V(\hat{p}^{2})=V(\hat{q}^{2})$, we obtain

\begin{equation}
V(P) = \frac{h^{2}f^{2}}{8 \tau^{2}} V(\hat{q}^{2}).
\end{equation}
If a Gaussian random variable has expected value of zero, then the variance of its square is twice the square of its variance:

\begin{equation}
V(P) = \frac{h^{2}f^{2}}{8 \tau^{2}} 2 V^{2}(\hat{q}),
\end{equation}
which leads to

\begin{equation}
V(\hat{q})=\frac{2\tau}{hf} \sqrt{V(P)} .
\end{equation}
Using \eqref{VP_RIN} we obtain the expression for the quadrature variance caused by the RIN of the signal laser:

\begin{align}
V_{\text{RIN,sig}}(\hat{q}) & = \frac{2\tau}{hf} \sqrt{ \text{RIN}_{\text{sig}} \braket{P_{\text{sig}}}^{2} B } \notag \\
& = \frac{2\tau}{hf} \braket{P_{\text{sig}}} \sqrt{ \text{RIN}_{\text{sig}} B } \notag \\
& = \frac{2\tau}{hf} \frac{hf \braket{n_{\text{sig}}}}{\tau} \sqrt{ \text{RIN}_{\text{sig}} B } \notag \\
& = 2 \braket{n_{\text{sig}}} \sqrt{ \text{RIN}_{\text{sig}} B } .
\end{align}
The above expression describes the quadrature variance directly at the transmitter, hence at the channel input. However, when the excess noise is defined with respect to the channel output (as in this article), it needs to be multiplied by a factor $T$, the total channel transmittance. Moreover, according to \eqref{nvsVmod}, the mean photon number $\braket{n_{\text{sig}}}$ is $V_{\text{mod}}/2$, which yields the final result:

\begin{equation}
\boxed{
\xi_{\text{RIN,sig}}= TV_{\text{mod}} \sqrt{\text{RIN}_{\text{sig}} B } .
}
\end{equation}

\section{Relative Intensity Noise -- Local Oscillator}

When a signal beam is mixed with a local oscillator at a balanced beamsplitter the difference of the number operators at both output ports is proportional to the local-oscillator amplitude times the signal's quadrature components \eqref{homodynedetectionlaw}:

\begin{equation}
\Delta \hat{n} = | \alpha_{\text{LO}} | (\hat{q}\cos \theta +\hat{p} \sin \theta ).
\end{equation}
Obviously, a power fluctuation of the local oscillator will introduce an additional contribution to the excess noise. For simplicity we assume $\theta=0$, hence

\begin{equation}
\Delta \hat{n} = | \alpha_{\text{LO}} | \hat{q} .
\end{equation}
The variance of $\Delta \hat{n}$ is the variance of a product of two independent random variables which can be written as

\begin{align} \label{RINLO1}
V(\Delta \hat{n}) & = V(|\alpha_{\text{LO}}| \hat{q}) \notag \\
& = \braket{|\alpha_{\text{LO}}|^{2}} \braket{\hat{q}^{2}} - \braket{|\alpha_{\text{LO}}|}^{2} \braket{\hat{q}}^{2} \notag \\
& = \braket{|\alpha_{\text{LO}}|^{2}} \braket{\hat{q}^{2}} \notag \\
& \eqcolon \braket{|\alpha_{\text{LO}}|^{2}} V_{\neg \text{RIN,LO}}(\hat{q}) \notag \\
& = \left( V(|\alpha_{\text{LO}}|) + \braket{|\alpha_{\text{LO}}|}^{2} \right) V_{\neg \text{RIN,LO}}(\hat{q}) ,
\end{align}
where we used $\braket{\hat{q}}=0$ and the subscript $\neg \text{RIN,LO}$ indicates that $V_{\neg \text{RIN,LO}}(\hat{q})$ describes the variance of $\hat{q}$ without taking the local-oscillator's RIN into account. In order to find an expression for the variance caused by the LO's RIN we rewrite $V(\Delta \hat{n})$ in a different way, assuming a \emph{constant} LO amplitude $\alpha_{\text{LO}}$ and the RIN noise being already contained in $V(\hat{q})$:

\begin{align} \label{RINLO2}
V(\Delta \hat{n}) & = V(|\alpha_{\text{LO}}| \hat{q}) \notag \\
& = |\alpha_{\text{LO}}|^{2} V_{\text{tot}}(\hat{q}) \notag \\
& = |\alpha_{\text{LO}}|^{2} \left( V_{\neg \text{RIN,LO}}(\hat{q}) + V_{\text{RIN,LO}}(\hat{q}) \right) .
\end{align}
Equalising \eqref{RINLO1} and \eqref{RINLO2} yields

\begin{align}
\left( V(|\alpha_{\text{LO}}|) + \braket{|\alpha_{\text{LO}}|}^{2} \right) V_{\neg \text{RIN,LO}}(\hat{q}) & = |\alpha_{\text{LO}}|^{2} \left( V_{\neg \text{RIN,LO}}(\hat{q}) + V_{\text{RIN,LO}}(\hat{q}) \right) \notag \\
V(|\alpha_{\text{LO}}|) V_{\neg \text{RIN,LO}}(\hat{q}) & = |\alpha_{\text{LO}}|^{2} V_{\text{RIN,LO}}(\hat{q}),
\end{align}
and therefore

\begin{equation} \label{VRINLO1}
V_{\text{RIN,LO}}(\hat{q}) = \frac{V(|\alpha_{\text{LO}}|)}{|\alpha_{\text{LO}}|^{2}} V_{\neg \text{RIN,LO}}(\hat{q}).
\end{equation}
The expression $V(|\alpha_{\text{LO}}|)$ is clearly determined by the RIN. In order to understand the relation between $V(|\alpha_{\text{LO}}|)$ and the RIN, note that the mean photon number is given by the amplitude squared:

\begin{equation}
\braket{n}=|\alpha|^{2}
\end{equation}
with the derivative

\begin{equation}
\frac{\partial \braket{n}}{\partial |\alpha|}=2|\alpha|,
\end{equation}
from which the relation between the deviations and variances of $\braket{n}$ and $|\alpha|$ and follow:

\begin{subequations}
\begin{align}
\delta \braket{n} & = 2|\alpha| \ \delta |\alpha| \\
V(\braket{n}) & = 4 |\alpha|^{2} V(|\alpha|) .
\end{align}
\end{subequations}
Using this result, $V(|\alpha|)$ can be reexpressed as follows:

\begin{align}
V(|\alpha|) & = \frac{1}{4 |\alpha|^{2}} V(\braket{n}) \notag \\
& =\frac{hf}{4 \braket{P}\tau} V(\braket{n}) \notag \\
& =\frac{hf}{4 \braket{P}\tau} V\left( \frac{\tau}{hf} P \right) \notag \\
& =\frac{hf}{4 \braket{P}\tau} \frac{\tau^{2}}{h^{2}f^{2}} V(P) \notag \\
& =\frac{\tau}{4 \braket{P} hf} V(P) \notag \\
& =\frac{\tau}{4 \braket{P} hf} \text{RIN}{\braket{P}^{2}} B \notag \\
& =\frac{\tau}{4 hf} \text{RIN}{\braket{P}} B .
\end{align}
Inserting this result into \eqref{VRINLO1} yields

\begin{align}
V_{\text{RIN,LO}}(\hat{q}) & = \frac{1}{|\alpha_{\text{LO}}|^{2}} \ \frac{\tau}{4 hf} \text{RIN}_{\text{LO}} \ {\braket{P_{\text{LO}}}} \ B \ V_{\neg \text{RIN,LO}}(\hat{q}) \notag \\
& = \frac{hf}{\braket{P_{\text{LO}}} \tau} \ \frac{\tau}{4 hf} \text{RIN}_{\text{LO}} \ {\braket{P_{\text{LO}}}} \ B \ V_{\neg \text{RIN,LO}}(\hat{q}),
\end{align}
which leads to the final expression

\begin{equation}
\boxed{
\xi_{\text{RIN,LO}} = \frac{1}{4} \text{RIN}_{\text{LO}} \ B \ V_{\neg \text{RIN,LO}}(\hat{q}) .
}
\end{equation}

\section{Modulation Noise}

In this section we will model the excess noise caused by Alice's modulation of a coherent state.\footnote{Note that in this section we will only derive the excess noise caused by modulation-voltage noise of the signal generator. Further noise sources such as the quantisation error of the DAC, phase deviations of the $q/p$-modulator, the ripple in the electro-optic modulation and detection responses are not investigated at this point.} We assume the following experimental setup:

\begin{itemize}
\item A signal generator -- say, a digital-to-analog converter (DAC) or, alternatively, an arbitrary-waveform generator (AWG) -- translates a certain signal-bit information into a voltage $U_{\text{DAC}}$ with a specific deviation $\delta U_{\text{DAC}}$.
\item This voltage and its deviation are amplified by a factor of $g$ to the voltage required to drive the $q/p$-modulator. The amplified voltage is

\begin{equation} \label{eq_noisyampU}
U=g(U_{\text{DAC}} + \delta U_{\text{DAC}}).
\end{equation}
\item A pulsed laser emits a coherent state $\ket{\tilde{\alpha}}$ which is subsequently injected into the $q/p$-modulator.
\item Depending on the voltage $U$ applied to the modulator, the quadratures of $\ket{\tilde{\alpha}}$ are adjusted according to Alice's choice of random numbers. The modulator consists of a Mach-Zehnder interferometer (MZI) with another nested MZI in each arm to which the phase rotations $\pm \phi_{1}$ and $\pm \phi_{2}$ are applied respectively. The output state of the modulator is a coherent state $\ket{\tilde{\alpha}'}$ with the eigenvalue\footnote{See Appendix~\ref{ch_modulator} for the principles of a $q/p$-modulator and a derivation of its output state.}

\begin{equation} \label{eq_tildealpha'}
\tilde{\alpha}'=\frac{1}{2} \tilde{\alpha} (\cos \varphi_{1} + i \cos \varphi_{2}).
\end{equation}
The initial input state $\ket{\tilde{\alpha}}$ can be arbitrary. However, for simplicity we assume the eigenvalue to be real-valued: $\tilde{\alpha} \in \mathbb{R}$. Under this assumption the real and imaginary part of $\tilde{\alpha}'$, hence the $q$- and $p$-quadrature are

\begin{subequations}
\begin{align}
\tilde{q}' & =\frac{1}{2} \tilde{\alpha} \cos \varphi_{1}, \\
\tilde{p}' & =\frac{1}{2} \tilde{\alpha} \cos \varphi_{2}.
\end{align}
\end{subequations}
\item Followed by the modulator, an attenuator damps the coherent state by a factor of $\sqrt{t}$ such that its photon number corresponds to the modulation variance $\braket{n}=V_{\text{mod}}/2$. The attenuated state is then

\begin{equation} \label{eq_attenalpha}
\ket{\alpha}= \ket{\sqrt{t} \tilde{\alpha}'}.
\end{equation}
\item Alice sends $\ket{\alpha}$ to Bob who receives the state $\ket{\sqrt{T} \alpha}$, where $T$ is the total transmission (including channel transmission, detection- and coupling efficiency).
\end{itemize}
In order to derive the noise parameter $\xi_{\text{mod}}$ we go back to the expression for the $q$-quadrature of the modulated state $\ket{\tilde{\alpha}'}$:

\begin{equation}
\tilde{q}' =\frac{1}{2} \tilde{\alpha} \cos \varphi_{1} .
\end{equation}
The phase rotation $\varphi_{1}$ is equal to the product of the applied voltage $U$ and some conversion factor $\nu$ which describes in rad/V the amount of phase rotation per applied voltage:

\begin{equation} \label{eq_nu}
    \varphi_{1} = U \nu.
\end{equation}
This conversion factor can be expressed in terms of the entity $U_{\pi}$ which describes the voltage required to achieve a phase rotation of $\pi$. In our setup, where the applied voltage is the amplified DAC voltage $U=gU_{\text{DAC}}$, the quadrature $\tilde{q}'$ can be reexpressed in the following way:

\begin{align}
\tilde{q}' & = \frac{1}{2} \tilde{\alpha} \cos (U \nu) \notag \\
& = \frac{1}{2} \tilde{\alpha} \cos \left( U \frac{\pi}{U_{\pi}} \right) \notag \\
& = \frac{1}{2} \tilde{\alpha} \cos \left( g \ U_{\text{DAC}} \frac{\pi}{U_{\pi}} \right) .
\end{align}
However, when a noisy voltage, amplified by a factor $g$ as shown in \eqref{eq_noisyampU}, is applied the modulated quadrature will become noisy as well:

\begin{align}
\tilde{q}' + \delta \tilde{q}' & =\frac{1}{2} \tilde{\alpha} \cos \left( g[U_{\text{DAC}} + \delta U_{\text{DAC}} ] \frac{\pi}{U_{\pi}} \right) \notag \\
& = \frac{1}{2} \tilde{\alpha} \cos \left( \pi g \ \frac{U_{\text{DAC}}}{U_{\pi}} + \pi g \ \frac{\delta U_{\text{DAC}}}{U_{\pi}} \right).
\end{align}
We define $A \coloneq \pi g \ U_{\text{DAC}} /U_{\pi}$ and $\lambda \coloneq \pi g \ \delta U_{\text{DAC}} / U_{\pi}$ and make a Taylor expansion of the above equation around $\lambda=0$:

\begin{align}
\left( \tilde{q}' + \delta \tilde{q}' \right) (\lambda) & =\frac{1}{2} \tilde{\alpha} \cos \left( A + \lambda \right) \notag \\
& = \frac{1}{2} \tilde{\alpha} \cos A - \frac{\tilde{\alpha}}{2} \lambda \sin A - \frac{\tilde{\alpha}}{4} \lambda^{2} \cos A + \mathcal{O} (\lambda^{3}) .
\end{align}
By cancelling $\tilde{q}'=1/2 \ \tilde{\alpha} \cos A$ on both sides we obtain the expression for the quadrature noise:

\begin{align} \label{eq_quadrDACnoise}
\delta \tilde{q}' (\lambda) & = - \frac{\tilde{\alpha}}{2} \lambda \sin A - \frac{\tilde{\alpha}}{4} \lambda^{2} \cos A + \mathcal{O} (\lambda^{3})
\end{align}
The sine and cosine function obey of course $-1 \leq \sin A, \cos A \leq 1$ for any argument $A$. Therefore we obtain an upper bound for the quadrature modulation noise

\begin{align} \label{eq_upperboundquadrDACnoise}
|\delta \tilde{q}'| & \leq \frac{|\tilde{\alpha}|}{2} \lambda + \frac{|\tilde{\alpha}|}{4} \lambda^{2} \notag \\
& = \frac{|\tilde{\alpha}|}{2} \pi g \ \frac{\delta U_{\text{DAC}}}{U_{\pi}} + \frac{|\tilde{\alpha}|}{4} \left( \pi g \ \frac{\delta U_{\text{DAC}}}{U_{\pi}} \right)^{2} \notag \\
& = \frac{1}{2} |\tilde{\alpha}| \left( \pi g \ \frac{\delta U_{\text{DAC}}}{U_{\pi}} + \frac{1}{2} \left[ \pi g \ \frac{\delta U_{\text{DAC}}}{U_{\pi}} \right]^{2} \right) .
\end{align}
Damping by the attenuator and the channel yields

\begin{align}
|\delta q| & \leq \frac{1}{2} \sqrt{T} \sqrt{t} |\tilde{\alpha}| \left( \pi g \ \frac{\delta U_{\text{DAC}}}{U_{\pi}} + \frac{1}{2} \left[ \pi g \ \frac{\delta U_{\text{DAC}}}{U_{\pi}} \right]^{2} \right) .
\end{align}
The modulation-caused variance of the quadrature is then

\begin{equation}
V_{\xi,\text{mod}} (q) \leq \frac{1}{4} T t |\tilde{\alpha}|^{2} \left( \pi g \ \frac{\delta U_{\text{DAC}}}{U_{\pi}} + \frac{1}{2} \left[ \pi g \ \frac{\delta U_{\text{DAC}}}{U_{\pi}} \right]^{2} \right)^{2} .
\end{equation}
If we compute the square of the modulated state $\tilde{\alpha}'$ \eqref{eq_tildealpha'},

\begin{align}
| \tilde{\alpha}' |^{2} & = \left( \frac{1}{2} \tilde{\alpha} \cos \varphi_{1} \right)^{2} + \left( \frac{1}{2} \tilde{\alpha} \cos \varphi_{2} \right)^{2}  && \hspace{-70pt} \text{for } \tilde{\alpha} \in \mathbb{R} \notag \\
& \leq \frac{1}{2}  \tilde{\alpha} ^{2}  && \hspace{-70pt} \text{for } \varphi_{1},\varphi_{2} \in \{0,\pi \} ,
\end{align}
we see that the square of the input state $\tilde{\alpha}$ and the square of the modulated state $\tilde{\alpha}'$ are related by a factor of 2. Moreover, due to \eqref{eq_attenalpha} we can write  $t | \tilde{\alpha} ' |^{2}=|\alpha|^{2}$, where $\alpha$ represents the state sent from Alice to Bob. Using this and the fact that $|\alpha|^{2} = \braket{n} = V_{\text{mod}} / 2$ we arrive at

\begin{align}
V_{\xi,\text{mod}} (q) & \leq \frac{1}{2} T t |\alpha'|^{2} \left( \pi g \ \frac{\delta U_{\text{DAC}}}{U_{\pi}} + \frac{1}{2} \left[ \pi g \ \frac{\delta U_{\text{DAC}}}{U_{\pi}} \right]^{2} \right)^{2} \notag \\
& = \frac{1}{2} T |\alpha|^{2} \left( \pi g \ \frac{\delta U_{\text{DAC}}}{U_{\pi}} + \frac{1}{2} \left[ \pi g \ \frac{\delta U_{\text{DAC}}}{U_{\pi}} \right]^{2} \right)^{2} \notag \\
& = \frac{1}{4} T V_{\text{mod}} \left( \pi g \ \frac{\delta U_{\text{DAC}}}{U_{\pi}} + \frac{1}{2} \left[ \pi g \ \frac{\delta U_{\text{DAC}}}{U_{\pi}} \right]^{2} \right)^{2} .
\end{align}
The above expression describes the variance of the quadrature \emph{component} $q$. Knowing that $V_{\xi}(\hat{q})=4V_{\xi}(q)$ we obtain the variance of the quadrature \emph{operator} $\hat{q}$:

\begin{equation} \label{eq_modnoise1}
\boxed{
V_{\xi,\text{mod}} (\hat{q}) = \xi_{\text{mod}} \leq T V_{\text{mod}} \left( \pi g \ \frac{\delta U_{\text{DAC}}}{U_{\pi}} + \frac{1}{2} \left[ \pi g \ \frac{\delta U_{\text{DAC}}}{U_{\pi}} \right]^{2} \right)^{2} .
}
\end{equation}
Note that in the case of QPSK modulation this expression simplifies drastically: In this scheme all four points in phase space can be obtained by phase rotations of $\varphi_{1},\varphi_{2} \in \{0,\pi \}$. Therefore the voltage gain factor will adjust the DAC voltage such that it matches $U_{\pi}$, the voltage required for a phase rotation of $\pi$:

\begin{equation}
g=\frac{U_{\pi}}{U_{\text{DAC}}}.
\end{equation}
Inserting into \eqref{eq_modnoise1} yields the final expression for the excess noise caused by modulation voltage noise in the QPSK scheme:\footnote{Note that in the case of an applied phase $\pi$ (hence $g=U_{\pi} /U_{\text{DAC}}$) the variable $A$ in \eqref{eq_quadrDACnoise} will become $\pi$, therefore erasing the term proportional to $\sin A$ in the error $\delta \tilde{q}'$. In this case the error will only depend on the second and higher orders of $\lambda=\pi \ \delta U_{\text{DAC}} / U_{\text{DAC}}$. However, in order to take other modulator models into account (e.g. $\alpha'=1/2 \alpha [\cos \varphi_{1} + i \sin \varphi_{2}]$ instead of $\alpha'=1/2 \alpha [\cos \varphi_{1} + i \cos \varphi_{2}]$) we stick to the very generous upper bound \eqref{eq_upperboundquadrDACnoise} which is valid for any modulation scheme, any phase angle $\varphi$ and both quadratures $q$ and $p$.}

\begin{align}
\boxed{
\xi_{\text{mod,QPSK}} \leq T V_{\text{mod}} \left( \pi  \frac{\delta U_{\text{DAC}}}{U_{\text{DAC}}} + \frac{\pi^{2}}{2} \left[ \frac{\delta U_{\text{DAC}}}{U_{\text{DAC}}} \right]^{2} \right)^{2} .
}
\end{align}

\section{Phase-Recovery Noise}

Coherent detection of the quantum signal requires a well-known phase- and frequency relation between the transmitter laser and the local oscillator. Moreover, the transmitter laser will carry a specific amount of phase noise $\delta \varphi$. This phase noise as well as phase offsets between signal laser and LO can be compensated using a strong reference signal sent by Alice, the pilot tone~\cite{qi2015generating, soh2015self, kleis2017continuous, laudenbach2017pilot}. The pilot tone carries a well-known phase $\omega t$ with a fixed phase relation to the signal pulse. Bob performs heterodyne detection of the pilot tone, measuring its $q$- and $p$-quadrature. If he makes his measurements periodically in a rate of exactly $\omega$, he can determine the deviation from a fixed and temporarily constant reference phase. Any such measured drift from this phase is used to readjust the measured phase of the quantum signal accordingly. Since relative phase deviations of the signal laser and LO can be compensated as accurately as the quadratures of the pilot tone can be measured, the remaining phase noise after recovery can be described in terms of the variance of the pilot-tone quadratures. However, the pilot-tone is influenced by various noise sources, very similar to the signal itself:

\begin{equation}
V(\hat{q}_{\text{PT}}) = V_{0} + \xi_{\text{PT}} = 1 + \xi_{\text{PT,modul}} + \xi_{\text{PT,phase}} +\xi_{\text{PT,Ram}} +\xi_{\text{PT,det}} +\xi_{\text{PT,ADC}} ,
\end{equation}
where the pilot tone's phase noise is determined by the accuracy of the sampling time:

\begin{equation}
\delta \varphi = \omega \delta t,
\end{equation}
In order to translate this angle deviation into a variance of the quadratures, we note that

\begin{equation}
q_{\text{PT}}=|\alpha_{\text{PT}}| \cos(\omega t).
\end{equation}
Hence a deviation of $\delta t$ will cause the quadrature to deviate according to

\begin{equation}
q_{\text{PT}} + \delta q_{\text{PT}} = |\alpha_{\text{PT}}| \cos(\omega t + \omega \delta t).
\end{equation}
Expanding around $\omega \delta t=0$ yields

\begin{align}
q_{\text{PT}} + \delta q_{\text{PT}} = |\alpha_{\text{PT}}| \cos(\omega t) + |\alpha_{\text{PT}}| \left(- \omega \delta t \sin (\omega t) - \frac{1}{2} (\omega \delta t)^{2} \cos (\omega t) \right) + \mathcal{O} \left( ( \omega \delta t)^{3} \right).
\end{align}
Therefore the error in the quadrature reads

\begin{align}
\delta q_{\text{PT}} =  - |\alpha_{\text{PT}}| \left( \omega \delta t  \sin (\omega t) + \frac{1}{2} (\omega \delta t)^{2} \cos (\omega t) \right) + \mathcal{O} \left( ( \omega \delta t)^{3} \right)  .
\end{align}
Using the fact that the modulus of sine and cosine is always smaller than or equal to one, we obtain the upper bound

\begin{align}
| \delta q_{\text{PT}} | \leq  |\alpha_{\text{PT}}| \left(\omega \delta t + \frac{1}{2} (\omega \delta t)^{2}  \right),
\end{align}
which, when squared, yields the variance

\begin{equation}
V(q_{\text{PT}}) \leq n_{\text{PT}} \left( \omega \delta t + \frac{1}{2} (\omega \delta t)^{2} \right)^{2}.
\end{equation}
However, this is only the variance of the quadrature \emph{component} $q$. In order to get $\xi_{\text{PT,phase}}$, hence the variance of the quadrature \emph{operator}, we need to multiply by a factor four:

\begin{equation}
V(\hat{q}_{\text{PT}}) = \xi_{\text{PT,phase}} \leq 4 n_{\text{PT}} \left( \omega \delta t + \frac{1}{2} (\omega \delta t)^{2} \right)^{2} .
\end{equation}
The other noise components of the pilot tone, $\xi_{\text{PT,modul}}$, $\xi_{\text{PT,Ram}}$, $\xi_{\text{PT,det}}$ and $\xi_{\text{PT,ADC}}$, are computed exactly as the ones of the actual quantum signal, described in the other sections of this chapter. When the total noise of the pilot tone $V(\hat{q}_{\text{PT}})\eqcolon V_{\text{PT}}$ is determined, it needs to be related to the phase-recovery noise (PR noise) of the quantum signal by putting them in proportion to their mean photon numbers:

\begin{equation} \label{eq_quPTratio}
\frac{V_{\text{PT}}}{\braket{n_{\text{PT}}}} = \frac{\xi_{\text{PR}}}{\braket{n_{\text{qu}}}},
\end{equation}
where $\braket{n_{\text{qu}}}$ represents the mean photon number of the quantum signal. The above equation is true for a case where one single pilot-tone measurement is used to recover the phase of one signal state. In experiment it is common to sample several, say $N$, pilot-tone measurements, each with a photon number $\braket{n_{\text{PT}}}$, and then take the average. We will briefly discuss how numerous measurements influence $V_{\text{PT}}$ and the above equation. Say the quadratures of the pilot tone carry, by virtue of various noise components, the variance $V_{\text{PT}}=V(\hat{q}_{\text{PT}})$. A number of $N$ subsequent measurements of the quadrature $\hat{q}_{\text{PT}}$ will yield the average

\begin{equation}
\braket{\hat{q}_{\text{PT}}}=\frac{1}{N} \sum_{i} \hat{q}_{\text{PT}}^{i} ,
\end{equation}
The variance of this mean value is computed as follows:

\begin{align}
V(\braket{\hat{q}_{\text{PT}}}) & = V \left( \frac{1}{N} \sum_{i=1}^{N} \hat{q}_{\text{PT}}^{i} \right) \notag \\
& = \frac{1}{N^{2}}  V \left( \sum_{i=1}^{N} \hat{q}_{\text{PT}}^{i} \right) \notag \\
& = \frac{1}{N^{2}} \sum_{i=1}^{N} V \left( \hat{q}_{\text{PT}}^{i}  \right) .
\end{align}
However, the variance of a single measurement of $\hat{q}_{\text{PT}}^{i}$ is $V_{\text{PT}}$, so we can rewrite the above expression to obtain

\begin{align}
V(\braket{\hat{q}_{\text{PT}}}) & = \frac{1}{N^{2}} \sum_{i=1}^{N} V_{\text{PT}} = \frac{1}{N^{2}} N V_{\text{PT}} = \frac{1}{N} V_{\text{PT}} .
\end{align}
Therefore, sampling over $N$ pilot-tone states with each $\braket{n_{\text{PT}}}$ mean photons transforms \ref{eq_quPTratio} to

\begin{equation}
\frac{V_{\text{PT}}}{N \braket{n_{\text{PT}}}} = \frac{\xi_{\text{PR}}}{\braket{n_{\text{qu}}}}.
\end{equation}
Rearranging the above equation and using the relation $\braket{n}=V_{\text{mod}}/2$ yields the final expression for the phase noise of the quantum signal:

\begin{align}
\boxed{
\xi_{\text{PR}} = \frac{1}{2} V_{\text{mod}} \frac{V_{\text{PT}}}{N\braket{n_{\text{PT}}}} = \frac{1}{2} V_{\text{mod}} \frac{1 + \xi_{\text{PT}}}{N\braket{n_{\text{PT}}}} .
}
\end{align}

\section{Raman Noise} \label{sec_xiRam}

When the quantum channel is wavelength-multiplexed with a classical channel, this classical channel will generate very broadband noise photons by Raman scattering. We assume the state represented by the Raman photons to be a thermal state.
\\ \\
In shot-noise units the photon-number operator is given by\footnote{Find a derivation of the number operator in terms of quadrature operators in SNU in Appendix~\ref{ch_coherent}.}

\begin{equation}
\hat{n}=\frac{1}{4}(\hat{q}^{2}+\hat{p}^{2}) - \frac{1}{2} .
\end{equation}
When the quadratures have a mean of zero ($\braket{\hat{q}}=\braket{\hat{p}}=0$) then their variance is given by the mean of their square: $V(\hat{q})=\braket{\hat{q}^{2}}$ (and analogous for $\hat{p}$). This allows us to write the expectation value of the number operator as

\begin{align}
\braket{ \hat{n} } & =\frac{1}{4} \left( \braket{ \hat{q}^{2} } + \braket{ \hat{p}^{2} } \right) - \frac{1}{2} \notag \\
& = \frac{1}{4} \left( V(\hat{q}) + V(\hat{p}) \right) - \frac{1}{2}
\end{align}
Further assuming that the $q$- and $p$-quadrature carry the same total variance $V(\hat{q})=V(\hat{p})\eqcolon V$, this expression simplifies to

\begin{equation}
\braket{ \hat{n}}=\frac{1}{2}( V - 1) .
\end{equation}
Let the above expression describe the total photon number of an incoming pulse, contaminated by Raman photons. The total photon number $\braket{ \hat{n}_{\text{tot}} }$ will then be the sum of attenuated signal photons and Raman photons; and the total variance $V_{\text{tot}}$ will be the sum of the damped modulation variance, one shot noise and the variance of the Raman noise:

\begin{align}
\braket{ \hat{n}_{\text{tot}}} & =\frac{1}{2}( V_{\text{tot}} - 1) \notag \\
T \braket{ \hat{n}_{A} } + \braket{ \hat{n}_{\text{Ram}} } & = \frac{1}{2}( TV_{\text{mod}} + 1 + V_{\text{Ram}} - 1) .
\end{align}
Knowing that $\braket{\hat{n}_{A}} = V_{\text{mod}}/2$ we obtain the simple result

\begin{equation} \label{eq_ramanphotonnumber}
\braket{ \hat{n}_{\text{Ram}} } = \frac{1}{2} V_{\text{Ram}} = \frac{1}{2} \xi_{\text{Ram}} .
\end{equation}
The mean photon number can be reexpressed as optical power times integration time divided by energy per photon:

\begin{equation} \label{eq_photonnumber}
\braket{ \hat{n} } = \frac{P \ \tau}{hf}.
\end{equation}
When the Raman noise is measured in terms of spectral noise density $N_{\text{Ram}}$ in units of dBm/nm, the linear spectral density is

\begin{align}
p_{\text{Ram}} & = 10^{\frac{N_{\text{Ram}}}{10}} \ \frac{\text{mW}}{\text{nm}} \notag \\
& = 10^{\frac{N_{\text{Ram}}}{10}} \times 10^{6} \frac{\text{W}}{\text{m}} .
\end{align}
Therefore the Raman noise power in linear scale is

\begin{equation}
P_{\text{Ram}}= \Delta \lambda \ 10^{N_{\text{Ram}}/10} \times 10^{6} \ \frac{\text{W}}{\text{m}} ,
\end{equation}
with $\Delta \lambda$ being the filter bandwidth (e.g.\ $\Delta \lambda \approx \SI{8}{\pico\metre}$ at $B=\SI{1}{\giga\hertz}$ and $f=\SI{193.4}{THz}$ for $\lambda=\SI{1550}{\nano\metre}$). Inserting the above expression into \eqref{eq_photonnumber} and using \eqref{eq_ramanphotonnumber}, we arrive at the final expression for the Raman noise in SNU:\footnote{Note that in case of a polarisation-multiplexed reference signal a polarised beamsplitter (PBS) is used to separate the quantum signal from the reference signal. Assuming unpolarised Raman photons, this BS will divide the optical Raman power, and hence the excess noise $\xi_{\text{Ram}}$, in half. However, due to mixing of the quantum signal with the local oscillator, giving rise to sum- and difference frequency terms, the signal's bandwidth $B$ is increased by a factor $2$, therefore compensating the effect of the PBS and leaving the expression \eqref{eq_xiram} unchanged.}

\begin{equation} \label{eq_xiram}
\boxed{
\xi_{\text{Ram}}= 2 \ \frac{\Delta \lambda \ 10^{N_{\text{Ram}}/10} \ \tau}{hf}  \times 10^{6} .
}
\end{equation}

\section{Common-Mode Rejection Ratio} \label{sec_CMRRnoise}

In a typical balanced homodyne receiver two PIN diodes convert, by their responsivity $\rho$, the optical input power into electric currents $I_{1}$ and $I_{2}$ whose difference $\Delta I$ is proportional to the quadrature of the input signal. The current difference is subsequently amplified to a measurable voltage $U$ by a large transimpedance $g$. However, a realistic differential amplifier will not only amplify the current difference but, to a small portion, also their mean value:

\begin{equation}
U= g (I_{1}-I_{2}) + g_{s} \frac{1}{2} (I_{1}+I_{2}) .
\end{equation}
The ratio of differential gain $g$ and common-mode gain $g_{s}$ is referred to as common-mode rejection ratio (CMRR):

\begin{equation}
\text{CMRR}=\left| \frac{g}{g_{s}} \right| .
\end{equation}
Favourably $g_{s}$ is very small and the CMRR therefore very high. In order to model the effect of the CMRR, we go to the expression for the photon numbers at the output ports of the homodyne beamsplitter \eqref{homodyningn1n2}:

\begin{subequations}
\begin{align}
\hat{n}_{1} & =\hat{a}_{1}^{\dagger}\hat{a}_{1}= \frac{1}{2} (\hat{a}^{\dagger} + \alpha_{\text{LO}}^{*})(\hat{a} + \alpha_{\text{LO}})=\frac{1}{2} ( \hat{a}^{\dagger}\hat{a} + \alpha_{\text{LO}}^{*}\alpha_{\text{LO}} + \alpha_{\text{LO}}\hat{a}^{\dagger} + \alpha_{\text{LO}}^{*}\hat{a} ), \\
\hat{n}_{2} & =\hat{a}_{2}^{\dagger}\hat{a}_{2}= \frac{1}{2} (\hat{a}^{\dagger} - \alpha_{\text{LO}}^{*})(\hat{a} - \alpha_{\text{LO}})=\frac{1}{2} ( \hat{a}^{\dagger}\hat{a} + \alpha_{\text{LO}}^{*}\alpha_{\text{LO}} - \alpha_{\text{LO}}\hat{a}^{\dagger} - \alpha_{\text{LO}}^{*}\hat{a} ).
\end{align}
\end{subequations}
Taking the difference of the two operators yields \eqref{homodynedetectionlaw}

\begin{equation}
\Delta \hat{n} = | \alpha_{\text{LO}} | (\hat{q}\cos \theta +\hat{p} \sin \theta ),
\end{equation}
where $\theta$ is the phase of the LO. On the other hand, adding the two operators yields

\begin{align}
\hat{n}_{1}+\hat{n}_{2} & =\hat{a}^{\dagger}\hat{a}+\alpha_{\text{LO}}^{*}\alpha_{\text{LO}} \notag \\
& = \hat{n}_{\text{sig}}+\braket{n_{\text{LO}}} ,
\end{align}
which is independent of the LO phase $\theta$. A photon number $n$ corresponds to the optical power by $P=hfn/\tau$. So the amplified voltage, given the above photon numbers and a PIN responsivity $\rho$, can be written as

\begin{align}
\hat{U}= \frac{hf}{\tau} \rho \left( g |\alpha_{\text{LO}}| (\hat{q}\cos \theta +\hat{p} \sin \theta) + \frac{g_{s}}{2} (\hat{n}_{\text{sig}}+\braket{n_{\text{LO}}}) \right).
\end{align}
Again, we simplify to the case $\theta=0$. Furthermore, we express $g_{s}$ in terms of $g$ and the CMRR, so we obtain

\begin{align}
\hat{U}= \frac{hf}{\tau} \rho g \left( |\alpha_{\text{LO}}| \hat{q} + \frac{1}{2 \text{CMRR}} (\hat{n}_{\text{sig}}+\braket{n_{\text{LO}}}) \right).
\end{align}
Since for any random variable $X$: $V(X+\text{const.})=V(X)$, the CMRR does \emph{not} contribute to the variance of the measurement outcome if the photon numbers $\braket{\hat{n}_{\text{sig}}}$ and $\braket{n_{\text{LO}}}$ are constant.\footnote{Note that, even if the CMRR term does not contribute to the excess noise, it still produces a voltage offset which has to be subtracted from the measurement before post-processing.} However, if $\braket{\hat{n}_{\text{sig}}}$ and/or $\braket{n_{\text{LO}}}$ are random variables themselves, then the CMRR term \emph{does} contribute to the excess noise and the variance of the total voltage is given by

\begin{equation}
V(\hat{U})= \frac{h^{2}f^{2}}{\tau^{2}} \rho^{2} g^{2} V(|\alpha_{\text{LO}}| \hat{q}) + \frac{h^{2}f^{2}}{\tau^{2}} \rho^{2} g^{2} \frac{1}{4 \text{CMRR}^{2}} \left( V(\hat{n}_{\text{sig}}) + V(\braket{n_{\text{LO}}}) \right) .
\end{equation}
For now we want to focus on the CMRR contribution only, so we write

\begin{equation}
V_{\text{CMRR}} (\hat{U})= \frac{h^{2}f^{2}}{\tau^{2}} \rho^{2} g^{2} \frac{1}{4 \text{CMRR}^{2}} \left( V(\hat{n}_{\text{sig}}) + V(\braket{n_{\text{LO}}}) \right)
\end{equation}
We assume the fluctuations of the photon numbers to be dominantly caused by relative intensity noise (RIN) of the signal and local oscillator respectively. Therefore we reexpress the variances of the photon numbers in terms of optical laser powers which can then be written in terms of RIN:

\begin{align} \label{VUCMRR}
V_{\text{CMRR}} (\hat{U}) & = \frac{h^{2}f^{2}}{\tau^{2}} \rho^{2} g^{2} \frac{1}{4 \text{CMRR}^{2}} \left( \frac{\tau^{2}}{h^{2}f^{2}} V(P_{\text{sig}}) + \frac{\tau^{2}}{h^{2}f^{2}} V(P_{\text{LO}}) \right) \notag \\
& =  \frac{h^{2}f^{2}}{\tau^{2}} \rho^{2} g^{2} \frac{1}{4 \text{CMRR}^{2}} \left( \frac{\tau^{2}}{h^{2}f^{2}} P_{\text{sig}}^{2} \text{RIN}_{\text{sig}} B + \frac{\tau^{2}}{h^{2}f^{2}} P_{\text{LO}}^{2} \text{RIN}_{\text{LO}}  B \right) \notag \\
& =  \frac{h^{2}f^{2}}{\tau^{2}} \rho^{2} g^{2} \frac{1}{4 \text{CMRR}^{2}} \left( \braket{n_{\text{sig}}}^{2} \text{RIN}_{\text{sig}} B + \frac{\tau^{2}}{h^{2}f^{2}} P_{\text{LO}}^{2} \text{RIN}_{\text{LO}}  B \right) \notag \\
& =  \frac{h^{2}f^{2}}{\tau^{2}} \rho^{2} g^{2} \frac{1}{4 \text{CMRR}^{2}} \left( \frac{V_{\text{mod}}^{2}}{4} \text{RIN}_{\text{sig}} B + \frac{\tau^{2}}{h^{2}f^{2}} P_{\text{LO}}^{2} \text{RIN}_{\text{LO}}  B \right) 
\end{align}
Back-transforming the voltage variance through the receiver setup into a variance of the quadrature operator $\hat{q}$ yields

\begin{align}
V(U) & =g^{2} V(\Delta I)=g^{2}\rho^{2} V(\Delta P) = g^{2}\rho^{2} \frac{h^{2}f^{2}}{\tau^{2}} V(\Delta n) \notag \\
& = g^{2}\rho^{2} \frac{h^{2}f^{2}}{\tau^{2}} |\alpha_{\text{LO}}|^{2} V(\hat{q}) \notag \\
& = g^{2}\rho^{2} \frac{h^{2}f^{2}}{\tau^{2}} \braket{n_{\text{LO}}} V(\hat{q}) \notag \\
& = g^{2}\rho^{2} \frac{hf}{\tau} P_{\text{LO}} V(\hat{q}),
\end{align}
and therefore

\begin{equation}
V(\hat{q})= \frac{\tau}{hf \rho^{2} g^{2} P_{\text{LO}}} V(U).
\end{equation}
By inserting \eqref{VUCMRR} we obtain

\begin{align}
V(\hat{q}) & = \frac{\tau}{hf \rho^{2} g^{2} P_{\text{LO}}}  \frac{h^{2}f^{2}}{\tau^{2}} \rho^{2} g^{2} \frac{1}{4 \text{CMRR}^{2}} \left( \frac{V_{\text{mod}}^{2}}{4} \text{RIN}_{\text{sig}} B + \frac{\tau^{2}}{h^{2}f^{2}} P_{\text{LO}}^{2} \text{RIN}_{\text{LO}}  B \right) \notag \\
& = \frac{hf}{\tau P_{\text{LO}}} \frac{1}{4 \text{CMRR}^{2}} \left( \frac{V_{\text{mod}}^{2}}{4} \text{RIN}_{\text{sig}} B + \frac{\tau^{2}}{h^{2}f^{2}} P_{\text{LO}}^{2} \text{RIN}_{\text{LO}}  B \right),
\end{align}
which brings us to the expression:

\begin{align} \label{eq_xiCMRR}
\xi_{\text{CMRR}} = \frac{1}{4 \text{CMRR}^{2}} \left( \frac{hf V_{\text{mod}}^{2}}{4 \tau P_{\text{LO}}} \text{RIN}_{\text{sig}} B + \frac{\tau}{hf} P_{\text{LO}} \text{RIN}_{\text{LO}}  B \right).
\end{align}
Note that, by Equation~\eqref{eq_covmathet}, heterodyne detection will halve the transmittance \emph{and} the excess noise with all it's constituents:

\begin{equation}
\xi \longrightarrow \frac{1}{2} \xi = \frac{1}{2} (\xi_{\text{modul}} + \xi_{\text{RIN}} + \xi_{\text{phase}} + \xi_{\text{Raman}} + \xi_{\text{CMRR}} + \xi_{\text{det}} + \xi_{\text{ADC}} + \dots).
\end{equation}
However, the measured CMRR noise will (similar to the detection- and ADC noise) \emph{not} halve since there will be \emph{two} balanced receivers (and ADCs) in the measurement setup -- one at each output port of the beamsplitter. Due to our convention that the excess noise $\xi$ is defined at the channel output \emph{before} the heterodyning beamsplitter, and because Equation~\eqref{eq_xiCMRR} represents the CMRR noise \emph{per detector}, the CMRR-noise parameter has to doubled in case of heterodyne detection:

\begin{align}
\xi_{\text{CMRR}} (\text{het}) & = 2 \xi_{\text{CMRR}} (\text{hom}) .
\end{align}
Therefore we arrive at the final expression

\begin{align}
\boxed{
\xi_{\text{CMRR}} = \frac{\mu}{4 \text{CMRR}^{2}} \left( \frac{hf V_{\text{mod}}^{2}}{4 \tau P_{\text{LO}}} \text{RIN}_{\text{sig}} B + \frac{\tau}{hf} P_{\text{LO}} \text{RIN}_{\text{LO}}  B \right).
}
\end{align}
where $\mu=1$ in case of homodyne detection and $\mu=2$ in case of heterodyne detection.

\section{Detection Noise}

As derived in Appendix~\ref{ch_homodynedet}, the number-difference operator at the output ports of a homodyning beamsplitter is (in SNU) represented as

\begin{equation}
\Delta \hat{n} = |\alpha_{\text{LO}}| \ ( \hat{q} \cos \theta + \hat{p} \sin \theta ) ,
\end{equation}
with $\theta$ being the phase of the local oscillator. Consider WLOG a measurement of the $q$-quadrature, i.e.\ $\theta=0$. So the above expression simplifies to

\begin{equation}
\Delta \hat{n} = |\alpha_{\text{LO}}| \ \hat{q} .
\end{equation}
The variance of this operator is

\begin{align}
V(\Delta \hat{n}) & = |\alpha_{\text{LO}}|^{2} \ V(\hat{q}) \notag \\
& = \braket{n_{\text{LO}}} \ V(\hat{q}) \notag \\
& = \frac{P_{\text{LO}} \ \tau}{hf} \ V(\hat{q}) ,
\end{align}
where $P_{\text{LO}}$ is the optical local-oscillator power, $\tau$ is the LO's pulse duration, $h$ represents Planck's constant and $f$ the optical LO frequency. Since we assume all excess-noise components to be stochastically independent, we can regard the variance caused by detection noise to be independent from other noise sources and therefore express it on its own:

\begin{align} \label{eq_LOdeviation}
V_{\text{det}}(\Delta \hat{n}) & = \frac{P_{\text{LO}} \ \tau}{hf} \ V_{\text{det}}(\hat{q}) \notag \\
& = \frac{P_{\text{LO}} \ \tau}{hf} \ \xi_{\text{det}} .
\end{align}
Now consider a balanced homodyne receiver with a given voltage noise density at its output

\begin{equation}
u_{\text{det}}=\text{NEP} \rho \ g ,
\end{equation}
where NEP is the noise-equivalent power (in $\text{W}/\sqrt{\text{Hz}}$), $\rho$ is the responsivity of the PIN diodes (in A/W) and $g$ is the gain factor of the amplifier (in V/A). The unit of the voltage-noise density $u_{\text{det}}$ is therefore $\text{V}/\sqrt{\text{Hz}}$. At a given electronic bandwidth $B$ the receiver's voltage noise is then

\begin{equation}
U_{\text{det}}= u_{\text{det}} \ \sqrt{B} = \text{NEP} \rho \ g \ \sqrt{B} .
\end{equation}
The excess-noise parameter $\xi$, as it appears in the covariance matrix, refers to an additional variance at the channel output \emph{before} Bob's measurement apparatus. So in order to understand how the voltage noise at the receiver's output will affect the excess noise, we will translate it -- walking backwards through the receiver -- into a variance $\xi_{\text{det}}$ of the input signal. This is equivalent to a noisy signal with variance $\xi$ being detected by a noiseless receiver. The voltage output $U$ of a balanced receiver is the amplification of the current difference $\Delta I$ of the PIN diodes by the gain factor $g$. So a voltage noise $U_{\text{det}}$ relates to a certain deviation $\delta$ of the current difference:

\begin{equation}
\delta_{\text{det}}(\Delta I) = \frac{U_{\text{det}}}{g} = \text{NEP} \ \rho \ \sqrt{B} .
\end{equation}
The optical-power difference $\Delta P$ at the input of the diodes relates linearly to the current difference $\Delta I$ by the diode's responsivity $\rho$ and the same goes for their deviations:

\begin{equation}
\delta_{\text{det}}(\Delta P) = \frac{\delta_{\text{det}}(\Delta I)}{r} = \text{NEP} \ \sqrt{B} .
\end{equation}
The next step is to translate the deviation of the optical-power difference into a deviation of the photon-number difference:

\begin{equation}
\delta_{\text{det}}(\Delta n) = \frac{\delta_{\text{det}}(\Delta P) \ \tau}{hf} = \frac{ \text{NEP} \ \sqrt{B} \tau}{hf} .
\end{equation}
Using $V_{\text{det}}(\Delta n)=\delta_{\text{det}}^{2}(\Delta n)$, we obtain

\begin{equation}
V_{\text{det}}(\Delta n) = \frac{ \text{NEP}^{2} \ B \tau^{2}}{h^{2}f^{2}} .
\end{equation}
Equalising the above expression with \eqref{eq_LOdeviation} yields

\begin{equation}
\frac{ \text{NEP}^{2} \ B \tau^{2}}{h^{2}f^{2}} = \frac{P_{\text{LO}} \ \tau}{hf} \ \xi_{\text{det}} ,
\end{equation}
which allows us to find the required expression for the detection noise in shot-noise units:

\begin{equation}
\xi_{\text{det}} = \frac{ \text{NEP}^{2} \ B \ \tau}{hf \ P_{\text{LO}}} . 
\end{equation}
Since in case of heterodyne detection, there will be two noisy detectors -- one at each output of the heterodyning beamsplitter -- the measured detection noise does not halve by action of the beamsplitter. Therefore (as explained in Section~\ref{sec_CMRRnoise} on the CMRR noise) we get

\begin{align}
\xi_{\text{det}} (\text{het}) & = 2 \xi_{\text{det}} (\text{hom}) .
\end{align}
This yields the final expression

\begin{equation} \label{eq_xidet}
\boxed{
\xi_{\text{det}} = \mu  \ \frac{ \text{NEP}^{2} \ B \ \tau}{hf \ P_{\text{LO}}} , 
}
\end{equation}
where $\mu=1$ ($\mu=2$) for homodyne (heterodyne) detection.

Alternatively, the detection noise can be conveniently expressed in terms of the \emph{clearance} $C$ which is defined as the ratio of the experimental shot-noise variance and the variance caused by the receiver's electronic noise:

\begin{align}
C=\frac{V_{0}(\hat{q})+V_{\text{det}}(\hat{q})}{V_{\text{det}}(\hat{q})}.
\end{align}
The shot-noise variance scales linearly with the optical power of the local oscillator which is, however, confined by the saturation limit of the receiver's PIN diodes. Experimentally, the numerator of the above equation can be determined by measuring the quadrature variance of the LO when it is mixed with a vacuum input. The denominator is the remaining quadrature variance after the LO is disconnected from the receiver. In shot-noise units we have by definition $V_{0}(\hat{q})=1$ and the above equation becomes

\begin{align}
C=\frac{1+V_{\text{det}}(\hat{q})}{V_{\text{det}}(\hat{q})} = \frac{1+\xi_{\text{det}}}{\xi_{\text{det}}}.
\end{align}
The detection noise with respect to the clearance is therefore (taking into account the factor $\mu=2$ in case of heterodyne detection)

\begin{equation}
\boxed{
\xi_{\text{det}} = \mu  \ \frac{1}{C-1} . 
}
\end{equation}

\section{ADC Quantisation Noise}

An incoming signal pulse will be detected and amplified by Bob's balanced receiver where the voltage at the receiver output will be proportional to the measured quadrature. However, if the output voltage is quantised by an analog-to-digital converter, this ADC will introduce an additional error to the measured signal, therefore contributing to the excess noise $\xi$. Like in the previous chapter on detection noise we will back-transform the ADC noise to the channel output where the noise in the covariance matrix is defined. A given error caused by the ADC is equivalent to a certain voltage noise $\delta U$ at the receiver output, quantised by a noiseless ADC. Reusing the considerations of the previous section, we know that this voltage noise translates into a deviation of the photon number difference of

\begin{equation}
\delta_{\text{ADC}}(\Delta n) = \frac{\delta_{\text{ADC}}(\Delta P) \ \tau}{hf} = \frac{ \delta_{\text{ADC}} U \ \tau }{hf \ g \ \rho} ,
\end{equation}
where the subscript ADC is meant to indicate the source of the deviation. The variance of the photon number difference in terms of $\delta_{\text{ADC}}$ is then

\begin{equation}
V_{\text{ADC}}(\Delta n) = \frac{ V_{\text{ADC}} (U)\ \tau^{2} }{h^{2}f^{2} \ g^{2} \ \rho^{2}} .
\end{equation}
Similar to the previous section, we express the variance of the number-difference at the output port of the homodyning BS in terms of the excess-noise parameter we want to find:
\begin{align}
V_{\text{ADC}}(\Delta \hat{n}) = \frac{P_{\text{LO}} \ \tau}{hf} \ \xi_{\text{ADC}} .
\end{align}
Comparing the above two equations yields

\begin{equation}
\xi_{\text{ADC}}= \frac{ \ \tau }{hf  \ g^{2} \ \rho^{2} \ P_{\text{LO}}} \ V_{\text{ADC}} (U).
\end{equation}
We assume $V_{\text{ADC}}$ to consist of two parts: the intrinsic ADC voltage noise variance $V_{\text{ADC,intr}}$ and the quantisation error variance $V_{\text{ADC,quant}}$: This allows to reexpress the ADC-caused excess noise as

\begin{equation} \label{eq_xiADC1}
\xi_{\text{ADC}}= \frac{ \ \tau }{hf  \ g^{2} \ \rho^{2} \ P_{\text{LO}}} \left( V_{\text{ADC,quant}} +V_{\text{ADC,intr}} \right) .
\end{equation}
In order to derive the quantisation error variance $V_{\text{ADC,quant}}$ consider a continuous random variable $x$ falling into a specific interval $[-q/2,q/2]$. The probability of $x$ being smaller than $-q/2$ or greater than $q/2$ is, trivially, zero. On the other hand, the probability density function for $x$ in the interval $[-q/2,q/2]$ is $1/q$. So the overall probability density function is

\begin{equation}
    p(x)= 
\begin{cases}
    \frac{1}{q}& \text{if } -\frac{q}{2} < x < \frac{q}{2} \\
    0              & \text{otherwise} .
\end{cases}
\end{equation}
The variance of a random variable is well known:

\begin{equation}
V(x)=\braket{x^{2}}-\braket{x}^{2},
\end{equation}
where in our case the mean value of $x$ is zero: $\braket{x}=0$. So the variance of a uniformly distributed random variable, quantised in the interval $[-q/2,q/2]$, is

\begin{align} \label{eq_quanterror}
V(x)=\braket{x^{2}} & = \int_{-\infty}^{\infty} x^{2} p(x) \ dx =  \int_{-q/2}^{q/2} x^{2} \frac{1}{q} \ dx = \left[ \frac{x^{3}}{3q} \right]_{-q/2}^{q/2} = 2 \frac{q^{2}}{24} = \frac{q^{2}}{12} .
\end{align}
In the case of a random voltage, quantised by an ADC with full voltage range $R_{U}$ and $n$ bits (hence $2^{n}$ voltage intervals) one single interval is referred to as the least significant bit (LSB) and given by

\begin{equation}
\text{LSB}=\frac{R_{U}}{2^{n}} .
\end{equation}
Therefore, comparing with \eqref{eq_quanterror}, we obtain the voltage variance caused by the quantisation of an analog-to-digital converter

\begin{equation}
V_{\text{ADC,quant}} = \frac{1}{12} \text{LSB}^{2} = \frac{1}{12} \frac{R_{U}^{2}}{2^{2n}}.
\end{equation}
Reinserting into \eqref{eq_xiADC1} and considering the effect of having $two$ ADCs in case of heterodyne detection (see previous section) yields

\begin{equation} \label{eq_xiADC}
\boxed{
\xi_{\text{ADC}}= \mu \ \frac{ \ \tau }{hf  \ g^{2} \ \rho^{2} \ P_{\text{LO}}} \left( \frac{1}{12} \frac{R_{U}^{2}}{2^{2n}} +V_{\text{ADC,intr}} \right) .
}
\end{equation}

\chapter{Experimental Implications} \label{ch_experimentalimplications}

Any experimental implementation of CV-QKD (or DV-QKD for that matter) needs to overcome a number of technological challenges in order to establish a satisfying secure-key rate over a given distance~\cite{diamanti2016practical}. The fundamental equations and noise models derived in this article allow us to simulate various experimental setups, test specific hardware choices and identify bottlenecks prior to the experiment~\cite{laudenbach2016cvsim}. In this concluding section we try to obtain an estimate in which way actual experimental parameters influence the performance of a CV-QKD setup.

Ultimately, the secure-key rate is proportional to the symbol rate $f_{\text{sym}}$ times the secret fraction $r$ (i.e.\ the secure key per transmitted symbol)

\begin{equation}
K \sim f_{\text{sym}} \cdot r,
\end{equation}
where $r$ is defined as before:

\begin{equation} \label{eq_secfrac}
r=\beta I_{AB}-\chi_{EB}.
\end{equation}
A natural first approach in the attempt to increase the key rate would be to go to a higher symbol rate. In fact, this is where the compatibility of CV-QKD with the highly advanced and ultrafast telecom technology strikes. State-of-the-art waveform generators, $q/p$-modulators and coherent receivers render symbol rates in the order of magnitude $\sim \SI{10}{Gbaud}$ feasible. However, even arbitrarily high symbol rates cannot increase the achievable transmission distance since the amount of symbols per second will have no effect on the secure bits \emph{one} symbol can carry and therefore cannot compensate for $r=0$. Furthermore, high rates require high-bandwidth detectors which, by \eqref{eq_xidet}, will increase the detection noise linearly with the bandwidth $B$. As illustrated below in the present section, the detection noise already makes up the better part of the total noise in a realistic CV-QKD implementation. Therefore, under the strict assumption that the noise generated at the receiver is accessible to Eve, the setup is extremely sensitive to detection noise -- and consequently to the increase of bandwidth and symbol rate. On the other hand, when the detection noise is classified as trusted noise, fast and high-bandwidth receivers are less problematic. Experimental implementations of CV-QKD protocols demonstrated symbol rates in the regime of kbaud up to $\SI{250}{Mbaud}$~\cite{laudenbach2017pilot}.

\begin{figure}[h]
\centering
\subcaptionbox{}
    [0.49\linewidth]{\includegraphics[width=0.4\linewidth]{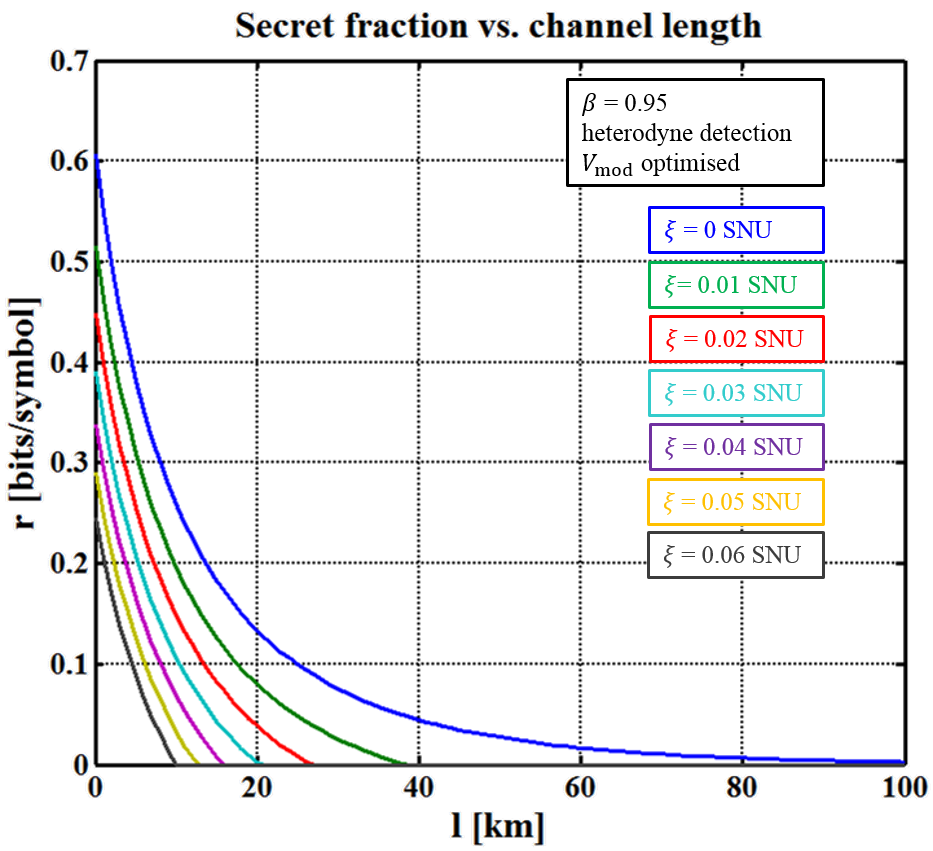}}
\subcaptionbox{}
    [0.49\linewidth]{\includegraphics[width=0.4\linewidth]{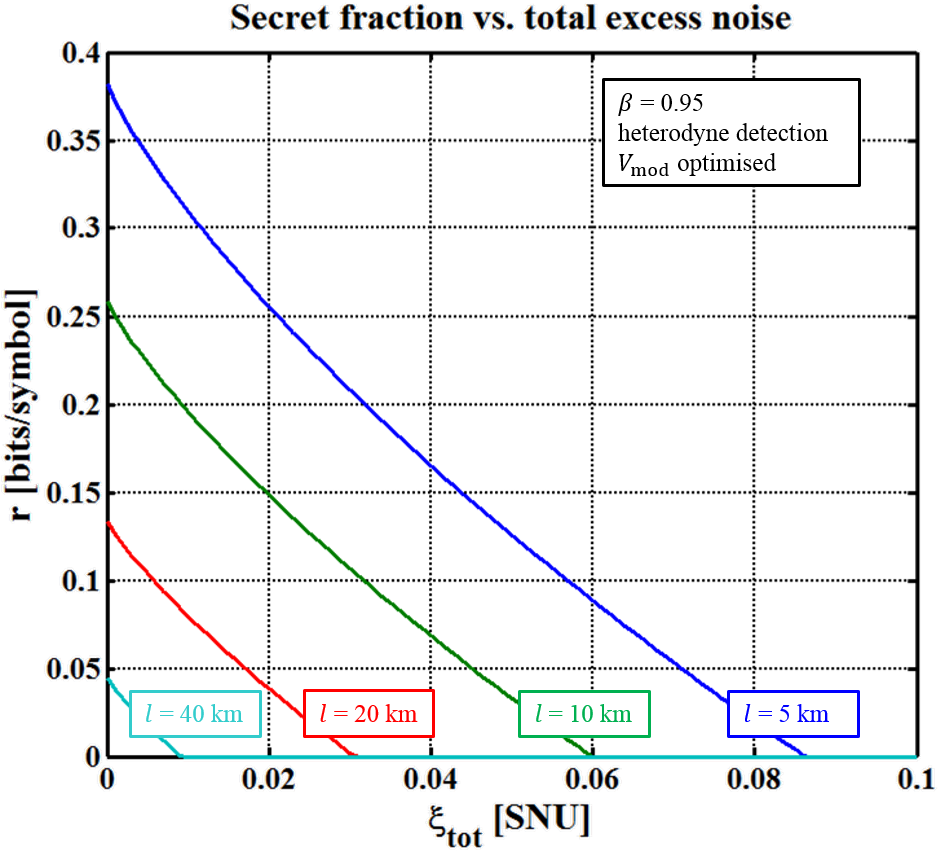}}
\caption{Secret fraction with respect to channel length and excess noise. The modulation variance $V_{\text{mod}}$ has been dynamically optimised to maximise the simulated key rate for each point in the graphs.}
\label{lengthnoise}
\end{figure}

The secret fraction \eqref{eq_secfrac} depends on merely four parameters, i.e.\ the modulation variance (directly related to the mean photon number per symbol), the transmittance, the excess noise and the reconciliation efficiency. Since the modulation variance $V_\text{mod}$ is basically variable and can (by optical attenuation) be easily adjusted to the given security requirements, the key parameters that primarily characterise a CV-QKD setup are the transmittance $T$, the excess noise $\xi$ and the reconciliation efficiency $\beta$. Figure~\ref{lengthnoise} illustrates the secure-key rate (i.e.\ secure key per symbol) with respect to transmittance and noise. As the indicated by Fig.~\ref{lengthnoise}(a), lower excess noise allows for higher loss tolerance and therefore longer transmission distances. Accordingly, as depicted by Fig.~\ref{lengthnoise}(b), the channel length defines an upper bound on the excess noise beyond which no secure key can be established. For instance, the graph shows that for a channel length of $l=\SI{40}{km}$ the threshold excess noise is lower than one percent of the shot noise which constitutes a considerable experimental challenge. Further investigations of the robustness of CV-QKD under the influence of channel loss and noise can be found in \cite{garcia2009continuous, lasota2017robustness}.

\begin{figure}
\centering
\includegraphics[width=0.95\linewidth]{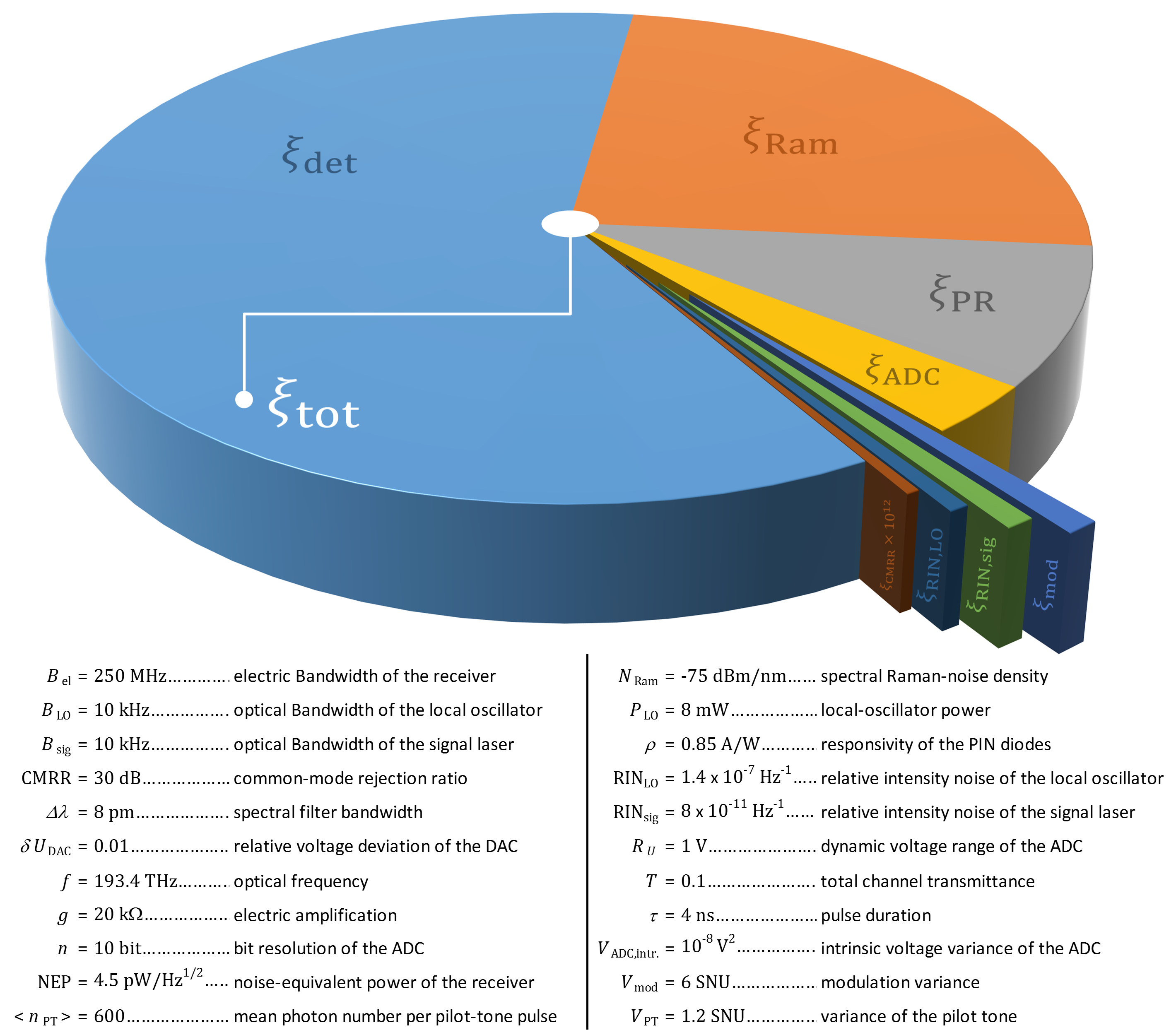}
\caption{Decomposition of the total excess noise in a typical CV-QKD setup. The values for each noise component were calculated using our analytical noise models as derived in Section~\ref{ch_noise} and the exemplary hardware parameters listed in the above table. In this particular example, the total excess noise amounts to $\xi_{\text{tot}} = \SI{0.0653}{SNU}$. With $\xi_{\text{det}} = \SI{0.0396}{SNU}$ the detection noise makes up about $\SI{60}{\%}$ of $\xi_{\text{tot}}$. Under the loose assumptions of trusted receiver devices (detector and ADC) the detection- and quantisation noise are not attributed to an eavesdropper. In this scenario the noise contributing to Eve's knowledge would only amount to $\xi_{\text{Eve}} = \SI{0.0236}{SNU}$ which is $\SI{36}{\%}$ of the total noise and therefore makes a more than considerable difference in terms of the Holevo bound and hence final secure-key rate. Apart from the detection noise the major components are the Raman noise $\xi_{\text{Ram}}$, the phase-recovery noise $\xi_{\text{PR}}$ and the quantisation noise $\xi_{\text{ADC}}$ caused by the analog-to-digital converter. According to our models, the modulation noise $\xi_{\text{mod}}$, the relative intensity noise of transmitter laser and LO, $\xi_{\text{RIN,sig}}$ and $\xi_{\text{RIN,LO}}$, as well as the noise caused by the common-mode-rejection ratio $\xi_{\text{CMRR}}$ do not contribute significantly to the total noise.}
\label{total_excess_noise}
\end{figure}

The excess noise $\xi$ is the figure of merit when it comes to appraising the experimental performance of a CV-QKD setup. It may have multiple origins which are discussed in all detail in Section~\ref{ch_noise}. Figure~\ref{total_excess_noise} depicts a representation of the total excess noise $\xi_{\text{tot}}$ corresponding to a typical exemplary setup and the decomposition into its constituents according to our noise models. The pie chart illustrates the dominance of the detection noise $\xi_{\text{det}}$ over the other noise sources. Other significant noise contributors are Raman scattering, caused by classical DWDM channels in the fibre, imperfect phase recovery and the quantisation of the measurement results by analog-to-digital converters. On the other hand, the modulation noise, the relative intensity noise as well as the common-mode-rejection ratio of the detectors have a rather minor effect.

\begin{figure}
\centering
\subcaptionbox{}
    [0.49\linewidth]{\includegraphics[width=0.4\linewidth]{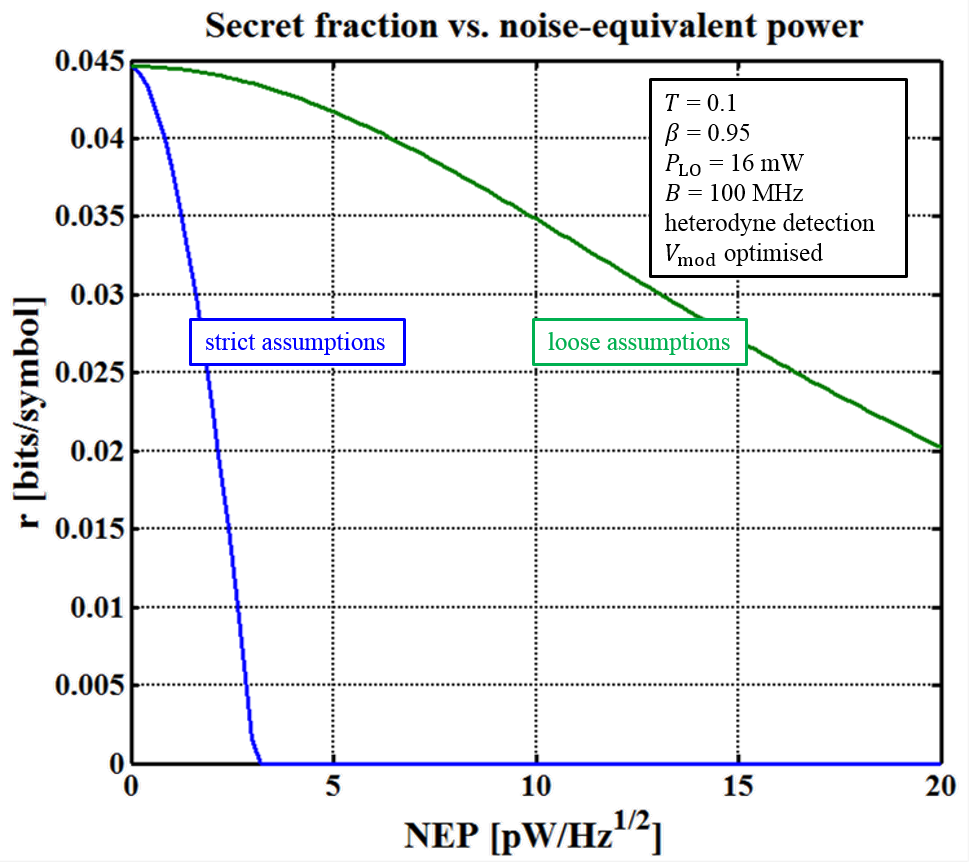}}
\subcaptionbox{}
    [0.49\linewidth]{\includegraphics[width=0.4\linewidth]{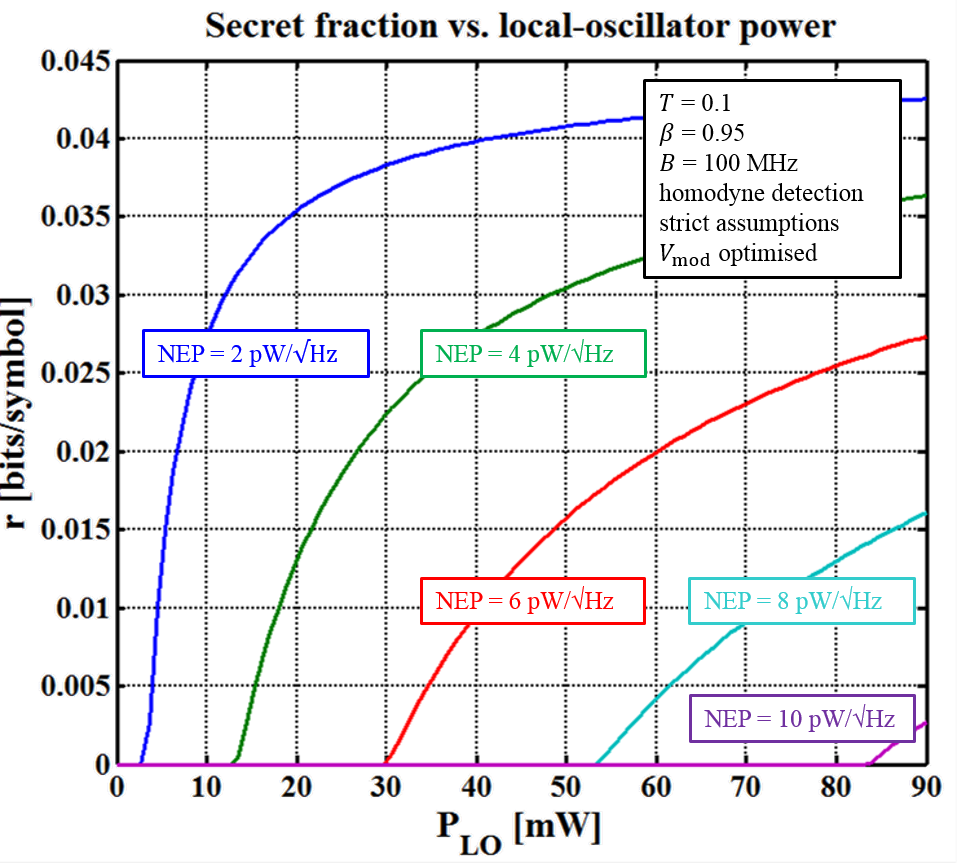}}
\caption{(a) Secret fraction vs.\ noise-equivalent power under strict (untrusted detector) and loose (trusted detector) assumptions. (b)~Secret fraction vs. local-oscillator power for different NEP values under strict assumptions. The detector's NEP can be compensated with a higher LO power, keeping the electronic noise small with respect to the shot noise. However, in practice this compensation is not always feasible due the saturation limit of conventional PIN diodes (usually in the order of $\sim \SI{10}{mW}$). For both plots all other noise sources were neglected.}
\label{detnoise}
\end{figure}

The relative magnitude of the detection noise $\xi_{\text{det}}$ undermines the performance discrepancy depending on whether  receiver-related noise sources are classified as trusted or not. Figure~\ref{detnoise}(a) emphasises to which extent the security assumptions determine the requirements on the balanced detectors: Under the strict assumption that the detection noise contributes to an eavesdropper's information, a low noise-equivalent power (NEP) is critical in order to establish a non-zero secure key (blue curve, $\text{NEP} < \SI{4}{pW/\sqrt{HZ}}$ for $T=0.1$ and $P_{\text{LO}}=\SI{8}{mW}$, neglecting all other noise sources). On the other hand, in the trusted-detector scenario the requirements on the NEP are far more relaxed (green curve) since it only influences the SNR but not the Holevo bound. According to our model~\eqref{eq_xidet}, the detection noise can be mitigated by increasing the local oscillator power. Figure~\ref{detnoise}(b) indicates the minimal LO power with respect to the NEP under strict assumptions. Most low-noise balanced detectors, however, suffer from the low optical saturation limit of the PIN diodes which confines the LO power to the order of magnitude $\sim \SI{10}{mW}$. Attempts to increase the performance of CV-QKD in the untrusted-receiver scenario must primarily go into the direction of decreasing the intrinsic detector noise (NEP) or increasing the saturation limit of the PIN diodes.

\begin{figure}
\centering
\subcaptionbox{}
    [0.49\linewidth]{\includegraphics[width=0.4\linewidth]{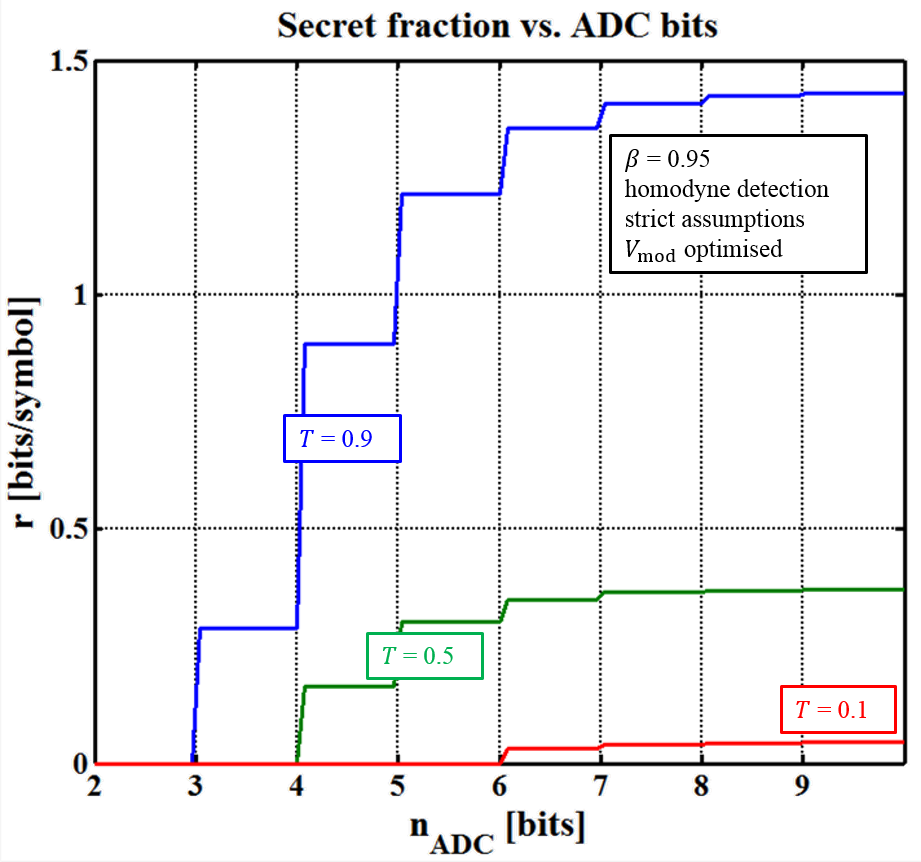}}
\subcaptionbox{}
    [0.49\linewidth]{\includegraphics[width=0.4\linewidth]{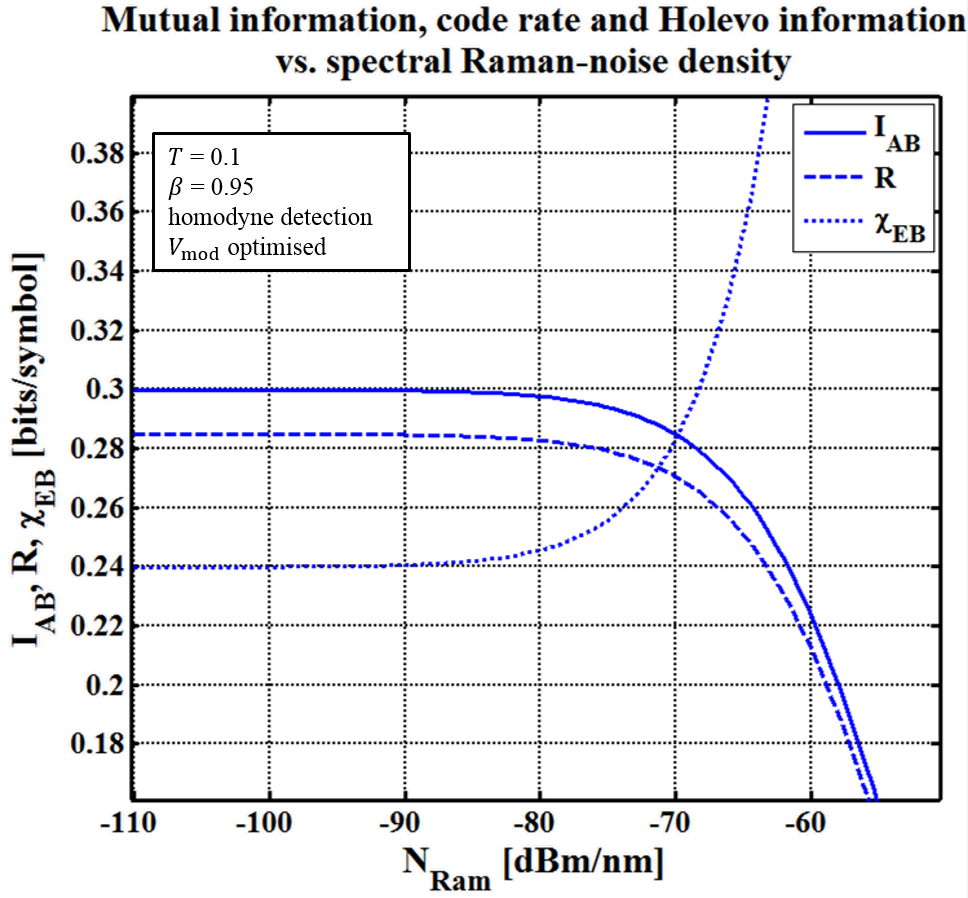}}
\caption{(a) Secret fraction $r$ vs. bit resolution $n_{\text{ADC}}$ of the analog-to-digital converter (ADC). The plot illustrates in which way the tolerable quantisation error depends on the total transmittance (and therefore on the total-noise threshold). For instance, at $T=0.9$ three ADC bits are sufficient to establish a non-zero key, while at $T=0.1$ at least six bits are required. (b) Mutual information $I_{AB}$, code rate $R=0.95 I_{AB}$ and Holevo information $\chi_{EB}$ with respect to the spectral Raman-noise density $N_{\text{Ram}}$. A secure key can only be established as long as $R=0.95 I_{AB} > \chi_{EB}$, in this configuration corresponding to a Raman-noise level of $N_{\text{Ram}} < \SI{-73}{dBm/nm}$ (one classical DWDM channel corresponding to $N_{\text{Ram}} \sim \SI{-90}{dBm/nm}$). For both plots all other noise sources were neglected.}
\label{morenoise}
\end{figure}

The quantisation noise caused by the analog-to-digital (ADC) converter can, under loose security assumptions, also be classified as trusted and therefore not be attributed to Eve. For the pie chart in Fig.~\ref{total_excess_noise} we assumed the ADC to have a resolution of $\SI{8}{bit}$. According to our model~\eqref{eq_xiADC}, the quantisation noise is proportional to the inverse of $2^{2n}$ where $n$ is the bit resolution of the ADC. Therefore, increasing $n$ by one bit reduces $\xi_{\text{ADC}}$ by a factor of $4$, increasing $n$ by two bits will improve $\xi_{\text{ADC}}$ by factor of $16$. We conclude that, unlike the detection noise, the quantisation noise does not constitute a serious impediment in practical CV-QKD implementations, even under strict security assumptions. This is also illustrated by Fig.~\ref{morenoise}(a) where we depict the secret fraction with respect to the ADC's bit resolution in the untrusted-receiver scenario. The graph shows that even for a lossy channel with $T=0.1$ a six-bit resolution is sufficient to establish a non-zero key.

As one of its main advantages, CV-QKD allows for the integration into existing telecom networks and does not require the deployment of dedicated dark fibres. However, the co-existence of the quantum channel with several wavelength-multiplexed classical channels will be a source of detrimental crosstalk. In particular, Raman scattering caused by the classical channel will increase the channel noise of the CV-QKD transmission. The Raman noise is modelled in Section~\ref{sec_xiRam}. Figure~\ref{morenoise}(b) depicts the mutual information, the code rate and the Holevo bound with respect to the spectral Raman-noise density.

\begin{figure}
\centering
\subcaptionbox{}
    [0.49\linewidth]{\includegraphics[width=0.4\linewidth]{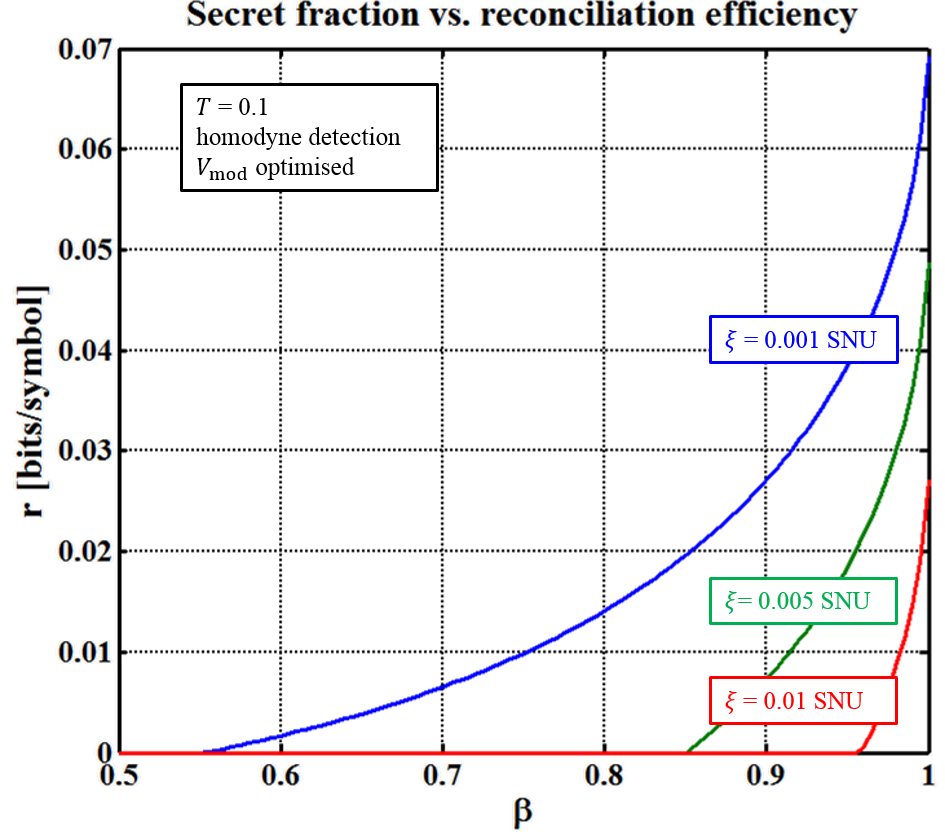}}
\subcaptionbox{}
    [0.49\linewidth]{\includegraphics[width=0.4\linewidth]{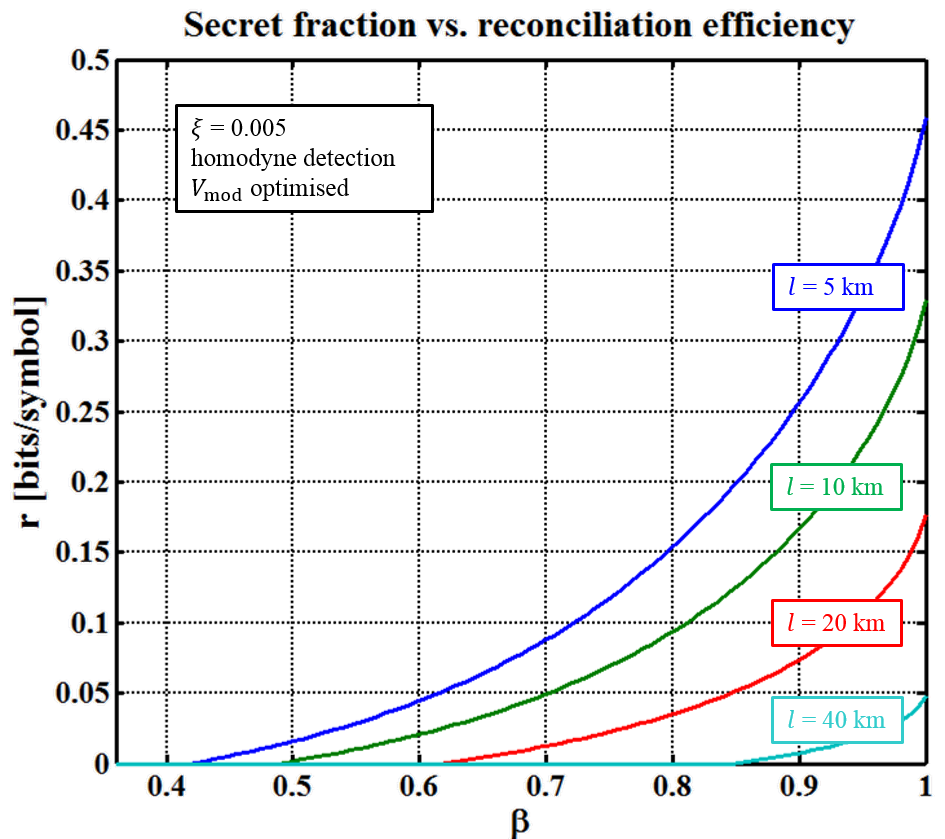}}
\caption{Secret fraction vs.\ reconciliation efficiency. The plots illustrate how crucial efficient post-processing becomes with increasing excess noise $\xi$ (a) and channel length (b). For instance, Fig.~(a) shows that for a total transmittance of $T=0.1$ and an excess-noise parameter of $\xi=\SI{0.01}{SNU}$ the reconciliation efficiency has to be higher than $0.96$.}
\label{beta}
\end{figure}

\begin{figure}
\centering
\subcaptionbox{}
    [0.49\linewidth]{\includegraphics[width=0.4\linewidth]{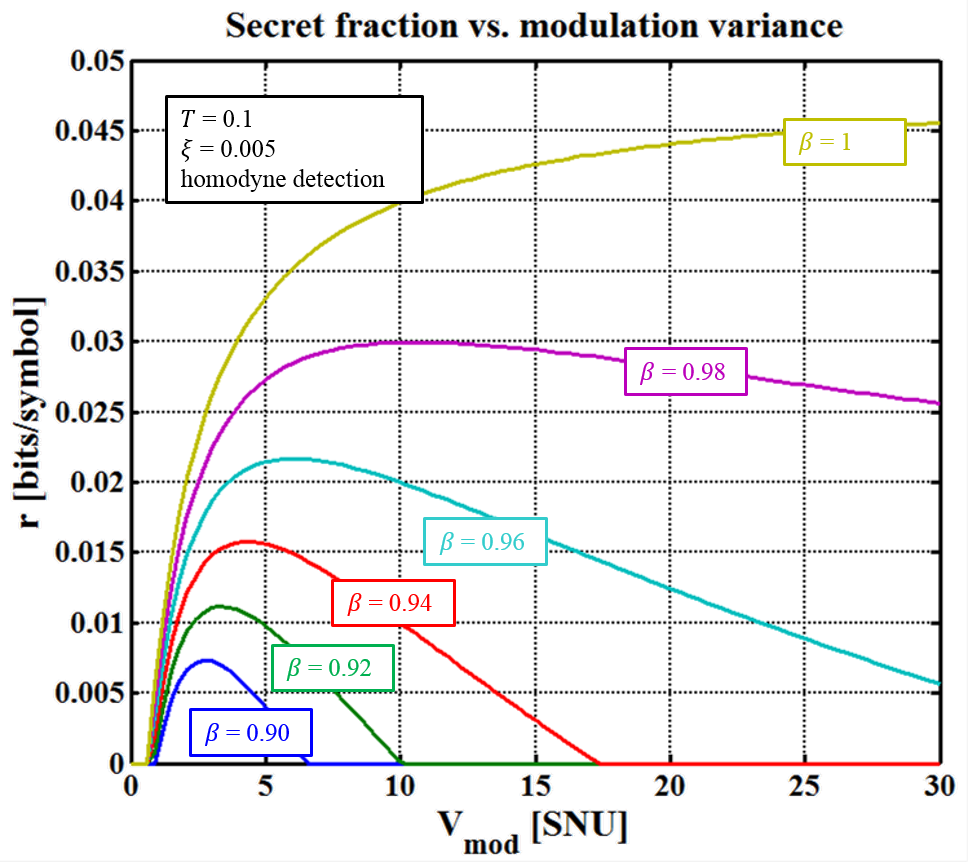}}
\subcaptionbox{}
    [0.49\linewidth]{\includegraphics[width=0.4\linewidth]{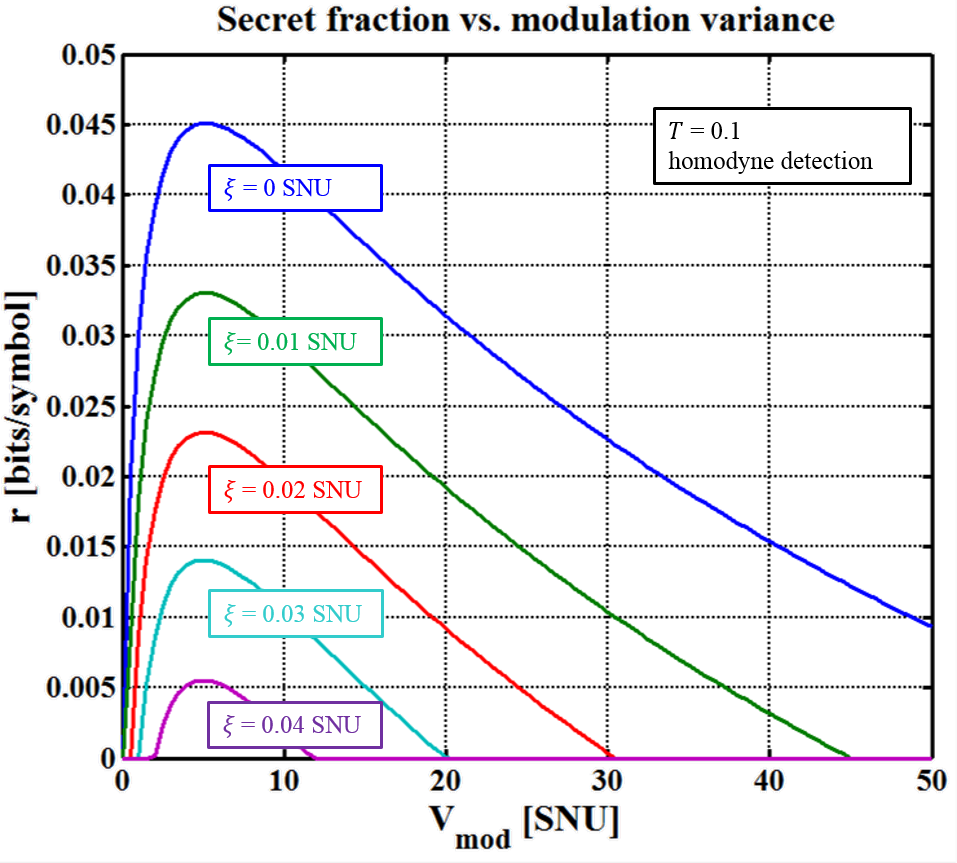}}
\caption{Secret fraction vs.\ modulation variance. The smaller the reconciliation efficiency $\beta$ (a) and the higher the excess noise $\xi$ (b), the smaller becomes the interval of the modulation variance in which a non-zero secure key can be established.}
\label{Vmod}
\end{figure}

Next to the excess noise which originates from various hardware imperfections, efficient software algorithms turn out to be a serious bottleneck in CV-QKD. While imperfect reconciliation $\beta < 1$ is detrimental for Alice's and Bob's mutual information, it does not affect Eve's information on the secret key~\eqref{eq_secfrac}. Figure~\ref{beta} depicts the secret fraction with respect to the reconciliation efficiency for various excess-noise parameters and channel lenghts. Figure~\ref{Vmod}(a) shows that in the idealised case of $\beta=1$ the modulation variance (and therefore the SNR) can be increased to an arbitrary level since Alice and Bob can use the full mutual information for their key and do not leave any information advantage to Eve.

Using variable optical attenuation the modulation variance $V_{\text{mod}}=2\braket{n}$ can be tuned arbitrarily and optimised to the other characteristic parameters $T$, $\xi$ and $\beta$. The dependence of the optimal modulation variance and its tolerance on the reconciliation efficiency and excess noise are illustrated in Fig.~\ref{Vmod}.

\chapter*{Acknowledgements}
\addcontentsline{toc}{chapter}{Acknowledgements}

We thank Martin Suda for proofreading and contributing valuable remarks. Moreover, we gratefully acknowledge the time and effort the anonymous reviewers have dedicated to profoundly study and improve this lengthy manuscript. Their recommendations and additional references significantly contributed to the quality, integrity and thoroughness of the final article.

\begin{appendices}

\chapter{Prerequisites on Coherent States} \label{ch_coherent}

Coherent states $\ket{\alpha}$ are eigenstates of the annihilation operator, hence

\begin{equation} \label{eq_coherentEVequation}
\hat{a} \ket{\alpha} = \alpha \ket{\alpha} ,
\end{equation}
where $\alpha$ is a complex number whose real and imaginary part can be understood as the quadrature components in phase space:

\begin{equation}
\alpha = q + ip .
\end{equation}
The ladder operators $\hat{a}$ and $\hat{a}^{\dagger}$ are uniquely defined by their action on a Fock state $\ket{n}$:
\begin{subequations}
\begin{align}
\hat{a}^{\dagger} \ket{n} & = \sqrt{n+1} \ket{n+1} , \\
\hat{a} \ket{n} & = \sqrt{n} \ket{n-1} , \\
\hat{a}^{\dagger} \hat{a} \ket{n} & \coloneq \hat{n} \ket{n} = n \ket{n} ,
\end{align}
\end{subequations}
where $\hat{n}$ is the boson-number operator and $n$ is its eigenvalue; and their commutation relations:

\begin{subequations}
\begin{align}
[\hat{a}_{j},\hat{a}^{\dagger}_{k}] & = \delta_{jk} \mathbb{1} , \\
[\hat{a}^{\dagger}_{j},\hat{a}^{\dagger}_{k}] & = [\hat{a}_{j},\hat{a}_{k}] = 0 .
\end{align}
\end{subequations}
Moreover, from the above relation immediately follows the result

\begin{align}\label{commut}
\hat{a} \hat{a}^{\dagger}&=\hat{n} + \mathbb{1} . 
\end{align}
The ladder operators can be reexpressed in terms of quadrature operators (given in shot-noise units):

\begin{subequations}
\begin{align}
\hat{a} & =\frac{1}{2} ( \hat{q} + i \hat{p}), \\
\hat{a}^{\dagger} & =\frac{1}{2} ( \hat{q} - i \hat{p}) ,
\end{align}
\end{subequations}
which immediately leads to

\begin{subequations}
\begin{align}
\hat{q} & = \hat{a} + \hat{a}^{\dagger} , \\
\hat{p} & = -i (\hat{a} - \hat{a}^{\dagger}) .
\end{align}
\end{subequations}
The two quadrature operators then fulfil the commutation relation

\begin{align}
[\hat{q},\hat{p}] & =-i \ [\hat{a} + \hat{a}^{\dagger},\hat{a} - \hat{a}^{\dagger}]\notag \\
& = -i \left( \underbrace{[\hat{a},\hat{a}]}_{=0} - \underbrace{[\hat{a},\hat{a}^{\dagger}]}_{=1} + \underbrace{[\hat{a}^{\dagger},\hat{a}]}_{=-1} - \underbrace{[\hat{a}^{\dagger},\hat{a}^{\dagger}]}_{=0} \right) = 2i .
\end{align}
In a coherent state both quadratures carry the same uncertainty (standard deviation), hence the same variance:

\begin{equation}
V(\hat{q})=V(\hat{p}).
\end{equation}
By definition of a coherent state the uncertainty product $\delta \hat{q} \delta \hat{p}$ is minimal. In order to derive the uncertainty of the quadrature operators of a coherent state, we begin with the definition of the variance:

\begin{equation} \label{eq_V}
V(\hat{q})=\braket{\hat{q}^{2}}-\braket{\hat{q}}^{2} .
\end{equation}
The expectation value $\braket{\hat{q}}$ can be obtained very quickly:

\begin{align} \label{eq_exp1}
\braket{\hat{q}} & = \braket{\alpha | \hat{q} | \alpha} = \braket{\alpha | \hat{a} + \hat{a}^{\dagger} | \alpha} = \braket{\alpha | \hat{a} | \alpha} + \braket{\alpha | \hat{a}^{\dagger} | \alpha} = \alpha \braket{\alpha | \alpha} + \alpha^{*} \braket{\alpha | \alpha} \notag \\ 
& = \alpha + \alpha^{*} = (q+ip) + (q-ip) = 2q .
\end{align}
To compute the expectation value of the squared operator we proceed similarly:\footnote{We can use the result \eqref{eq_exp2} to relate the modulation variance of quadrature \emph{operators} and \emph{components} in Gaussian modulated coherent-state CV-QKD. Taking $\braket{\hat{q}^{2}} = 4 q^{2} + 1$ and assuming $q$ to be a realization of a random variable $\randq$ that is distributed as $\randq \sim \mathcal{N} (0,\tilde{V}_{\text{mod}})$, we obtain $\braket{\hat{q}^{2}} = 4 \braket{\randq^{2}} + 1$ or $V(\hat{q}) = 4 \tilde{V}_{\text{mod}} + 1$.}

\begin{align} \label{eq_exp2}
\braket{\hat{q}^{2}} & = \braket{\alpha | \hat{q}^{2} | \alpha} = \braket{\alpha | ( \hat{a} + \hat{a}^{\dagger})^{2} | \alpha} \notag \\
& = \braket{\alpha | \hat{a}^{2} | \alpha} + \braket{\alpha | (\hat{a}^{\dagger})^{2}  | \alpha} + \braket{\alpha | \underbrace{\hat{a}\hat{a}^{\dagger}}_{=\hat{a}^{\dagger}\hat{a} + \mathbb{1}} | \alpha} + \braket{\alpha | \hat{a}^{\dagger}\hat{a} | \alpha} \notag \\
& = \alpha^{2} + (\alpha^{*})^{2} + \alpha^{*}\alpha  + 1+  \alpha^{*}\alpha \notag \\ 
& = q^{2}-p^{2}+2iqp + q^{2}-p^{2} - 2iqp + 2(q^{2}+p^{2}) + 1 \notag \\
& = 4 q^{2} + 1 .
\end{align}
Reinserting \eqref{eq_exp1} and \eqref{eq_exp2} into \eqref{eq_V} yields

\begin{equation}
V(\hat{q})=4 q^{2} + 1 -4 q^{2} = 1 ,
\end{equation}
which is the minimal uncertainty in shot-noise units, often referred to as shot noise. $V(\hat{p})$ is computed analogously yielding the same result. Thus the uncertainty relation in SNU is given by

\begin{equation}\label{eq:uncertaintyrelSNU}
\delta \hat{q} \delta \hat{p} \geq 1 .
\end{equation}
The expectation value of the number operator $\hat{n}$ is 

\begin{align}
\braket{\hat{n}}=\braket{\alpha | \hat{n} | \alpha } = \braket{\alpha | \hat{a}^{\dagger} \hat{a} | \alpha }  = \alpha^{*}\alpha \braket{\alpha | \alpha } = |\alpha|^{2} = q^{2} + p^{2} ,
\end{align}
or in terms of the quadrature operators

\begin{align} \label{expect_n}
\hat{n} & =  \hat{a}^{\dagger}\hat{a}  \notag \\
& = \frac{1}{4}  ( \hat{q} - i \hat{p}) ( \hat{q} + i \hat{p}) \notag \\
& = \frac{1}{4} \left( \hat{q}^{2} + \hat{p}^{2} + \underbrace{ i \ [\hat{q},\hat{p}] }_{=-2} \right) \notag \\
& = \frac{1}{4} \left( \hat{q}^{2} + \hat{p}^{2} \right) - \frac{1}{2}
\end{align}

with the expectation value
\begin{align}
\braket{\hat{n}} = \frac{1}{4} \left( \braket{ \hat{q}^{2} } + \braket{ \hat{p}^{2}} \right) - \frac{1}{2} .
\end{align}

\chapter{Prerequisites on Gaussian Quantum Information} \label{ch_gaussianquinfo}

\section{Displacement Vector and Covariance Matrix}

This section summarises, in an eclectic way, the basic theoretic tools required to analyse, interpret and transform Gaussian quantum states. We will restrict ourselves to what is essential for the analysis of continuous-variable QKD. These tools are selected from the arsenal of the Gaussian-quantum-information formalism which is comprehensively described in \cite{weedbrook2012gaussian}. In this formalism an $N$-mode Gaussian state is represented by a displacement vector $\mathbf{\hat{x}}$ of dimension $2N$ and a covariance matrix $\Sigma$ of dimension $2N \times 2N$. The displacement vector reads

\begin{equation}
\mathbf{\hat{x}} = \left( \hat{q}_{1} , \hat{p}_{1} , \hat{q}_{2} , \hat{p}_{2} , \dots \hat{q}_{N} , \hat{p}_{N} \right)^{T} ,
\end{equation}
and its elements fulfil the commutation relation

\begin{equation}
[\hat{x}^{j},\hat{x}^{k}] = 2i \Omega^{jk} ,
\end{equation}
where $\Omega$ is the $2N \times 2N$ matrix

\begin{equation} \label{eq_Omega}
\Omega=\bigoplus_{l=1}^{N}
\begin{pmatrix}
0 & 1 \\
-1 & 0
\end{pmatrix} .
\end{equation}
The elements of the covariance matrix are given by

\begin{align}
\Sigma^{ij} & = \frac{1}{2} \left\langle \left\lbrace \hat{x}^{j} - \langle \hat{x}^{j} \rangle ,\hat{x}^{k} - \langle \hat{x}^{k} \rangle \right\rbrace \right\rangle  \notag \\
& = \frac{1}{2} \left( \langle \hat{x}^{j}\hat{x}^{k} \rangle + \langle \hat{x}^{k}\hat{x}^{j} \rangle \right) - \langle \hat{x}^{j} \rangle \langle \hat{x}^{k} \rangle \eqcolon \mathbb{E} (\hat{x}^{j},\hat{x}^{k}) .
\end{align}
Note that along the diagonal we have

\begin{align}
\Sigma^{jj} = \langle (\hat{x}^{j})^{2} \rangle - \langle \hat{x}^{j} \rangle^{2} = V(\hat{x}^{j}).
\end{align}
In matrix notation we get

\begin{equation}
\Sigma = 
\begin{pmatrix}
V(\hat{q}_{1}) & \mathbb{E} (\hat{q}_{1},\hat{p}_{1}) & \mathbb{E} (\hat{q}_{1},\hat{q}_{2}) & \mathbb{E} (\hat{q}_{1},\hat{p}_{2}) & \cdots & \mathbb{E} (\hat{q}_{1},\hat{q}_{N}) & \mathbb{E} (\hat{q}_{1},\hat{p}_{N}) \\
  & V(\hat{p}_{1}) & \mathbb{E} (\hat{p}_{1},\hat{q}_{2}) & \mathbb{E} (\hat{p}_{1},\hat{p}_{2}) & \cdots & \mathbb{E} (\hat{p}_{1},\hat{q}_{N}) & \mathbb{E} (\hat{p}_{1},\hat{p}_{N}) \\
  &  & V(\hat{q}_{2}) & \mathbb{E} (\hat{q}_{2},\hat{p}_{2}) & \cdots & \mathbb{E} (\hat{q}_{2},\hat{q}_{N}) & \mathbb{E} (\hat{q}_{2},\hat{p}_{N}) \\
  &  &  & V(\hat{p}_{2}) & \cdots & \mathbb{E} (\hat{p}_{2},\hat{q}_{N}) & \mathbb{E} (\hat{p}_{2},\hat{p}_{N}) \\
  &  &  &  & \ddots & \vdots &\vdots \\
  &  \multicolumn{2}{l}{{\Large \text{symmetric}}} &  &  & V(\hat{q}_{N}) & \mathbb{E} (\hat{q}_{N},\hat{p}_{N}) \\
  &  &  &  &  &  & V(\hat{p}_{N}) \\
\end{pmatrix} .
\end{equation}
The covariance matrix describes the mutual correlations of the mode quadratures. If the cross terms $\mathbb{E} (\hat{x}^{j},\hat{x}^{k})$ are zero for the $q$- and $p$-quadratures of two modes, then these two modes are uncorrelated to each other, i.e.\ separable.

\section{Symplectic Operations} \label{sec_symplop}

Unitary operators in Hilbert-space notation correspond to \emph{symplectic} operators in the formalism of Gaussian quantum information. An operator, represented by a matrix $S$ is symplectic if the relation

\begin{equation}
S\Omega S^{T} = \Omega
\end{equation}
holds. A symplectic operation transforms a Gaussian state as follows:

\begin{subequations}
\begin{align}
\mathbf{\hat{x}} & \longrightarrow S \mathbf{\hat{x}} , \\
\Sigma & \longrightarrow S \Sigma S^{T} .
\end{align}
\end{subequations}
The most important symplectic operator in the realm of this paper is the beamsplitter (BS), represented as

\begin{equation}
\text{BS} =
\begin{pmatrix}
\sqrt{T} \mathbb{1}_{2} & \sqrt{1-T} \mathbb{1}_{2} \\
-\sqrt{1-T} \mathbb{1}_{2} & \sqrt{T} \mathbb{1}_{2} 
\end{pmatrix} .
\end{equation}

\section{Composite States}

In Hilbert-space representation, two states $A$ and $B$ constitute a total state by action of the direct product:

\begin{equation}
\rho_{\text{tot}}=\rho_{A} \otimes \rho_{B} .
\end{equation}
In Gaussian quantum information two states $A$ and $B$ with $M$ and $N$ modes respectively combine to a displacement vector of dimension $2(M+N)$:

\begin{equation}
\mathbf{\hat{x}}_{\text{tot}} = \binom{\mathbf{\hat{x}}_{A}}{\mathbf{\hat{x}}_{B}} ,
\end{equation}
and the covariance matrix describing the total state will be

\begin{equation}
\Sigma_{\text{tot}}= \Sigma_{A} \oplus \Sigma_{B} =
\begin{pmatrix}
\Sigma_{A} & 0 \\
0 & \Sigma_{B}
\end{pmatrix}
\end{equation}
with dimension $2(M+N) \times 2(M+N)$. Reversely, a state Gaussian state is separable if and only if it can be written as a direct sum in the above fashion. Otherwise, if there are non-zero off-diagonal terms in $\Sigma_{\text{tot}}$ between $\Sigma_{A}$ and $\Sigma_{B}$ (like it is the case for a TMSVS) we speak of an entangled state.

\section{Von Neumann entropy} \label{sec_vonneumann}

The von Neumann entropy of a Gaussian state is described in terms of the \emph{symplectic eigenvalues} of its covariance matrix. The symplectic eigenvalues of a matrix $\Sigma$ are defined as the modulus of the ordinary eigenvalues of the matrix

\begin{equation}
\tilde{\Sigma}= i \Omega \Sigma
\end{equation}
with $\Omega$ defined as in \eqref{eq_Omega}. Since only the modulus of the eigenvalues is considered, an $2N \times 2N$ covariance matrix has exactly $N$ symplectic eigenvalues. A Gaussian state with covariance matrix $\Sigma$ has a von Neumann entropy, given by

\begin{equation} \label{eq_vonneumann}
S=\sum_{i} g(\nu_{i})
\end{equation}
with

\begin{equation}
g(\nu)= \left( \frac{\nu+1}{2} \right) \log_{2} \left( \frac{\nu+1}{2} \right) - \left( \frac{\nu-1}{2} \right) \log_{2} \left( \frac{\nu-1}{2} \right)
\end{equation}
and $\nu_{i}$ being the symplectic eigenvalues of $\Sigma$.

\section{Partial Measurements} \label{sec_partial}

Consider an $N$-mode Gaussian state, described by a covariance matrix of dimension $2N \times 2N$ (two quadratures per mode). A measurement on one mode will modify the Gaussian state of the remaining modes, depending on their correlations to the measured one. Let $B$ be the mode on which a measurement is performed, $A$ be the $N-1$ remaining modes and $C$ the correlations between $B$ and $A$. Then the covariance matrix of the total state before the measurement can be written as:

\begin{equation} \label{eq_generalcovmat}
\Sigma =
\begin{pmatrix}
\Sigma_{A} & \Sigma_{C} \\
\Sigma_{C}^{T} & \Sigma_{B} 
\end{pmatrix} ,
\end{equation}
where $\Sigma_{A} \in \mathbb{R}^{2(N-1) \times 2(N-1)}$, $\Sigma_{B} \in \mathbb{R}^{2 \times 2}$ and $\Sigma_{C} \in \mathbb{R}^{2(N-1) \times 2}$. The impact of a partial measurement of mode $B$ on the covariance matrix of the other ones, $\Sigma_{A} \stackrel{B}{\longrightarrow} \Sigma_{A|B}$, depends on whether homodyne or heterodyne measurement is performed, i.e.\ on whether only one or both of $B$'s quadratures are measured. In case of \emph{homodyne detection} of $q$ or $p$, a partial measurement of $B$ transforms $A$ as follows:

\begin{equation} \label{eq_partialhom1}
\Sigma_{A|B} = \Sigma_{A} - \Sigma_{C} ( \Pi_{q,p} \Sigma_{B} \Pi_{q,p})^{-1} \Sigma_{C}^{T},
\end{equation}
where $\Pi_{q} = \diag (1,0)$ (in case $q$ is measured) and $\Pi_{p} = \text{diag} (0,1)$ (in case $p$ is measured). So

\begin{subequations} \label{eq_projectionhom}
\begin{align}
\Pi_{q} \Sigma_{B} \Pi_{q} =
\begin{pmatrix}
\Sigma_{B}^{11} & 0 \\
0 & 0
\end{pmatrix}
=
\begin{pmatrix}
V(\hat{q}_{B}) & 0 \\
0 & 0
\end{pmatrix}
\qquad q\text{-measurement,} \\
\Pi_{p} \Sigma_{B} \Pi_{p} =
\begin{pmatrix}
0 & 0 \\
0 & \Sigma_{B}^{22}
\end{pmatrix}
=
\begin{pmatrix}
0 & 0 \\
0 & V(\hat{p}_{B})
\end{pmatrix}
\qquad p\text{-measurement.}
\end{align}
\end{subequations}
The expression $^{-1}$ in \eqref{eq_partialhom1} indicates a pseudoinverse. However, if a matrix is diagonal, as it's the case for $\Pi \Sigma_{B} \Pi$ (for any $\Sigma_{B}$), the pseudoinverse is just an inversion of the elements: $\diag(a_{1},\dots,a_{n})^{-1}=\diag(a_{1}^{-1},\dots,a_{n}^{-1})$. Using this and \eqref{eq_projectionhom}, we can reexpress \eqref{eq_partialhom1} in the convenient form

\begin{equation} \label{eq_partialhom}
\boxed{
\Sigma_{A|B} = \Sigma_{A} - \frac{1}{V(q_{B},p_{B})} \Sigma_{C} \Pi_{q,p}  \Sigma_{C}^{T} .
}
\end{equation}
A \emph{heterodyne detection} of the mode $B$ will affect the other modes as follows:

\begin{equation} \label{eq_partialhet1}
\Sigma_{A|B} = \Sigma_{A} - \Sigma_{C} (\Sigma_{B}+\mathbb{1}_{2} )^{-1} \Sigma_{C}^{T} .
\end{equation}
The addition of the unit matrix $\mathbb{1}_{2}$ can be intuitively understood as a portion of shot-noise, added to the variances of $q$ and $p$ through the vacuum input of the balanced beamsplitter. This equation can be nicely derived using only some basic considerations and the relation for a partial homodyne measurement \eqref{eq_partialhom}: Consider a balanced beamsplitter, splitting the mode $B$ in half in order to allow for a subsequent simultaneous measurement of both quadratures $q$ and $p$. According to the methods described in Appendix~\ref{sec_symplop}, this BS transforms the covariance matrix \eqref{eq_generalcovmat} to

\begin{align}
\Sigma' & = 
\begin{pmatrix}
\mathbb{1}_{2} & 0 & 0 \\
0 & \frac{1}{\sqrt{2}} \mathbb{1}_{2} \ & \frac{1}{\sqrt{2}} \mathbb{1}_{2} \\
0 & -\frac{1}{\sqrt{2}} \mathbb{1}_{2} & \frac{1}{\sqrt{2}} \mathbb{1}_{2} \\
\end{pmatrix}
\begin{pmatrix}
\Sigma_{A} & \Sigma_{C} & 0 \\
\Sigma_{C}^{T} & \Sigma_{B} & 0 \\
0 & 0 & \mathbb{1}_{2} 
\end{pmatrix}
\begin{pmatrix}
\mathbb{1}_{2} & 0 & 0 \\
0 & \frac{1}{\sqrt{2}} \mathbb{1}_{2} \ & -\frac{1}{\sqrt{2}} \mathbb{1}_{2} \\
0 & \frac{1}{\sqrt{2}} \mathbb{1}_{2} & \frac{1}{\sqrt{2}} \mathbb{1}_{2} \\
\end{pmatrix} \notag \\
& =
\bordermatrix{
 & A & B_{1} & B_{2} \cr
A &\Sigma_{A} & \frac{1}{\sqrt{2}}\Sigma_{C} & -\frac{1}{\sqrt{2}}\Sigma_{C} \cr
B_{1} &\frac{1}{\sqrt{2}}\Sigma_{C} & \frac{\Sigma_{B}+\mathbb{1}_{2}}{2} & \frac{-\Sigma_{B}+\mathbb{1}_{2}}{2} \cr
B_{2} &-\frac{1}{\sqrt{2}}\Sigma_{C} & \frac{-\Sigma_{B}+\mathbb{1}_{2}}{2} & \frac{\Sigma_{B}+\mathbb{1}_{2}}{2} 
} 
\eqcolon
\begin{pmatrix}
\Sigma_{A} & \Sigma_{\gamma} & -\Sigma_{\gamma} \\
\Sigma_{\gamma} &  \Sigma_{\beta}  & \Sigma_{\delta} \\
-\Sigma_{\gamma} & \Sigma_{\delta} &  \Sigma_{\beta} 
\end{pmatrix} .
\end{align}
So we added an additional (vacuum) mode to the total system and performed a BS operation, mixing the vacuum mode with the mode $B$. Since $B$ is now split in half, we can perform two homodyne measurements, one on each half of the initial $B$. Say we are measuring the $q$-quadrature on mode $B_{2}$. According to \eqref{eq_partialhom}, Alice's mode, conditioned on a homodyne measurement on $B_{2}$, is represented by the covariance matrix

\begin{equation}
\Sigma_{A'}\coloneq\Sigma_{A|B_{2}} = \Sigma_{A} - \frac{1}{\Sigma_{\beta}^{11}} \Sigma_{\gamma} \Pi_{q}  \Sigma_{\gamma}^{T} .
\end{equation}
Now let a second partial measurement take place, i.e.\ let the quadrature $p$ be measured on the mode $B_{1}$. This will affect the $A$ mode a second time in a similar way:

\begin{equation}
\Sigma_{A''}\coloneq\Sigma_{A|B_{1}B_{2}} = \Sigma_{A'} - \frac{1}{\Sigma_{\beta}^{22}} \Sigma_{\gamma} \Pi_{p}  \Sigma_{\gamma}^{T} .
\end{equation}
Putting the above two equations together, we obtain

\begin{align}
\Sigma_{A|B_{1}B_{2}} & = \Sigma_{A} -\frac{1}{\Sigma_{\beta}^{11}} \Sigma_{\gamma} \Pi_{q}  \Sigma_{\gamma}^{T} - \frac{1}{\Sigma_{\beta}^{22}} \Sigma_{\gamma} \Pi_{p}  \Sigma_{\gamma}^{T},
\end{align}
and since $\Sigma_{\beta}=(\Sigma_{B}+\mathbb{1})/2$ and $\Sigma_{\gamma}=\Sigma_{C}/\sqrt{2}$,

\begin{align}
\Sigma_{A|B_{1}B_{2}} & = \Sigma_{A} - \frac{2}{\Sigma_{B}^{11}+1} \frac{1}{2} \Sigma_{C} \Pi_{q}  \Sigma_{C}^{T} - \frac{2}{\Sigma_{B}^{22}+1} \frac{1}{2} \Sigma_{C} \Pi_{p}  \Sigma_{C}^{T} \notag \\
& =  \Sigma_{A} - \Sigma_{C} \left( \frac{1}{\Sigma_{B}^{11}+1}  \Pi_{q} + \frac{1}{\Sigma_{B}^{22}+1} \Pi_{p} \right) \Sigma_{C}^{T} \notag \\
& =  \Sigma_{A} - \Sigma_{C} 
\begin{pmatrix}
\frac{1}{\Sigma_{B}^{11}+1} & 0 \\
0 & \frac{1}{\Sigma_{B}^{22}+1}
\end{pmatrix}
\Sigma_{C}^{T} \notag \\
& =  \Sigma_{A} - \Sigma_{C} 
\begin{pmatrix}
\Sigma_{B}^{11}+1 & 0 \\
0 & \Sigma_{B}^{22}+1
\end{pmatrix}^{-1}
\Sigma_{C}^{T} \notag \\
& = \Sigma_{A} - \Sigma_{C} (\Sigma_{B}+\mathbb{1}_{2} )^{-1} \Sigma_{C}^{T} ,
\end{align}
which coincides exactly with \eqref{eq_partialhet1}. When $B$ is a coherent state with $V(\hat{q}_{B})=V_{B}(\hat{p}_{B})\eqcolon V_{B}$, \eqref{eq_partialhet1} simplifies to

\begin{equation} \label{eq_partialhet}
\boxed{
\Sigma_{A|B} = \Sigma_{A} - \frac{1}{V_{B}+1} \Sigma_{C} \Sigma_{C}^{T} .
}
\end{equation}

\chapter{Derivation of the Covariance Matrix} 

In this chapter we will derive the initial covariance matrix of a two-mode squeezed vacuum state (TMSVS) which is in ket-notation represented as \cite{weedbrook2012gaussian}

\begin{align}
\ket{\Psi} = \left( \frac{2}{V+1} \right)^{1/2} \sum_{k=0}^{\infty} \left( \frac{V-1}{V+1} \right)^{k/2} \ket{k,k} .
\end{align}
The expectation values will be calculated according to

\begin{subequations}
\begin{align}
\langle \hat{x}_{i} \rangle & = \braket{\Psi | \hat{x}_{i} | \Psi} , \\
\langle \hat{x}_{i} \hat{x}_{j} \rangle & = \braket{\Psi | \hat{x}_{i}\hat{x}_{j} | \Psi} .
\end{align}
\end{subequations}
Some useful relations, helpful at deriving the elements of the covariance matrix, are

\begin{align} \label{EVzero}
\braket{\Psi | \hat{a}^{\dagger}_{*} | \Psi} = \braket{\Psi | \hat{a}_{*} | \Psi} = \braket{\Psi | (\hat{a}^{\dagger}_{*})^{2} | \Psi} = \braket{\Psi | \hat{a}_{*}^{2} | \Psi} = \braket{\Psi | \hat{a}^{\dagger}_{A} \hat{a}_{B} | \Psi} = \braket{\Psi | \hat{a}_{A} \hat{a}^{\dagger}_{B} | \Psi} = 0
\end{align}
(with $*=A,B$ respectively) due to exclusively orthogonal Fock states in the scalar product. These results turn out useful when we try for example to find the expectation value of the quadrature operators of the TMSVS:

\begin{subequations}
\begin{align} 
\langle \hat{q}_{*} \rangle & = \braket{\Psi | \hat{q}_{*} | \Psi} = \braket{\Psi | \hat{a}_{*} | \Psi} + \braket{\Psi | \hat{a}^{\dagger}_{*} | \Psi} = 0 \label{EVq} \\
\langle \hat{p}_{*} \rangle & = \braket{\Psi | \hat{p}_{*} | \Psi} = -i \left( \braket{\Psi | \hat{a}_{*} | \Psi}  - \braket{\Psi | \hat{a}^{\dagger}_{*} | \Psi} \right) = 0 \label{EVp}
\end{align}
\end{subequations}
We quickly make sure the normalisation condition is fulfilled:

\begin{align}
\braket{\Psi | \Psi} = \frac{2}{V+1} \underbrace{\sum_{k=0}^{\infty} \left( \frac{V-1}{V+1} \right)^{k} }_{\stackrel{\footnotemark}{=}\frac{V+1}{2}} = 1 .
\end{align}
\footnotetext{Note that $\sum_{k=0}^{\infty}q^{k}=1/(1-q)$ for $q<1$. So with $q=(V-1)/(V+1)=1-2/(V+1)$ we get $(1-q)^{-1}=(2/(V+1))^{-1}=(V+1)/2$.}

\section{Initial Covariance Matrix} \label{sec_covmatinit}

A TMSVS is described by a $4\times4$-matrix $\Sigma$. In the following sections we will derive its 16 components individually.

\subsection{$\Sigma_{11}$, $\Sigma_{22}$, $\Sigma_{33}$, $\Sigma_{44}$}

The diagonal terms are given by the variances of the quadrature operators $\hat{q}_{A}$, $\hat{p}_{A}$, $\hat{q}_{B}$ and $\hat{p}_{B}$. The variance of an operator can be expressed in terms of the expectation values of the operator and its square; and the expectation values are computed by reexpressing the quadrature operators in terms of ladder operators:

\begin{align}
\Sigma_{11} & = \langle \hat{q}_{A}^{2} \rangle - \underbrace{ \langle \hat{q}_{A} \rangle^{2}}_{\stackrel{\eqref{EVq}}{=} 0} = \langle \hat{q}_{A}^{2} \rangle = \left\langle \left( \hat{a}_{A} + \hat{a}^{\dagger}_{A} \right)^{2} \right\rangle = \notag \\
& = \underbrace{ \braket{\Psi | (\hat{a}^{\dagger}_{A})^{2} | \Psi} }_{\stackrel{\eqref{EVzero}}{=}0} + \braket{\Psi | \hat{a}^{\dagger}_{A} \hat{a}_{A} | \Psi} + \braket{\Psi | \underbrace{ \hat{a}_{A} \hat{a}^{\dagger}_{A} }_{\stackrel{\eqref{commut}}{=} \hat{a}_{A}^{\dagger}\hat{a}_{A} + \mathbb{1}} | \Psi}  + \underbrace{ \braket{\Psi | \hat{a}_{A}^{2} | \Psi} }_{=0} = 2 \braket{\Psi | \hat{a}^{\dagger}_{A} \hat{a}_{A} | \Psi} + \braket{\Psi|\Psi} \notag \\
& = \frac{4}{V+1} \underbrace{ \sum_{k=0}^{\infty} \left( \frac{V-1}{V+1} \right)^{k} k }_{\stackrel{\footnotemark}{=} 1/4(V-1)(V+1)} + 1 = V - 1 +1 = V
\end{align}
In analogous fashion we get $\Sigma_{22}=\Sigma_{33}=\Sigma_{44}=V$.

\footnotetext{Note that $\sum_{k=0}^{\infty}kq^{k}=q\partial_{q}\sum_{k=0}^{\infty}q^{k}=q\partial_{q}1/(1-q)=q/(1-q)^{2}$ for $q<1$. So with $q=(V-1)/(V+1)=1-2/(V+1)$ we get $q(1-q)^{-2}=(V-1)/(V+1)(2/(V+1))^{-2}=1/4(V-1)(V+1)$.}

\subsection{$\Sigma_{12}$, $\Sigma_{21}$, $\Sigma_{34}$, $\Sigma_{43}$}

We proceed similarly as in the previous section and compute the required expectation values of the quadrature operators by reexpressing them in terms of creation and annihilation operators:

\begin{align}
\Sigma_{12} = \Sigma_{21} &  = \frac{1}{2} \left( \langle \hat{q}_{A} \hat{p}_{A} \rangle +  \langle \hat{p}_{A} \hat{q}_{A} \rangle \right) - \langle \hat{q}_{A} \rangle \langle \hat{p}_{A} \rangle = \frac{1}{2} \left( \langle \hat{q}_{A} \hat{p}_{A} \rangle +  \langle \hat{p}_{A} \hat{q}_{A} \rangle \right)
\end{align}
with

\begin{subequations}
\begin{align}
\langle \hat{q}_{A} \hat{p}_{A} \rangle & = i \left( \underbrace{ \braket{\Psi | (\hat{a}^{\dagger}_{A})^{2} | \Psi} }_{\stackrel{\eqref{EVzero}}{=} 0 } - \braket{\Psi | \hat{a}^{\dagger}_{A} \hat{a}_{A} | \Psi} + \braket{\Psi | \underbrace{\hat{a}_{A} \hat{a}^{\dagger}_{A}}_{= \hat{a}_{A}^{\dagger}\hat{a}_{A} + \mathbb{1}} | \Psi} - \underbrace{ \braket{\Psi | \hat{a}_{A}^{2} | \Psi} }_{=0} \right) \notag \\
& = i \left( - \braket{\Psi | \hat{a}^{\dagger}_{A} \hat{a}_{A} | \Psi} + \braket{\Psi | \hat{a}^{\dagger}_{A} \hat{a}_{A} | \Psi} + \braket{\Psi | \Psi} \right) = i \\
\langle \hat{p}_{A} \hat{q}_{A} \rangle & = i \left(  \braket{\Psi | (\hat{a}^{\dagger}_{A})^{2} | \Psi} + \braket{\Psi | \hat{a}^{\dagger}_{A} \hat{a}_{A} | \Psi} - \braket{\Psi | \hat{a}_{A} \hat{a}^{\dagger}_{A} | \Psi} - \braket{\Psi | \hat{a}_{A}^{2} | \Psi} \right) \notag \\
& = i \left( \braket{\Psi | \hat{a}^{\dagger}_{A} \hat{a}_{A} | \Psi} - \braket{\Psi | \hat{a}^{\dagger}_{A} \hat{a}_{A} | \Psi} - \braket{\Psi | \Psi} \right) = -i
\end{align}
\end{subequations}
Thus we get
\begin{align}
\Sigma_{12} = \Sigma_{21} = 0
\end{align}
and analogously for $\Sigma_{34}=\Sigma_{43}=0$.

\subsection{$\Sigma_{13}$, $\Sigma_{31}$, $\Sigma_{24}$, $\Sigma_{42}$}

\begin{align}
\Sigma_{13}=\Sigma_{31} &  = \frac{1}{2} \left( \langle \hat{q}_{A} \hat{q}_{B} \rangle +  \langle \hat{q}_{B} \hat{q}_{A} \rangle \right) - \langle \hat{q}_{A} \rangle \langle \hat{q}_{B} \rangle = \langle \hat{q}_{A} \hat{q}_{B} \rangle \notag \\
& = \braket{\Psi | \hat{a}^{\dagger}_{A} \hat{a}^{\dagger}_{B} | \Psi} + \underbrace{ \braket{\Psi | \hat{a}^{\dagger}_{A} \hat{a}_{B} | \Psi} }_{=0} + \underbrace{\braket{\Psi | \hat{a}_{A} \hat{a}^{\dagger}_{B} | \Psi}}_{=0} +\braket{\Psi | \hat{a}_{A} \hat{a}_{B} | \Psi} ,
\end{align}
where

\begin{subequations}
\begin{align}
\braket{\Psi | \hat{a}^{\dagger}_{A} \hat{a}^{\dagger}_{B} | \Psi} & = \frac{2}{V+1} \left( \sum_{j=0}^{\infty} \left( \frac{V-1}{V+1} \right)^{j/2} \bra{j,j} \right) \left( \sum_{k=0}^{\infty} \left( \frac{V-1}{V+1} \right)^{k/2} (k+1) \ket{k+1,k+1} \right) \notag \\
& = \frac{2}{V+1} \underbrace{ \sum_{k=0}^{\infty} \left( \frac{V-1}{V+1} \right)^{k+1/2} (k+1) }_{\frac{1}{4} \left( \frac{V-1}{V+1} \right)^{1/2} (V+1)^{2} } \notag \\
& = \frac{1}{2} \left( \frac{V-1}{V+1} \right)^{1/2} (V+1) \notag \\
& = \frac{1}{2} \sqrt{V^{2}-1} , \\
\braket{\Psi | \hat{a}_{A} \hat{a}_{B} | \Psi} & \stackrel{\footnotemark}{=} \frac{2}{V+1} \left( \sum_{j=0}^{\infty} \left( \frac{V-1}{V+1} \right)^{j/2} \bra{j,j} \right) \left( \sum_{k=0}^{\infty} \left( \frac{V-1}{V+1} \right)^{k/2} k \ket{k-1,k-1} \right) \notag \\
& = \frac{2}{V+1} \underbrace{ \sum_{k=1}^{\infty} \left( \frac{V-1}{V+1} \right)^{k-1/2} k }_{\frac{1}{4} \left( \frac{V+1}{V-1} \right)^{1/2} (V+1)(V-1) } \notag \\
& = \frac{1}{2} \left( \frac{V+1}{V-1} \right)^{1/2} (V-1) \notag \\
& = \frac{1}{2} \sqrt{V^{2}-1} .
\end{align}
\end{subequations}
Therefore we obtain
\footnotetext{or alternatively $\braket{\Psi | \hat{a}_{A} \hat{a}_{B} | \Psi} \coloneq\ \braket{\Psi | \Psi'} = \braket{\Psi' | \Psi}^{*}=\braket{\Psi | \hat{a}^{\dagger}_{A} \hat{a}^{\dagger}_{B} | \Psi}^{*}=\frac{1}{2} \sqrt{V^{2}-1}$.}

\begin{align}
\Sigma_{13}=\Sigma_{31}=\frac{1}{2} \sqrt{V^{2}-1} + \frac{1}{2} \sqrt{V^{2}-1} = \sqrt{V^{2}-1} ,
\end{align}
and analogously

\begin{align}
\Sigma_{24}=\Sigma_{42} & = \frac{1}{2} \left( \langle \hat{p}_{A} \hat{p}_{B} \rangle +  \langle \hat{p}_{B} \hat{p}_{A} \rangle - 2 \langle \hat{p}_{A} \rangle \langle \hat{p}_{B} \rangle \right) = \langle \hat{p}_{A} \hat{p}_{B} \rangle \notag \\
& = - \braket{\Psi | \hat{a}^{\dagger}_{A} \hat{a}^{\dagger}_{B} | \Psi} + \underbrace{ \braket{\Psi | \hat{a}^{\dagger}_{A} \hat{a}_{B} | \Psi} }_{=0} + \underbrace{\braket{\Psi | \hat{a}_{A} \hat{a}^{\dagger}_{B} | \Psi}}_{=0} - \braket{\Psi | \hat{a}_{A} \hat{a}_{B} | \Psi} \notag \\
& = - \sqrt{V^{2}-1} .
\end{align}

\subsection{$\Sigma_{14}$, $\Sigma_{41}$, $\Sigma_{32}$, $\Sigma_{23}$}

\begin{align}
\Sigma_{14} = \Sigma_{41} & = \frac{1}{2} \left( \langle \hat{q}_{A} \hat{p}_{B} \rangle +  \langle \hat{p}_{B} \hat{q}_{A} \rangle - 2 \langle \hat{q}_{A} \rangle \langle \hat{p}_{B} \rangle \right) = \langle \hat{q}_{A} \hat{p}_{B} \rangle
\\
& = i \left( \underbrace{ \braket{\Psi | \hat{a}^{\dagger}_{A} \hat{a}^{\dagger}_{B} | \Psi} }_{=\sqrt{V^{2}-1}/2} - \underbrace{ \braket{\Psi | \hat{a}^{\dagger}_{A} \hat{a}_{B} | \Psi} }_{=0} + \underbrace{\braket{\Psi | \hat{a}_{A} \hat{a}^{\dagger}_{B} | \Psi}}_{=0} - \underbrace{ \braket{\Psi | \hat{a}_{A} \hat{a}_{B} | \Psi} }_{=\sqrt{V^{2}-1}/2}  \right) = 0
\end{align}
and similarly $\Sigma_{32}=\Sigma_{23}=0$.

\subsection{Final $\Sigma$}

Collecting all the components computed in the previous sections we obtain the final expression for the covariance matrix of a TMSVS:
\begin{align}
\Sigma_{AB}=
\begin{pmatrix}
V \mathbb{1}_{2} & \sqrt{V^{2}-1} \sigma_{z} \\ 
\sqrt{V^{2}-1} \sigma_{z} & V \mathbb{1}_{2} 
\end{pmatrix} .
\end{align}

\section{Channel Transmission}

In this section we will investigate how transmission of one TMSVS-mode through a lossy channel will affect the covariance matrix. First we rewrite the covariance matrix of a TMSVS:

\begin{align}
\Sigma_{AB}=
\begin{pmatrix}
V \mathbb{1}_{2} & \sqrt{V^{2}-1} \sigma_{z} \\ 
\sqrt{V^{2}-1} \sigma_{z} & V \mathbb{1}_{2}
\end{pmatrix} \eqcolon
\begin{pmatrix}
v & w  \\ 
w & v
\end{pmatrix} 
\end{align}
with $v=V \mathbb{1}_{2}$ and $w=\sqrt{V^{2}-1} \sigma_{z}$. The channel loss can be modelled with a beamsplitter of transmittance $T$. In order to take the vacuum input of the beamsplitter into account we add another mode to the covariance matrix. The additional vacuum mode is represented by a $2 \times 2$ unit matrix and we integrate it to our total state using direct summation:

\begin{align} \label{eq_Sigma_tot}
\Sigma_{\text{tot}}=\Sigma_{AB} \oplus \underbrace{\Sigma_{\text{vac}}}_{=\mathbb{1}_{2}} = 
\begin{pmatrix}
V \mathbb{1}_{2} & \sqrt{V^{2}-1} \sigma_{z} & 0 \\ 
\sqrt{V^{2}-1} \sigma_{z} & V \mathbb{1}_{2} & 0 \\
0 & 0 & \mathbb{1}_{2}
\end{pmatrix} =
\begin{pmatrix}
v & w & 0 \\ 
w & v & 0 \\
0 & 0 & \mathbb{1}_{2}
\end{pmatrix} .
\end{align}
The beamsplitter matrix in symplectic expression is

\begin{align}
\text{BS}=
\begin{pmatrix}
\sqrt{T} \mathbb{1}_{2} & \sqrt{1-T} \mathbb{1}_{2} \\
-\sqrt{1-T} \mathbb{1}_{2} & \sqrt{T} \mathbb{1}_{2}
\end{pmatrix} \eqcolon
\begin{pmatrix}
t & r  \\
-r & t
\end{pmatrix}
\end{align}
with $t=\sqrt{T}\mathbb{1}_{2}$, $r=\sqrt{1-T}\mathbb{1}_{2}$. In order for this matrix to match with the dimensions of our covariance matrix \eqref{eq_Sigma_tot}, we add a $2 \times 2$ unitary representing the action of the BS on Alice's mode:

\begin{align}
\text{BS}_{\text{tot}} = \mathbb{1}_{A} \oplus
\begin{pmatrix}
t & r  \\
-r & t
\end{pmatrix}_{B,\text{vac}} = 
\begin{pmatrix}
\mathbb{1}_{2} & 0 & 0 \\
0 & t & r  \\
0 & -r & t
\end{pmatrix}
\end{align}
In this expression the beamsplitter act on Bob's mode and the vacuum input but will leave Alice's mode unchanged. A Gaussian state, represented by a covariance matrix $\Sigma$, transforms by action of a symplectic operator $S$ according to

\begin{align}
\Sigma'=S\Sigma S^{T} ,
\end{align}
where the prime indicates the state \emph{after} channel transmission of Bob's mode. Therefore we obtain

\begin{align}
\Sigma_{\text{tot}}' & = \text{BS}_{\text{tot}}\Sigma_{\text{tot}} \text{BS}_{\text{tot}}^{T} \notag \\
&=
\begin{pmatrix}
\mathbb{1}_{2} & 0 & 0 \\
0 & t & r  \\
0 & -r & t
\end{pmatrix}
\begin{pmatrix}
v & w & 0 \\ 
w & v & 0 \\
0 & 0 & \mathbb{1}_{2}
\end{pmatrix}
\begin{pmatrix}
\mathbb{1}_{2} & 0 & 0 \\
0 & t & -r  \\
0 & r & t
\end{pmatrix} =
\begin{pmatrix}
v  & tw & -rw \\ 
tw & t^{2}v+r^{2} & -trv+tr \\
-rw & -trv+tr & r^{2}v+t^{2}
\end{pmatrix} ,
\end{align}
where Alice's and Bob's substate reads

\begin{align}
\Sigma'_{AB}=
\begin{pmatrix}
v & tw \\
tw & t^{2}v+r^{2}
\end{pmatrix} =
\begin{pmatrix}
V \mathbb{1}_{2} &  \sqrt{T}\sqrt{V^{2}-1} \sigma_{z} \\
 \sqrt{T}\sqrt{V^{2}-1} \sigma_{z} & (TV+1-T) \mathbb{1}_{2} 
\end{pmatrix} . 
\end{align}

\section{Channel Transmission including Excess Noise} \label{sec_covmatchannelnoise}

In order to model the insertion of excess noise, we proceed in a similar fashion as in the previous section. But this time the vacuum input of the channel-transmission beamsplitter is exchanged by a noisy input:

\begin{align}
\Sigma_{\text{tot}}=\Sigma_{AB} \oplus \underbrace{\Sigma_{\text{noise}}}_{\eqcolon N \mathbb{1}_{2}} = 
\begin{pmatrix}
V \mathbb{1}_{2} & \sqrt{V^{2}-1} \sigma_{z} & 0 \\ 
\sqrt{V^{2}-1} \sigma_{z} & V \mathbb{1}_{2} & 0 \\
0 & 0 & N \mathbb{1}_{2}
\end{pmatrix} \eqcolon
\begin{pmatrix}
v & w & 0 \\ 
w & v & 0 \\
0 & 0 & n
\end{pmatrix} ,
\end{align}
where we assume the quadratures of the noise input $\Sigma_{\text{noise}}$ to carry the same variance $N$. Action of the beamsplitter on the total covariance matrix (TMSVS plus noise input) yields

\begin{align}
\Sigma_{\text{tot}}' & = \text{BS}_{\text{tot}}\Sigma_{\text{tot}} \text{BS}_{\text{tot}}^{T} \notag \\
&=
\begin{pmatrix}
\mathbb{1}_{2} & 0 & 0 \\
0 & t & r  \\
0 & -r & t
\end{pmatrix}
\begin{pmatrix}
v & w & 0 \\ 
w & v & 0 \\
0 & 0 & n
\end{pmatrix}
\begin{pmatrix}
\mathbb{1}_{2} & 0 & 0 \\
0 & t & -r  \\
0 & r & t
\end{pmatrix} =
\begin{pmatrix}
v  & tw & -rw \\ 
tw & t^{2}v+r^{2}n & -trv+trn \\
-rw & -trv+trn & r^{2}v+t^{2}n
\end{pmatrix} .
\end{align}
So the variance of the quadratures in Bob's mode reads

\begin{align}
t^{2}v+r^{2}n = (TV+[1-T]N)\mathbb{1}_{2} .
\end{align}
If we define the noise input $N$ as follows:

\begin{align}
N=1+\frac{\xi}{1-T},
\end{align}
we obtain

\begin{align}
TV+(1-T)N = T(V-1) + 1 + \xi ,
\end{align}
which yields the covariance matrix in its final form:

\begin{align}
\Sigma'_{AB}=
\begin{pmatrix}
V \mathbb{1}_{2} &  \sqrt{T}\sqrt{V^{2}-1} \sigma_{z} \\
 \sqrt{T}\sqrt{V^{2}-1} \sigma_{z} & (T[V-1]+1+\xi) \mathbb{1}_{2}
\end{pmatrix} .
\end{align}

\section{Bob Heterodyning} \label{sec_bobhet}

In order for Bob to measure both quadrature components simultaneously, he needs to insert a balanced beamsplitter to split the signal in half. The second input of the beamsplitter adds an additional vacuum mode to the system:

\begin{align}
\Sigma'_{\text{tot}}=\Sigma'_{AB} \oplus \Sigma_{\text{vac}} = 
\begin{pmatrix}
V \mathbb{1}_{2} & \sqrt{T}\sqrt{V^{2}-1} \sigma_{z} & 0 \\ 
\sqrt{T}\sqrt{V^{2}-1} \sigma_{z} & (T[V-1]+1+\xi) \mathbb{1}_{2} & 0 \\
0 & 0 & \mathbb{1}_{2}
\end{pmatrix} .
\end{align}
A balanced BS acts on Bob's mode and the vacuum mode, not on Alice; therefore it is described by

\begin{align}
\text{BS}_{\text{tot}} = \mathbb{1}_{A} \oplus \frac{1}{\sqrt{2}}
\begin{pmatrix}
\mathbb{1}_{2} & \mathbb{1}_{2}  \\
-\mathbb{1}_{2} & \mathbb{1}_{2}
\end{pmatrix}_{B,\text{vac}} = 
\begin{pmatrix}
\mathbb{1}_{2} & 0 & 0 \\
0 & \mathbb{1}_{2} /\sqrt{2} & \mathbb{1}_{2} /\sqrt{2}  \\
0 & -\mathbb{1}_{2} /\sqrt{2} & \mathbb{1}_{2} /\sqrt{2}
\end{pmatrix} .
\end{align}
The action of the BS operator on the total state, including Bob's additional vacuum mode, yields

\begin{tiny}
\begin{align}
\Sigma''_{AB} & = \text{BS}_{\text{tot}}\Sigma'_{\text{tot}}\text{BS}_{\text{tot}}^{T} \notag \\
& = \bordermatrix{            & \hat{q}_{A} & \hat{p}_{A} & \hat{q}_{B_{1}} & \hat{p}_{B_{1}} & \hat{q}_{B_{2}} & \hat{p}_{B_{2}} \cr
                \hat{q}_{A}     & V &  0  & \sqrt{\frac{T}{2}(V^{2}-1)} & 0 & -\sqrt{\frac{T}{2}(V^{2}-1)} & 0 \cr
                \hat{p}_{A}     & 0  &  V & 0 & -\sqrt{\frac{T}{2}(V^{2}-1)} & 0 & \sqrt{\frac{T}{2}(V^{2}-1)} \cr
                \hat{q}_{B_{1}} & \sqrt{\frac{T}{2}(V^{2}-1)} & 0 & \frac{T}{2}(V-1)+1+\frac{\xi}{2} & 0 & -\frac{1}{2}(T(V-1)+\xi) & 0 \cr
                \hat{p}_{B_{1}} & 0  &   -\sqrt{\frac{T}{2}(V^{2}-1)}  & 0 & \frac{T}{2}(V-1)+1+\frac{\xi}{2} & 0 & -\frac{1}{2}(T(V-1)+\xi) \cr
                \hat{q}_{B_{2}} & -\sqrt{\frac{T}{2}(V^{2}-1)}  &   0  & -\frac{1}{2}(T(V-1)+\xi) & 0 & \frac{T}{2}(V-1)+1+\frac{\xi}{2} & 0 \cr
                \hat{p}_{B_{2}} & 0  &  \sqrt{\frac{T}{2}(V^{2}-1)}  & 0 & -\frac{1}{2}(T(V-1)+\xi) &0 & \frac{T}{2}(V-1)+1+\frac{\xi}{2}} \notag \\
& = 
\bordermatrix{            & \hat{q}_{A} & \hat{p}_{A} & \hat{q}_{B_{1}} & \hat{p}_{B_{1}} & \hat{q}_{B_{2}} & \hat{p}_{B_{2}} \cr
                \hat{q}_{A}     & V_{\text{mod}}+1 &  0  & \sqrt{\frac{T}{2}(V_{\text{mod}}^{2}+2V_{\text{mod}})} & 0 & -\sqrt{\frac{T}{2}(VV_{\text{mod}}^{2}+2V_{\text{mod}})} & 0 \cr
                \hat{p}_{A}     & 0  &  V_{\text{mod}}+1 & 0 & -\sqrt{\frac{T}{2}(V_{\text{mod}}^{2}-2)} & 0 & \sqrt{\frac{T}{2}(V_{\text{mod}}^{2}+2V_{\text{mod}})} \cr
                \hat{q}_{B_{1}} & \sqrt{\frac{T}{2}(V_{\text{mod}}^{2}+2V_{\text{mod}})} & 0 & \frac{T}{2}V_{\text{mod}}+1+\frac{\xi}{2} & 0 & -\frac{1}{2}(TV_{\text{mod}}+\xi) & 0 \cr
                \hat{p}_{B_{1}} & 0  &   -\sqrt{\frac{T}{2}(V_{\text{mod}}^{2}+2V_{\text{mod}})}  & 0 & \frac{T}{2}V_{\text{mod}}+1+\frac{\xi}{2} & 0 & -\frac{1}{2}(TV_{\text{mod}}+\xi) \cr
                \hat{q}_{B_{2}} & -\sqrt{\frac{T}{2}(V_{\text{mod}}^{2}+2V_{\text{mod}})}  &   0  & -\frac{1}{2}(TV_{\text{mod}}+\xi) & 0 & \frac{T}{2}V_{\text{mod}}+1+\frac{\xi}{2} & 0 \cr
                \hat{p}_{B_{2}} & 0  &  \sqrt{\frac{T}{2}(V_{\text{mod}}^{2}+2V_{\text{mod}})}  & 0 & -\frac{1}{2}(TV_{\text{mod}}+\xi) &0 & \frac{T}{2}V_{\text{mod}}+1+\frac{\xi}{2}} \notag \\
& = 
\bordermatrix{            & A & B_{1} & B_{2} \cr
                A         & (V_{\text{mod}}+1) \mathbb{1}_{2} & \sqrt{\frac{T}{2}(V_{\text{mod}}^{2}+2V_{\text{mod}})} \sigma_{z} & -\sqrt{\frac{T}{2}(V_{\text{mod}}^{2}+2V_{\text{mod}})}\sigma_{z} \cr
                B_{1}     & \sqrt{\frac{T}{2}(V_{\text{mod}}^{2}+2V_{\text{mod}})} \sigma_{z}  &  (\frac{T}{2}V_{\text{mod}}+1+\frac{\xi}{2})\mathbb{1}_{2} & -\frac{1}{2}(TV_{\text{mod}}+\xi)\mathbb{1}_{2} \cr
                B_{2} & -\sqrt{\frac{T}{2}(V_{\text{mod}}^{2}+2V_{\text{mod}})}\sigma_{z} & -\frac{1}{2}(TV_{\text{mod}}+\xi)\mathbb{1}_{2} & (\frac{T}{2}V_{\text{mod}}+1+\frac{\xi}{2})\mathbb{1}_{2} } .
\end{align}
\end{tiny}

\chapter{Quantum Homodyne Detection} \label{ch_homodynedet}

This chapter describes the quantum theory of homodyne detection. The derivation is leaning towards \cite{leonhardt1997measuring}. Consider a quantum coherent state with quadrature operators $\hat{q}$ and $\hat{p}$. In shot-noise units its annihilation operator is given by

\begin{equation} \label{eq_a}
\hat{a}=\frac{1}{2} \left( \hat{q} + i \hat{p} \right) .
\end{equation}
Furthermore, consider a classical optical field, represented by

\begin{equation} \label{eq_alpha}
\alpha_{\text{LO}} = q + ip = |\alpha_{\text{LO}}| \ e^{i\theta} , 
\end{equation}
where the subscript LO stands for local oscillator. Let $\hat{a}$ and $\alpha_{\text{LO}}$ each enter a balanced beamsplitter, represented by

\begin{equation}
\text{BS} = \frac{1}{\sqrt{2}}
\begin{pmatrix}
1 &  1 \\
1 & -1
\end{pmatrix} .
\end{equation}
The action of the beamsplitter on the quantum field $\hat{a}$ and the classical field $\alpha_{\text{LO}}$ yields

\begin{equation}
\text{BS} \binom{\hat{a}}{\alpha_{\text{LO}}} = \binom{ \frac{1}{\sqrt{2}} (\hat{a} + \alpha_{\text{LO}}) }{ \frac{1}{\sqrt{2}} (\hat{a} - \alpha_{\text{LO}}) } \eqcolon \binom{\hat{a}_{1}}{\hat{a}_{2}} ,
\end{equation}
where $\hat{a}_{1}$ and $\hat{a}_{2}$ represent the annihilation operators at the respective output ports of the beamsplitter. The photon-number operators at each output of the BS are easily evaluated by

\begin{subequations} \label{homodyningn1n2}
\begin{align} 
\hat{n}_{1} & =\hat{a}_{1}^{\dagger}\hat{a}_{1}= \frac{1}{2} (\hat{a}^{\dagger} + \alpha_{\text{LO}}^{*})(\hat{a} + \alpha_{\text{LO}})=\frac{1}{2} ( \hat{a}^{\dagger}\hat{a} + \alpha_{\text{LO}}^{*}\alpha_{\text{LO}} + \alpha_{\text{LO}}\hat{a}^{\dagger} + \alpha_{\text{LO}}^{*}\hat{a} ), \\
\hat{n}_{2} & =\hat{a}_{2}^{\dagger}\hat{a}_{2}= \frac{1}{2} (\hat{a}^{\dagger} - \alpha_{\text{LO}}^{*})(\hat{a} - \alpha_{\text{LO}})=\frac{1}{2} ( \hat{a}^{\dagger}\hat{a} + \alpha_{\text{LO}}^{*}\alpha_{\text{LO}} - \alpha_{\text{LO}}\hat{a}^{\dagger} - \alpha_{\text{LO}}^{*}\hat{a} ).
\end{align}
\end{subequations}
Subtraction of the two yields the number-difference operator:

\begin{equation}
\Delta \hat{n} = \hat{n}_{1} - \hat{n}_{2} = \alpha_{\text{LO}}\hat{a}^{\dagger} + \alpha_{\text{LO}}^{*}\hat{a} .
\end{equation}
Substituting \eqref{eq_a} and \eqref{eq_alpha} we arrive at

\begin{align} \label{homodynedetectionlaw}
\Delta \hat{n} & = |\alpha_{\text{LO}}| \ (\hat{a}^{\dagger} e^{i\theta} + \hat{a} e^{-i\theta} ) = \notag \\
& = |\alpha_{\text{LO}}| \ \frac{1}{2} ( [ \hat{q} - i \hat{p} ] e^{i\theta} + [ \hat{q} + i \hat{p} ] e^{-i\theta} ) = \notag \\
& = |\alpha_{\text{LO}}| \ \frac{1}{2} ( \hat{q} \underbrace{[ e^{i\theta} + e^{-i\theta} ]}_{2\cos\theta} + i \hat{p} \underbrace{[-e^{i\theta} + e^{-i\theta} ]}_{-2 i \sin \theta } ) = \notag \\
& = |\alpha_{\text{LO}}| \ ( \hat{q} \cos \theta + \hat{p} \sin \theta ) .
\end{align}
So depending on the local-oscillator phase $\theta$, the number-difference operator is either proportional to the $\hat{q}$ or $\hat{p}$ quadrature operator:

\begin{subequations}
\begin{align}
\Delta \hat{n} & = |\alpha_{\text{LO}}| \ \hat{q}  \qquad \text{for } \theta = 0, \\
\Delta \hat{n} & = |\alpha_{\text{LO}}| \ \hat{p}  \qquad \text{for } \theta = \frac{\pi}{2} .
\end{align}
\end{subequations}

\chapter{Principles of a $q/p$-Modulator} \label{ch_modulator}

\begin{figure}
\centering
\includegraphics[width=0.8\linewidth]{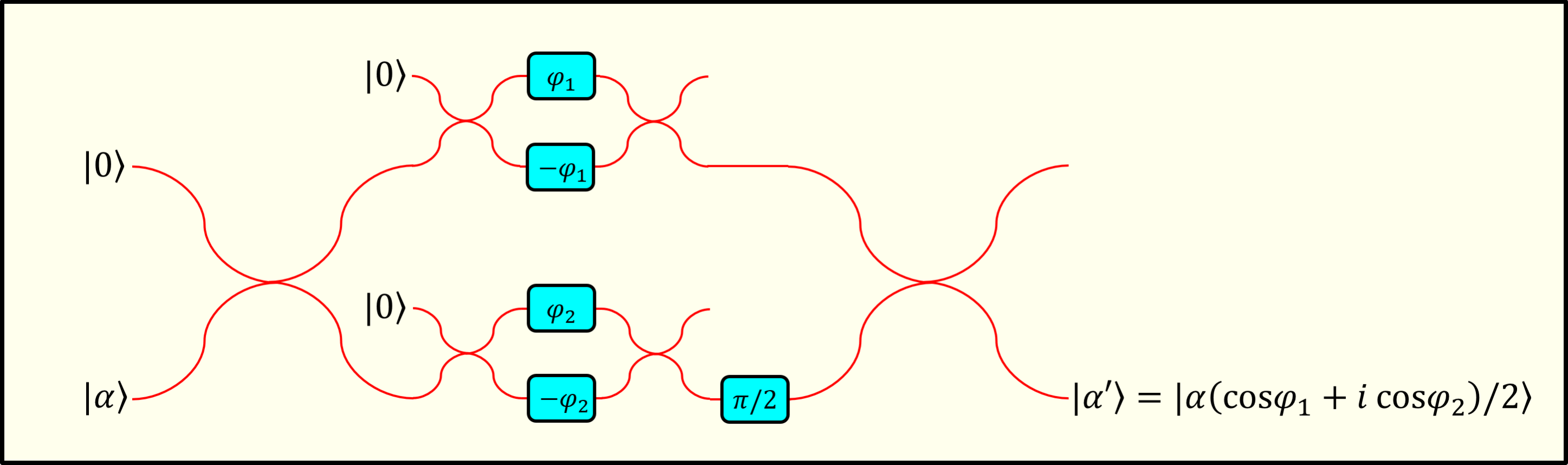}
\caption{Possible schematic of a $q/p$-modulator.}
\label{fig_modulator}
\end{figure}

In our model, sketched in Figure~\ref{fig_modulator}, we assume our modulator to consist of a double Mach-Zehnder interferometer (MZI), hence an MZI with another nested MZI in each arm. The working principle of the modulator can be described by the following steps:

\begin{itemize}
\item
A coherent state $\alpha$ and a vacuum enter the ports of a balanced beamsplitter (BS). The output states are
\begin{equation}
\frac{1}{\sqrt{2}}
\begin{pmatrix}
1 & 1 \\
1 & -1
\end{pmatrix}
\binom{\alpha}{0}
= \frac{1}{\sqrt{2}} \binom{\alpha}{\alpha} ,
\end{equation}
and each output corresponds to one arm of the outer MZI.
\item
Each arm of the MZI is again mixed with the vacuum by another balanced BS yielding the state

\begin{equation}
\left(
\frac{1}{\sqrt{2}}
\begin{pmatrix}
1 & 1 \\
1 & -1
\end{pmatrix}
\otimes \frac{1}{\sqrt{2}}
\begin{pmatrix}
1 & 1 \\
1 & -1
\end{pmatrix}
\right)
\left(
\frac{1}{\sqrt{2}} \binom{\alpha}{0} \otimes \frac{1}{\sqrt{2}} \binom{\alpha}{0}
\right)
= \frac{1}{2}
\begin{pmatrix}
\alpha \\
\alpha \\
\alpha \\
\alpha
\end{pmatrix} ,
\end{equation}
where the upper two components correspond two the respective arms of the first inner MZI and the lower two to the second inner MZI.

\item
Now a voltage is applied to the nonlinear waveguides in the four arms of the inner MZI's. This voltage, by the electro-optic effect, causes a drift of the waveguide's refractive index, therefore changing the phase to rotate proportionally to the applied voltage. We apply a phase rotation of $+\varphi_{1}$ and $-\varphi_{1}$ respectively to the two arms of the first inner MZI and a phase rotation of $+\varphi_{2}$ and $-\varphi_{2}$ to the arms of the second one:

\begin{equation}
\frac{1}{2}
\begin{pmatrix}
\alpha e^{i\varphi_{1}} \\
\alpha e^{-i\varphi_{1}} \\
\alpha e^{i\varphi_{2}} \\
\alpha e^{-i\varphi_{2}}
\end{pmatrix} .
\end{equation}

\item
The arms of the inner two MZI's are now combined again using balanced beamsplitters, yielding

\begin{equation}
\frac{1}{\sqrt{8}}
\begin{pmatrix}
\alpha (e^{i\varphi_{1}}+e^{-i\varphi_{1}}) \\
\alpha (e^{i\varphi_{1}}-e^{-i\varphi_{1}}) \\
\alpha (e^{i\varphi_{2}}+e^{-i\varphi_{2}}) \\
\alpha (e^{i\varphi_{2}}-e^{-i\varphi_{2}})
\end{pmatrix} 
=
\frac{2}{\sqrt{8}}
\begin{pmatrix}
\alpha \cos \varphi_{1} \\
\alpha i \sin \varphi_{1} \\
\alpha \cos \varphi_{2} \\
\alpha i \sin \varphi_{2}
\end{pmatrix} .
\end{equation}
We use only one output port of each beamsplitter, omitting the second and the fourth component of the above state vector. So we are left with

\begin{equation}
\frac{1}{\sqrt{2}}
\begin{pmatrix}
\alpha \cos \varphi_{1} \\
\alpha \cos \varphi_{2} \\
\end{pmatrix} .
\end{equation}

\item
Now, as the arms of the inner MZI's are recombined, we add a phase of $\pi /2$ to the second arm of the outer MZI:

\begin{equation}
\frac{1}{\sqrt{2}}
\begin{pmatrix}
\alpha \cos \varphi_{1} \\
\alpha e^{i\frac{\pi}{ 2}} \cos \varphi_{2} \\
\end{pmatrix} 
=
\frac{1}{\sqrt{2}}
\begin{pmatrix}
\alpha \cos \varphi_{1} \\
\alpha i \cos \varphi_{2} \\
\end{pmatrix} 
.
\end{equation}

\item
Finally we recombine the two arms of the outer MZI and obtain

\begin{equation}
\frac{1}{2}
\begin{pmatrix}
\alpha ( \cos \varphi_{1} + i \cos \varphi_{2} ) \\
\alpha ( \cos \varphi_{1} - i \cos \varphi_{2} )
\end{pmatrix} .
\end{equation}
We omit one of the two BS outputs and are left with the final state

\begin{equation}
\alpha' = \frac{1}{2} \alpha ( \cos \varphi_{1} + i \cos \varphi_{2} ) .
\end{equation}

\end{itemize}

\chapter{SI-, Natural- and Shot-Noise Units} \label{ch_units}

Coherent states can entirely be described in terms of annihilation and creation operators which fulfil the well-known relations listed in Appendix~\ref{ch_coherent}. However, depending on notational or computational convenience, one is free to define in which way the quadrature operators $\hat{q}$ and $\hat{p}$ are given with respect to the ladder operators. Depending on the definition of choice, one obtains according expressions for the commutator $[\hat{q},\hat{p}]$, the photon number operator $\hat{n}$ and minimal uncertainty product $\delta\hat{q}\delta\hat{p}$. A direct comparison of shot-noise units \cite{weedbrook2012gaussian} (SNU, used in this paper), natural units \cite{leonhardt1997measuring} (NU) and SI-units \cite{schleich2011quantum} can be found in Table \ref{tab_units}.

\begin{table} [b!]
\centering
\begin{tabular}{|>{\centering\arraybackslash} m{1in}|>{\centering\arraybackslash} m{1.6in}|>{\centering\arraybackslash} m{1.6in}|>{\centering\arraybackslash} m{1.6in}|}
\hline 
  & \textbf{SNU} & \textbf{NU} & \textbf{SI} \\ 
\hline \hline
$\hat{a}=$ & $\frac{1}{2}(\hat{q}+i\hat{p} )$ & $\frac{1}{\sqrt{2}}(\hat{q}+i\hat{p} )$ & $\frac{1}{\sqrt{2\hbar\omega}}(\omega\hat{q}+i\hat{p} )$ \\ 
\hline 
$\hat{a}^{\dagger}=$ & $\frac{1}{2}(\hat{q}-i\hat{p} )$ & $\frac{1}{\sqrt{2}}(\hat{q}-i\hat{p} )$ & $\frac{1}{\sqrt{2\hbar\omega}}(\omega\hat{q}-i\hat{p} )$ \\ 
\hline 
$\hat{q}=$ & $\hat{a}+\hat{a}^{\dagger}$ & $\frac{1}{\sqrt{2}} ( \hat{a}+\hat{a}^{\dagger} )$ & $\sqrt{\frac{\hbar}{2\omega}} ( \hat{a}+\hat{a}^{\dagger} )$ \\ 
& $\left( = \sqrt{2} \ \hat{q}_{\text{NU}} = \sqrt{\frac{2\omega}{\hbar}} \hat{q}_{\text{SI}} \right)$ & $\left( = \frac{1}{\sqrt{2}} \hat{q}_{\text{SNU}} = \sqrt{\frac{\omega}{\hbar}} \hat{q}_{\text{SI}} \right)$ & $\left( = \sqrt{\frac{\hbar}{2\omega}} \hat{q}_{\text{SNU}} = \sqrt{\frac{\hbar}{\omega}} \hat{q}_{\text{NU}} \right)$\\
\hline 
$\hat{p}=$ & $-i(\hat{a}-\hat{a}^{\dagger})$ & $\frac{-i}{\sqrt{2}} ( \hat{a}-\hat{a}^{\dagger} )$ & $-i \sqrt{\frac{\hbar \omega}{2}} ( \hat{a}-\hat{a}^{\dagger} )$ \\ 
& $ \left( = \sqrt{2} \ \hat{p}_{\text{NU}} = \sqrt{\frac{2}{\hbar\omega}} \hat{p}_{\text{SI}} \right)$ & $ \left( = \frac{1}{\sqrt{2}} \hat{p}_{\text{SNU}} = \frac{1}{\sqrt{\hbar\omega}} \hat{p}_{\text{SI}} \right)$ & $ \left( = \sqrt{\frac{\hbar\omega}{2}} \hat{p}_{\text{SNU}} = \sqrt{\hbar\omega} \hat{p}_{\text{NU}} \right)$ \\
\hline
$[\hat{q},\hat{p}] =$ & $2i$ & $i$ & $i\hbar$ \\ 
\hline 
$\hat{n}=$ & $\frac{1}{4}(\hat{q}^{2}+\hat{p}^{2}) - \frac{1}{2}$ & $\frac{1}{2}(\hat{q}^{2}+\hat{p}^{2})- \frac{1}{2}$ & $\frac{1}{2\hbar\omega}(\omega^{2}\hat{q}^{2}+\hat{p}^{2}) - \frac{1}{2}$ \\ 
\hline 
$\delta\hat{q}\delta\hat{p}\geq$ & $1$ & $\frac{1}{2}$ & $\frac{\hbar}{2}$ \\ 
\hline 
\end{tabular}
\caption{Comparison of shot-noise units (SNU), natural units (NU) and SI-units.}
\label{tab_units}
\end{table}

\end{appendices}


%


\printbibliography
\addcontentsline{toc}{chapter}{References}

\end{document}